\documentstyle[preprint,aps,epsfig]{revtex}

\def\plb#1#2#3#4{#1, Phys. Lett. {\bf #2B}, #3 (#4)}
\def\npb#1#2#3#4{#1, Nucl. Phys. {\bf B#2}, #3 (#4)}
\def\prd#1#2#3#4{#1, Phys. Rev. {\bf D#2}, #3 (#4)}
\def\prl#1#2#3#4{#1, Phys. Rev. Lett. {\bf #2}, #3 (#4)}

\def\rep#1#2#3#4{#1, Phys. Rep. {\bf #2}, #3 (#4)}

\def\beq{\begin{equation}}
\def\eeq{\end{equation}}
\def\bea{\begin{eqnarray}}
\def\eea{\end{eqnarray}}
\def\ba{\begin{array}}
\def\ea{\end{array}}
\def\bec{\begin{center}}
\def\ec{\end{center}}
\def\nl{\nonumber\\}

\begin{document}
\draft

\preprint{KAIST-TH 2002/09}

\title{ MSSM Higgs sector CP violation
at photon colliders: Revisited }

\author{Saebyok ${\rm Bae^1}$\thanks{Former address: Department
of Physics, KAIST, Daejeon, Korea (sbae@mail.kaist.ac.kr)},
Byungchul ${\rm Chung}^2$\thanks{E-mail address:
crash@muon.kaist.ac.kr} and P. ${\rm Ko^2}$\thanks{E-mail address:
pko@muon.kaist.ac.kr}}
\address{$^1$Center for Gifted Students and
$^2$Department of Physics, \\
Korea Advanced Institute of Science and Technology,
\\ Daejeon 305-701, Korea \\
}
\maketitle

\begin{abstract}
We present a comprehensive analysis on the MSSM Higgs sector
CP violation at photon colliders including the chargino contributions
as well as the contributions of other charged particles.
The chargino loop contributions can be important for the would-be CP odd
Higgs production at photon colliders. Polarization asymmetries are
indispensable in determining the CP properties of neutral Higgs bosons.
\end{abstract}




\newpage
\narrowtext
\tighten
\section{Introduction}
\label{sec:intro}

The discovery of Higgs boson(s) at the current/future colliders is one of
the most important goals of high energy particle physics experiments.
Its (non)discovery would be crucial for testing our present understanding of
the origin of electroweak symmetry breaking (EWSB) and the subsequent
generation of masses of electroweak (EW) gauge bosons and chiral fermions
in the Standard Model (SM). This would be also true of the Minimal
Supersymmetric Standard Model (MSSM), which is the most popular
candidate for the new physics beyond the SM.

The Higgs sector in the MSSM possesses three neutral Higgs
particles: two CP-even neutral scalars ($h$ and $H$), one CP-odd
neutral scalar ($A$), and a pair of charged Higgs scalars
($H^\pm$) \cite{Gunion}. The tree-level MSSM Higgs potential does
not allow spontaneous CP violation unlike general two-Higgs
doublet model. Even if one include the one-loop corrected
effective potential for the Higgs sector, the spontaneous CP
violation \cite{Maekawa} can not be realistic, because the
resulting lightest neutral Higgs boson should be far less than the
current lower limit on the Higgs boson \cite{Pomarol}. Still,
there are many new explicitly CP violating complex parameters in
the soft supersymmetry (SUSY) breaking sector of the MSSM
Lagrangian, and some of them can have large phases (without
conflict with the electron/neutron electric dipole moment (EDM)
constraints), and thus can lead to some observable consequences in
various CP violating phenomena in $K$ and $B$ decays \cite{Ko} and
electroweak baryogenesis \cite{Carena}, etc. Especially, the
complex phases of the stop and sbottom trilinear couplings
$A_{t,b}$ and the Higgsino mass parameter $\mu$ can cause the
mixing between CP-odd and CP-even neutral Higgs bosons in the
neutral Higgs sector via loop corrections in the MSSM, namely, the
Higgs sector CP violation \cite{PiWa}.

In most phenomenological studies of the MSSM, the large SUSY CP violating
(CPV) phases were usually neglected, since they may lead to large EDMs
of electron and neutron, or $\epsilon_K$, depending on whether they are
flavor preserving (FP) or flavor changing (FC). The SUSY CPV phases are
assumed to be very small, so that the only source of CP violation would be
the single Kobayashi-Maskawa (KM) phase in the CKM mixing matrix in the
charged weak current of down type quarks. In this case, the SUSY effects
on $K$ and $B$ phenomenology are minimal in the sense that deviations from
the SM predictions are quite small. However one can consider large
FP SUSY CPV phases, since one can avoid the EDM constraints in basically
three different ways:
\begin{itemize}
\item Decoupling (Effective SUSY Model): The 1st/2nd generation
sfermions are heavy (and degenerate to some extent) enough, so
that the SUSY CP and $\epsilon_K$ problems are evaded. Only third
generation sfermions and gauginos have to be lighter than $O(1)$
TeV in order that one solves the gauge hierarchy problem by SUSY
\cite{Dimopoulos}. In this case, the SUSY CPV phases need not be
zero, and they can lead to substantial deviations from the SM
cases, especially for the third generation. In this scenario, $B$
factories may be able to probe the SUSY CPV phases from direct
asymmetry in $B \rightarrow X_s \gamma$ and the lepton
forward-backward asymmetry in $B \rightarrow X_s l^+ l^-$, etc.

\item Cancellation: Various contributions to electron/neutron EDMs may
cancel one another, leading to the net results which are consistent with
experimental lower bounds \cite{Ibrahim}. In this case, many of the SUSY
CPV phases can be $O(1)$ as in the decoupling scenario. However this scenario
is tightly constrained when the data on the mercury ($^{199}$Hg) atom EDM
is included \cite{mercury}.

\item Non-universal Scenario:
$|A_e|, |A_{u,c}|, |A_{d,s}| \lesssim 10^{-3} |\mu|$ to evade $e/n$ EDM's,
but $A_t , A_b , A_{\tau}$ can have large CP violating phases \cite{Chang}.
However there is a strong two-loop Barr-Zee type constraint for large
$\tan\beta$. Therefore large CPV phases can be allowed in this scenario
and decoupling scenario only for $\tan \beta \lesssim 20-30$.
\end{itemize}

The reliable determinations of the neutral Higgs sector CP-violation in
the MSSM can be achieved by observing the CP-properties of all the three
neutral Higgs particles directly.
Higgs bosons can be produced in $\gamma \gamma$ collisions via one-loop
diagrams in which all the possible charged particles participate. The
$s$-channel resonance productions of neutral Higgs bosons in $\gamma \gamma$
collisions have been considered as crucial tools of studying
the CP properties of Higgs particles \cite{{GuHa},{Grza}}. Because the
polarizations of the colliding photons can strongly govern both the
$\gamma \gamma$ luminosity spectrum and the cross sections, obtaining the
highly polarizated photon beams is important to Higgs boson detections.
This is possible by Compton backscattering of laser photons off the linear
collider electron and positron beams which can produce high luminosity
$\gamma \gamma$ collisions with a wide spectrum of $\gamma \gamma$ center of
mass energy \cite{Exp_pp}.

In particular, one can observe CP violating effects through the $s$-channel
resonance for CP-odd neutral Higgs particle production in the linear collider.
Due to the mixing effect between the CP-odd and CP-even neutral Higgs bosons,
there are the additional loop contributions of charged scalars and vectors
to the would-be CP-odd neutral Higgs $H_2$ production in $\gamma \gamma$
collision, resulting in the enhanced production cross section.
In Ref.~\cite{SYChoi}, the CP violation of the neutral Higgs sector at a
photon collider was studied using the $s$-channel resonance production
cross sections and the polarization asymmetries of Higgs particles for
$3 \le \tan \beta \le 10$. In the loop diagrams relevant to $\gamma \gamma
\rightarrow$ neutral Higgs bosons, the contributions of charginos were
neglected by assuming that they were heavy enough to be decoupled from the
productions of the Higgs bosons. However, charginos are not much heavier
than the lighter stop in many SUSY breaking scenarios,
and their effects should
be included in a realistic analysis. The current lower limit on the lighter
chargino mass from LEP II experiment is only $M_{\tilde{\chi}^-_1}
> 103~ (83.6)$ GeV for $m_{\tilde{\nu}} > (<) \, 300$ GeV in the minimal
supergravity scenario \cite{Diaz}. It is even less stringent in the AMSB
scenario: $M_{\tilde{\chi}^-_1} > 45$ GeV.
Therefore we include the chargino contributions to
$\gamma\gamma\rightarrow H_{k=1,2,3}$, and investigated their effects when
other parameters are fixed as Ref.~\cite{SYChoi}.

In this work, we investigate the neutral Higgs productions at
$\gamma\gamma$ collisions, including the chargino loop contributions as well
as other charged particles in the MSSM, and study the CP properties of
the MSSM Higgs sector. This paper is organized as follows.
In Section II, we review briefly the loop-induced CP violation and the
mixing of CP-even and CP-odd Higgs bosons in the neutral Higgs sector of the
MSSM. In Section III, we derive the cross sections for the Higgs
productions in $\gamma\gamma$ collisions and the polarization asymmetries
in terms of two form factors appearing in the
$\gamma\gamma \rightarrow H_{k=1,2,3}$ amplitudes.
In Section IV, we present detailed numerical analyses and discuss the
potential importance of chargino loop contributions to the CP violation in
$\gamma\gamma \rightarrow H_k$. The formulae for the chargino and stop mass
matrices, their eigenvalues and the corresponding mixing matrices are given
in Appendix A. The interaction Lagrangians relevant to $\gamma\gamma
\rightarrow H_k$ are recapitulated for both convenience and completeness.

\section{The Neutral Higgs Sector in the MSSM}

The MSSM Lagrangian possesses many new CP violating phases in the
soft SUSY breaking terms in addition to the KM phase in the CKM
matrix element. Using Peccei-Quinn $U(1)_{\rm PQ}$ and $U(1)_{\rm
R}$ symmetries, we can redefine some parameters to be real. We
will work in the basis where $B\mu$ and the wino mass parameter
${M}_2$ are real.  In the MSSM, the Higgs potential is
CP-conserving at the tree level and only the soft terms (and the
usual CKM mixing matrix) can have CP violating phases. However,
CPV phases in soft terms can induce CP violation in the effective
potential of Higgs bosons through quantum corrections involving
squarks and other SUSY particles in the loop. The effective
potential of the Higgs fields at the one-loop level\footnote{We
follow the notations of the recent third paper of Ref.
\cite{PiWa}.} can be written as \bea \label{VHiggs} {\cal V}_ {\rm
Higgs}^{eff} &=& \mu_{1}^2 \Phi_{1}^{\dagger}\Phi_{1}+\mu_{2}^2
\Phi_{2}^{\dagger}\Phi_{2} +(m_{12}^2 \Phi_{1}^{\dagger} \Phi_{2}
+ {\rm h.c.}) \nonumber
\\
&&+
\lambda_{1}(\Phi_{1}^{\dagger}\Phi_{1})^2+\lambda_{2}(\Phi_{2}^{\dagger}
\Phi_{2})^2
+\lambda_{3}(\Phi_{1}^{\dagger} \Phi_{1})(\Phi_{2}^{\dagger} \Phi_{2})
+\lambda_{4}(\Phi_{1}^{\dagger} \Phi_{2})(\Phi_{2}^{\dagger} \Phi_{1})
\nonumber
\\
&&+
\lambda_{5}(\Phi_{1}^{\dagger}\Phi_{2})^2+\lambda_{5}^* (\Phi_{2}^{\dagger}
\Phi_{1})^2
+\lambda_{6}(\Phi_{1}^{\dagger} \Phi_{1})(\Phi_{1}^{\dagger} \Phi_{2})
+\lambda_{6}^*(\Phi_{1}^{\dagger} \Phi_{1})(\Phi_{2}^{\dagger} \Phi_{1})
\nonumber
\\
&&+
\lambda_{7}(\Phi_{2}^{\dagger} \Phi_{2})(\Phi_{1}^{\dagger} \Phi_{2})
+\lambda_{7}^*(\Phi_{2}^{\dagger} \Phi_{2})(\Phi_{2}^{\dagger} \Phi_{1}).
\eea
The fields $\Phi_i$ $(i=1,2)$ are the scalar components of the Higgs
superfields, with $\Phi_2 (\Phi_1)$ giving masses to the up-type (down-type)
fermions. In the MSSM, one has $\lambda_i =0$ ($i=5,6,7$) at tree level so
that there is not Higgs sector CP violations in the MSSM. But these couplings
are generated at one loop level and can be complex if $\mu, A_t$ possess CPV
phases. Also the Higgs bilinear couplings $m_{12}^2$ \cite{Pil} can be
complex by quantum corrections. For small $\tan \beta \sim O(1)$, where the
stop contributions are dominant over sbottom or chargino contributions to
the Higgs sector CP violations \cite{PiWa,Pil,IbNa},
one has, for example \cite{Bae},
\bea
\label{m12}
m_{12}^2
&\simeq&
B\mu + \frac{1}{16 \pi^2} h_t^2 A_t \mu \left[
\frac{m_{{\tilde t}_2}^2+m_{{\tilde t}_1}^2}
{4(m_{{\tilde t}_2}^2-m_{{\tilde t}_1}^2)}
\ln{\frac{m_{{\tilde t}_2}^2}{m_{{\tilde t}_1}^2}}
- \frac{1}{2} \right],
\eea
where the top Yukawa coupling
$h_t=\frac{\sqrt{2} m_t(\bar{m_t})}{ v \sin\beta }$, and
$m_{{\tilde t}_i}$ $(i=1, 2)$ are the masses of the lighter and heavier stops.
The contributions of the 1st and 2nd generation squarks are negligible because
of their small Yukawa couplings. The mixing of two CP-even Higgs bosons is
denoted by the real parameter $B \mu$, whereas the $h_t^2 A_t \mu$ term with
the complex $A_t$ trilinear coupling generates the mixings among all two
CP-even and one CP-odd  neutral Higgs bosons. Therefore, the quadratic term
of the Higgs fields with the coefficient $m_{12}^2$ plays an important role
in the Higgs mixing. If $h_t^2 A_t \mu$ terms are much less than $B \mu$, and
$\mu_1^2 \sim \mu_2^2 \sim B \mu$, we can expect that the
scalar-scalar mixing is much larger than the scalar-pseudoscalar mixing.
For large $\tan \beta \gtrsim 30$, the contribution of the chargino sector
can dominate those of the stop and sbottom sectors in the mixing between
the CP-even and CP-odd Higgs bosons \cite{IbNa}. The same is true of other
quartic couplings $\lambda_{5,6,7}$, whose imaginary parts vanish in the
CP conserving limit (or at tree level) in the MSSM.
One has to keep in mind that there is a strong constraint from two-loop
Barr-Zee type $e/n$ EDM constraints for large $\tan \beta$
($40 \lesssim \tan \beta \lesssim 60$).
Therefore, we will choose rather low $\tan \beta \lesssim 20$ and allow
maximal CPV phase in the $\mu$ and  $A_t$ parameters.

Since the electroweak gauge symmetry is broken spontaneously into
$U(1)_{\rm em}$, two Higgs doublets can be written as
\bea
\label{phi1}
\Phi_{1}
=
\left(
\begin{array}{c}
\phi_{1}^{+}
\\
(v_{1}+\phi_{1}+ia_{1})/\sqrt{2}
\end{array}
\right),~~~~~
\Phi_{2}
=
e^{i \xi}
\left(
\begin{array}{c}
\phi_{2}^{+}
\\
(v_{2}+\phi_{2}+ia_{2})/\sqrt{2}
\end{array}
\right), \eea where the VEVs $v_i$ are real. The relative phase
$\xi$, which is renormalization-scheme dependent,\footnote{Refer
to the third paper of Ref.~\cite{PiWa}.} is determined from the
minimum energy conditions of the Higgs potential \cite{PiWa},
i.e., the vanishing tadpole conditions $T_{\phi}={\partial{\cal
V}_{\rm Higgs}^{eff}}/{\partial \phi}=0$. It turns out $\xi$ is
very small in the $\overline{MS}$ scheme, and will be ignored in
the numerical analysis. Because the electroweak symmetry is
spontaneously broken to $U(1)_{\rm em}$, three Goldstone bosons
are eaten by $W^{\pm}, Z^0$ gauge bosons, and one ends up with two
charged Higgs and three neutral Higgs bosons. The $3 \times 3$
$({\rm mass})^2$ mass matrix ${\cal M}_N^2$ for three neutral
Higgs bosons is a real symmetric matrix, and is diagonalized by a
$3 \times 3$ orthogonal matrix $O$: \bea \label{Oij} O^{T} {\cal
M}^2_{N} O= \mbox{diag} (M_{H_{1}}^2, \, M_{H_{2}}^2, \,
M_{H_{3}}^2 ), \eea where $M_{H_{3}} \geq M_{H_{2}} \geq
M_{H_{1}}$. The corresponding mass eigenstates, $H_{i}$ $(i=1,2,3)
\,$, are defined from the weak eigenstates as \beq
\label{s-psmixing} (a, \phi_1, \phi_2)^T =O(H_1, H_2, H_3)^T. \eeq

\section{ Neutral Higgs Boson Productions at photon colliders}
\label{sec:nh}


Both within the SM and the MSSM,
the neutral Higgs decays into two gluons ($gg$) or two photons
($\gamma\gamma$) have been interesting subjects. The inverse of the former
process is a main production mechanism for the neutral Higgs bosons at hadron
colliders if the Higgs bosons have intermediate masses. The latter is an
important mode for tagging the neutral Higgs bosons at hadron colliders. Its
inverse process is the mechanism for neutral Higgs productions in the
$\gamma \gamma$ collision which can be run at next linear colliders (NLC).

The reactions $g g  \rightarrow H_k$ ($k=1,2,3$) are generated by the
(s)quark loops, and have been already discussed by two groups in the presence
of the MSSM Higgs sector CP violation \cite{ggh}. We have calculated these
processes and confirmed their results, although we do not reproduce them here.
The case for $\gamma\gamma \rightarrow H_k$ is more complicated than the
previous case ($g g \rightarrow H_k $), since one has to include all
the charged particle ($W^{\pm}$, $H^{\pm}$ and charginos) contributions
as well as the (s)quark loop contributions. It is straightforward to perform
the loop integrations. The only thing to take into account is the various
mixing components for charginos and neutral Higgs bosons. We present the
chargino mass matrix ${\cal M_C}$, its mass eigenvalues
$M_{\tilde{\chi}^-_1}$,  $M_{\tilde{\chi}^-_2}$ and two mixing matrices $U$
and $V$: $U^* {\cal M_C} V^{\dagger} = {\rm diag}( M_{\tilde{\chi}^-_1},
M_{\tilde{\chi}^-_2})$ (see Appendix A for explicit expressions).

The interaction Lagrangian between the charginos and three neutral
Higgs bosons is \beq {\cal L}( H_j \tilde{\chi}^+_k
\tilde{\chi}^-_l ) = H_j \overline{\tilde{\chi}^-_k} \left[ {\rm
Re}(\kappa^j_{kl}) + i \gamma^5 {\rm Im}(\kappa^j_{kl}) \right]
\tilde{\chi}^-_l, \eeq (with $j=1,2,3$ and $k,l=1,2$) where
\begin{equation}
\kappa^j_{kl} = - \frac{g}{\sqrt{2}} \left[ e^{+i\xi} U_{k1} V_{l2}
(O_{3, j} +i\cos\beta \, O_{1, j})
+ U_{k2} V_{l1}(O_{2, j} +i \sin\beta \, O_{1, j}) \right].
\end{equation}
In this work, it suffices to keep $\kappa_{kk}^j$ only, since we
consider the chargino loop contribution to $\gamma\gamma
\rightarrow H_i$. Note that there are two CP violating phases
($\xi$ and $\theta_\mu = $ arg($\mu$)) in the couplings
$\kappa^j_{kk}$. Also note that the $H_j -\tilde{\chi}^+_k
-\tilde{\chi}^-_k$ couplings arise from the Higgs-gaugino-Higgsino
couplings in the current basis. Thus the chargino loop effects
will be maximized if the wino-Higgsino mixing is large. This
requires $\mu \approx M_2$. In our study, however, we are
interested in large $\mu$ parameter (which we fix to $\mu = 1.2$
TeV) in order to have large CP mixing between CP-even and CP-odd
Higgs bosons from the stop loop. Then the charginos become too
heavy to be relevant to $\gamma\gamma \rightarrow H_i$. For a
smaller wino mass parameter $M_2 = 150$ GeV, the wino-Higgsino
mixing becomes smaller, but the lighter chargino mass becomes also
very light, and the loop function will be enhanced. The net result
turns out that the light chargino loop effects are important for
the reaction $\gamma\gamma \rightarrow H_i$ even if the lighter
chargino is dominantly a wino state ($M_2 \ll |\mu|$).

The amplitudes for $\gamma ( k_1, \epsilon_1 ) + \gamma ( k_2,
\epsilon_2 ) \rightarrow H_i (q)$ (with $i=1,2,3$) can be defined
in terms of two form factors $A_i (s)$ and $B_i (s)$ as follows in
a model independent way (we closely follow the convention of
Ref.~\cite{SYChoi} in the following): \beq {\cal M}(\gamma \gamma
\rightarrow H_i) = M_{H_i}  \frac{\alpha}{4\pi} \left\{ A_i(s)
\left[ \epsilon_1 \cdot \epsilon_2 - \frac{2}{s} (\epsilon_1 \cdot
k_2)(\epsilon_2 \cdot k_1) \right] - B_i(s) \frac{2}{s} \,
\epsilon_{\mu \nu \alpha \beta} \epsilon_1^\mu \epsilon_2^\nu
k_1^\alpha k_2^\beta \right\}, \eeq where $s \equiv ( k_1 + k_2
)^2 = M_{H_k}^2$. Including the chargino loop contributions, the
CP-even form factors $A_i$ at $s=M_{H_i}^2$ are \beq
A_i(s=M_{H_i}^2) = \sum_{f=t,b}A_i^f +
\sum_{\tilde{f}_j=\tilde{t}_{1,2}, \tilde{b}_{1,2}}
A_i^{\tilde{f}_j} +A_i^{H^{\pm}} + A_i^{W^{\pm}} + \sum_{j=1,2}
A_i^{\tilde{\chi}^\pm_j}, \eeq The CP-even functions $A_i^f$,
$A_i^{\tilde{f}_j}$, $A_i^{H^{\pm}}$, and $A_i^{W^{\pm}}$ are
given in Ref.~\cite{SYChoi}. We confirmed their results and
reproduced them and the related form factor loop functions in
Tables~1 and 2 for completeness. The chargino contribution to $A$
form factor is \beq A_i^{\tilde{\chi}^\pm_j}= 2 {\rm Re} (
\kappa^i_{jj} )  \frac{M_{H_i}}{M_{\tilde{\chi}^-_j}}
F_{sf}(\tau_{i \tilde{\chi}^\pm_j}), \eeq where $\tau_{i X}=
M_{H_i}^2/ 4 M_{X}^2$. The form factor $F_{sf}(\tau)=\tau^{-1}
\left[ 1+ (1-\tau^{-1}) f(\tau) \right]$ (and other loop functions
defined in Table~II) depends on the scaling function $f(\tau)$
\cite{Gunion}: \beq \label{f} f(\tau) =-\frac{1}{2} \int_0^1
\frac{{\rm d}y}{y} \log \left[1-4\tau y(1-y) \right]
=\left\{\ba{cl} {\rm arcsin}^2(\sqrt{\tau}) & {\rm for}~ \tau \le 1\\
-\frac{1}{4} \left[\log \left(
\frac{\sqrt{\tau}+\sqrt{\tau-1}}{\sqrt{\tau}-\sqrt{\tau-1}}\right)
-i\pi\right]^2 & {\rm for}~ \tau \ge 1. \ea \right. \eeq On the
other hand, the CP-odd form factor $B_i$ have contributions only
from the fermion loops and not from the boson loops: \beq
B_i(s=M_{H_i}^2) = \sum_{f=t, b} B_i^f + \sum_{j=1,2}
B_i^{\tilde{\chi}^\pm_j}, \eeq where $B_i^f$ are given in
\cite{SYChoi} (see also Tables~1 and 2), and the chargino
contributions are \beq \label{B_i} B_i^{\tilde{\chi}^\pm_j} = -2
{\rm Im} ( \kappa^i_{jj} ) \frac{M_{H_i}}{M_{\tilde{\chi}^-_j}}
F_{pf}(\tau_{i \tilde{\chi}^\pm_j}), \eeq where
$F_{pf}(\tau)=\tau^{-1}f(\tau)$. Therefore, when a CP-odd Higgs
boson $A$ is produced in $\gamma \gamma$ collision in the
CP-conserving limit, only fermion loops (not boson loops)
contributes to the production reaction.

It is also convenient to define two helicity amplitudes ${\cal M}_{\pm\pm}$ by
\beq
{\cal M}_{\lambda_1 \lambda_2} = - M_{H_k} \frac{\alpha}{4\pi}
\left\{ A_k (s) \delta_{\lambda_1 \lambda_2} + i \lambda_1  B_k (s)
\delta_{\lambda_1 \lambda_2} \right\},
\eeq
where $\lambda_{1,2}=\pm $ are photon helicities.
Then, in the narrow-width approximation, the partonic cross
sections of the $s$-channel Higgs productions \cite{SYChoi} are
\beq
\sigma(\gamma \gamma \rightarrow H_i) = \frac{\pi}{4M_{H_i}^4}
\left( \left|{\cal M}_{++} \right|^2 + \left| {\cal M}_{--} \right|^2
\right) \delta(1-M_{H_i}^2/s)
\equiv \hat{\sigma}_0(H_i) \delta(1-M_{H_i}^2/s).
\eeq
By using the amplitudes of $\gamma \gamma \rightarrow H_i$ at
$s=M_{H_i}^2$, we can also obtain the unpolarized decay rates of the
neutral Higgs bosons  into two photons,
\beq
\Gamma(H_i \rightarrow \gamma \gamma) = \frac{\alpha^2}{256 \pi^3} M_{H_i}
\left(
\left| A_i(s=M_{H_i}^2) \right|^2
+ \left| B_i(s=M_{H_i}^2) \right|^2 \right).
\eeq

The Higgs sector CP violation can be measured in the following
three polarization asymmetries ${\cal A}_a$ ($a=1,2,3$)
\cite{Grza} which are defined in terms of two independent helicity
amplitudes: \bea \label{asy1} {\cal A}_1 &=& \frac{ \left| {\cal
M}_{++} \right|^2 - \left| {\cal M}_{- -} \right|^2} {\left| {\cal
M}_{++} \right|^2 + \left| {\cal M}_{- -} \right|^2} = \frac{2{\rm
Im} (A_i(s)B_i(s)^*)}{\left| A_i(s) \right|^2 + \left| B_i(s)
\right|^2} ,
\\
\label{asy2} {\cal A}_2 &=& \frac{2{\rm Im}({\cal M}_{- -}^* {\cal
M}_{++})} {\left| {\cal M}_{++} \right|^2 + \left| {\cal M}_{- -}
\right|^2} =\frac{2{\rm Re}(A_i(s) B_i(s)^*)}{\left |A_i(s)
\right|^2 + \left| B_i(s) \right|^2} ,
\\
\label{asy3} {\cal A}_3 &=& \frac{2{\rm Re}({\cal M}_{- -}^*{\cal
M}_{++})} {\left| {\cal M}_{++} \right|^2 + \left| {\cal M}_{- -}
\right|^2} =\frac{\left |A_i(s) \right|^2 - \left| B_i(s)
\right|^2} {\left| A_i(s) \right|^2 + \left| B_i(s) \right|^2} ,
\eea In the CP-conserving limit, one of the form factors $A_i$ and
$B_i$ must vanish, so that ${\cal A}_1={\cal A}_2=0$, and ${\cal
A}_3=+1(-1)$ for a pure CP-even (CP-odd) Higgs scalar.
>From the definition of the function $f(\tau)$ in Eq.~(\ref{f}), we find
that the form factors $A_i$ and $B_i$ may be complex,
when the Higgs masses $M_{H_i}$ are two times larger than the particle mass
in the loop. This will induce rich structures in the polarization
asymmetries ${\cal A}_a$ as functions of Higgs masses and other
SUSY parameters in the presence of Higgs sector CP violation.

\section{Numerical Analyses}

The CP violation in the neutral Higgs sector through the stop loop
with the complex $A_t$ parameter always appear in the combination
of ${\rm arg}(A_t \mu)$. In the following numerical analyses, we
assume that the $\mu$ parameter is real and positive, in order to
simplify the discussions. For the complex $\mu$ parameter, the
chargino mass marix will contain CPV phase, thereby there would be
additional CP violating effects in the chargino loop contributions
to $\gamma\gamma \rightarrow H_i$. However, this CP violating
effect is independent of the CP violation in the neutral Higgs
sector through the mixing between the CP-even and the CP-odd Higgs
bosons. Since our focus in this work is to examine the reaction
$\gamma\gamma \rightarrow H_i$ in the presence of Higgs sector CP
violation through the mixing, we ignore complex phase in the
chargino sector. We also assume $A_t = A_b$ for simplicity even if
these couplings are independent in general. The CP violating phase
${\rm arg}(A_t)$ is varied between $0$ and $2\pi$. Also we choose
the same parameters as Ref.~\cite{SYChoi} (except for the wino
mass parameter $M_2$) in our numerical analyses in order to
investigate the chargino contributions more clearly; \bea
|A_t|=|A_b| = 0.4~{\rm TeV},~~ \mu = 1.2~{\rm TeV}, ~~ M_2 =
150~{\rm GeV},~~M_{\rm SUSY} = 0.5~{\rm TeV}. \label{param} \eea
Using these parameter set, we investigate in detail $\hat \sigma_0
(\gamma \gamma \rightarrow H_i)$ and ${\cal A}_a (H_i)$ for two
different values of $ \tan\beta =3$ and $\tan \beta = 10$ as
functions of each Higgs boson mass ($M_{H_k}$) and CP violating
phase ${\rm arg}(A_t)$ with/without chargino loop contributions.
As discussed in Section II, we do not consider a very large $\tan
\beta$ case, since the $A_t$ phase is strongly constrained by the
two-loop Barr-Zee type contributions to the EDMs of electron and
neutron. Note that the chargino contributions to the Higgs mixing
are negligible \cite{IbNa} for our choice of $\tan\beta = 3$ and
$\tan\beta = 10$.

\begin{figure}
\centerline{\epsfxsize=8cm \epsfbox{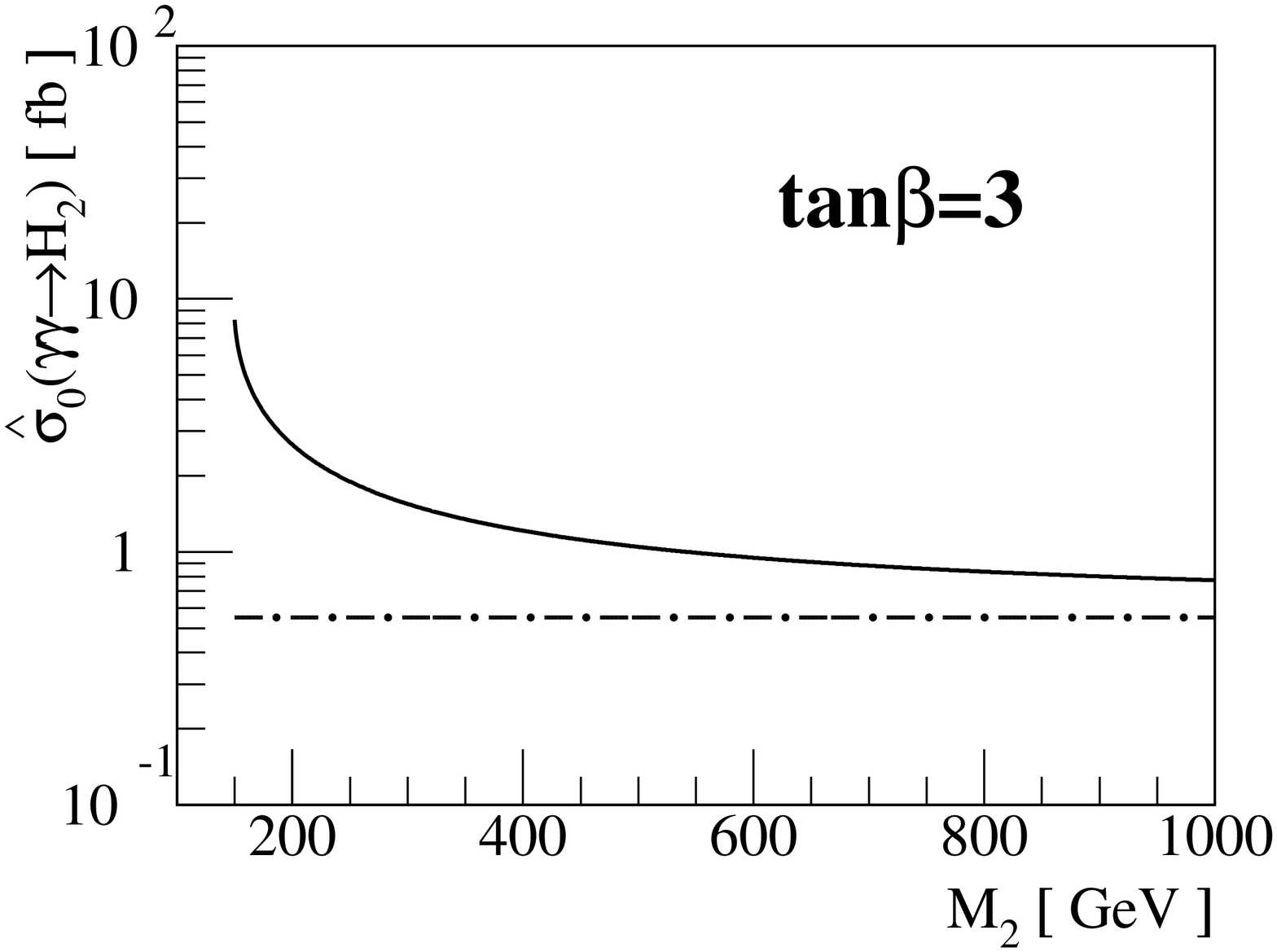} \epsfxsize=8cm
\epsfbox{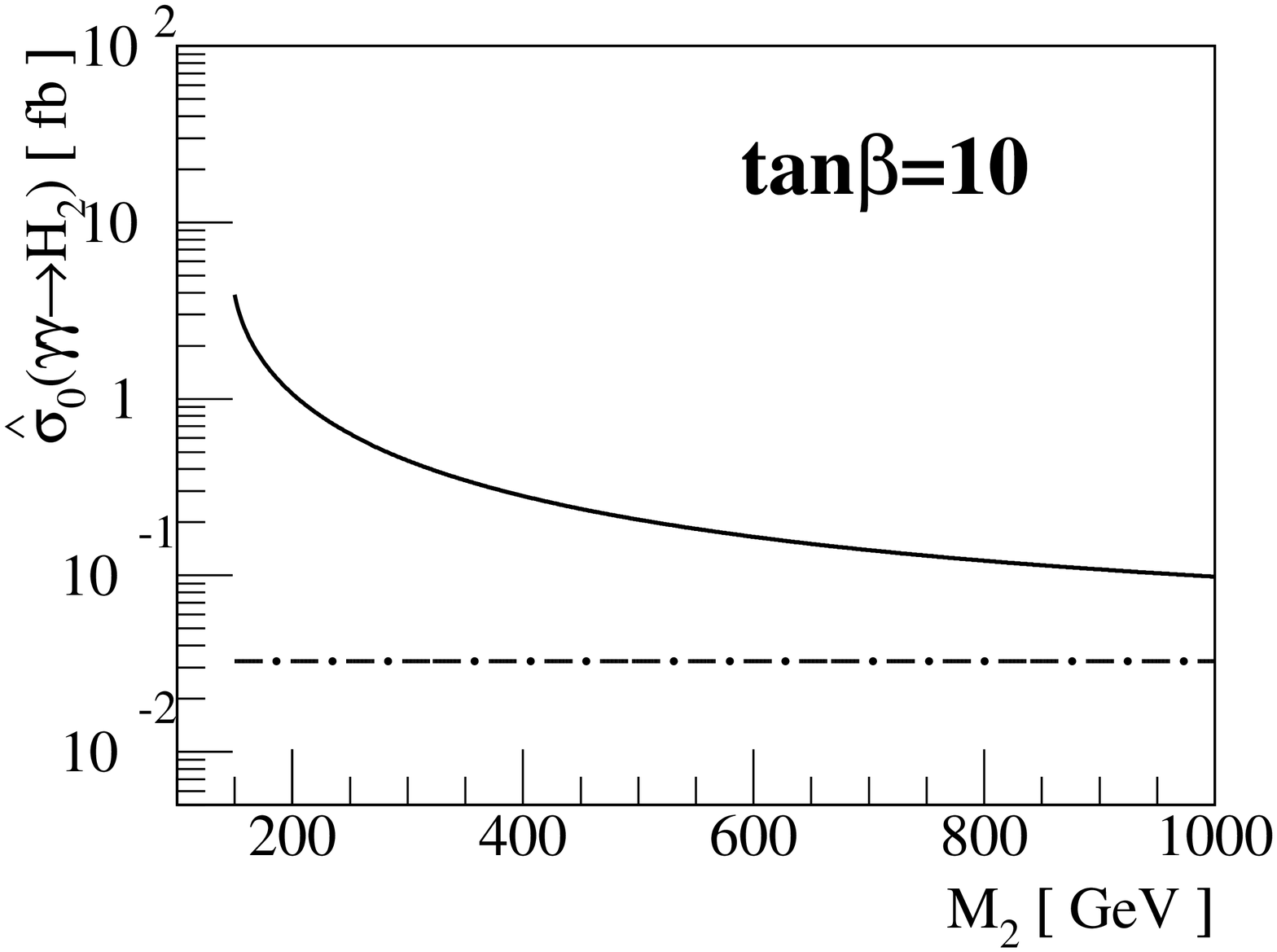}} ~~~~ \caption{ The cross sections for
$\gamma \gamma \rightarrow H_2$ with chargino loop contributions
(solid curves) and without chargino loop contribution (dash-dotted
curves) in unit of fb as functions of $M_2$ with $\tan\beta=3$ and
$\tan\beta=10$. We choose $|A_t| = 0.4$ TeV, $M_{H^+} = 300$ GeV
and ${\rm arg}(A_t)=0^\circ$. The left (right) figure is for
$\tan\beta=3$ ($\tan\beta=10$).} \label{ccc}
\end{figure}

It turns out the Higgs sector CP violation is most prominent in
the would-be CP-odd Higgs boson $H_2$ production at photon
colliders. Therefore we first discuss the production of the
would-be CP-odd Higgs scalar. In Fig.~\ref{ccc}, we show the
production cross section for $\gamma \gamma \rightarrow H_2$ as a
function of $M_2$ in the CP conserving limit (${\rm
arg}(A_t)=0^\circ$) for $\tan\beta = 3$ (on the left side) and
$\tan\beta = 10$ (on the right side), respectively. In both cases,
we assumed $\mu = 1.2$ TeV, and we set $M_{H^+} = 300$ GeV so that
$M_{H_2} = 291~(290)$ GeV for $\tan\beta = 3 ~(10)$, respectively.
The solid (dashed) curve represents the case with (without)
chargino contributions. For ${\rm arg}(A_t)=0^\circ$ (thick solid
curve), $H_2$ will be the pure CP-odd state ($A$) for our
parameter set (\ref{param}), since we can neglect the effects of
charginos on the Higgs mixing due to $\tan \beta \lesssim 20$
\cite{IbNa}. In this case, $\hat \sigma_0 (\gamma \gamma
\rightarrow H_2)$ has only the fermion loop contributions, since
the couplings of $H_2$ to the sfermion pairs, the charged
Higgs-boson and $W$-boson pairs vanish in the CP conserving limit.
The cross section for $\gamma \gamma \rightarrow H_2$ without
chargino loop contributions is independent of $M_2$ (the
horizontal dash-dotted lines), and are quite small ($\lesssim 1$
fb). The bottom-quark contribution is negligible compared to the
top-quark contribution for two reasons: (i) the small $b$ quark
mass and (ii) the smaller electric charge of $b$ quark (note that
the $\gamma \gamma \rightarrow H_2$ amplitude depends on $e_q^2$).
For our choice of parameters, the bottom quark contribution turns
out to get significant only for $\tan\beta \geq 10$, and can be
safely neglected for $\tan\beta \lesssim 10$. On the other hand,
the cross section for $\gamma \gamma \rightarrow H_2$ is enhanced
almost by an order of magnitude when the chargino loop
contributions are included. The chargino loop contributions to
$\gamma \gamma \rightarrow H_2$ can not be ignored at all, if
charginos are not very heavy.  This is true even if we set $M_2
\ll |\mu|$ so that the wino-Higgsino mixing is not large. Still
the lighter charginos are light enough ($M_2 = 150$ GeV for our
parameter set) and the loop contribution is important. Also
because of the $1/\tan\beta$ suppression factor for the top loop,
the chargino loop contribution becomes more important for larger
$\tan\beta$. Finally, as the $M_2$ increases, the lighter chargino
becomes heavier and the chargino loop contribution decreases
rather quickly due to the decoupling theorem. Since the chargino
mass arises dominantly from SUSY breaking rather than from
electroweak symmetry breaking, the decoupling of the chargino loop
contribution is more effective than the top loop contribution.
Also, the couplings  ${\rm Im}(\kappa^2_{jj})$ decrease more
quickly as functions of $\tan\beta$ compared to the loop functions
as $M_2$ increases. Therefore, the difference between the cross
sections for $\tan\beta=3$ and $\tan\beta = 10$ increases as $M_2$
increases.

\begin{figure}
\centerline{\epsfxsize=8cm \epsfbox{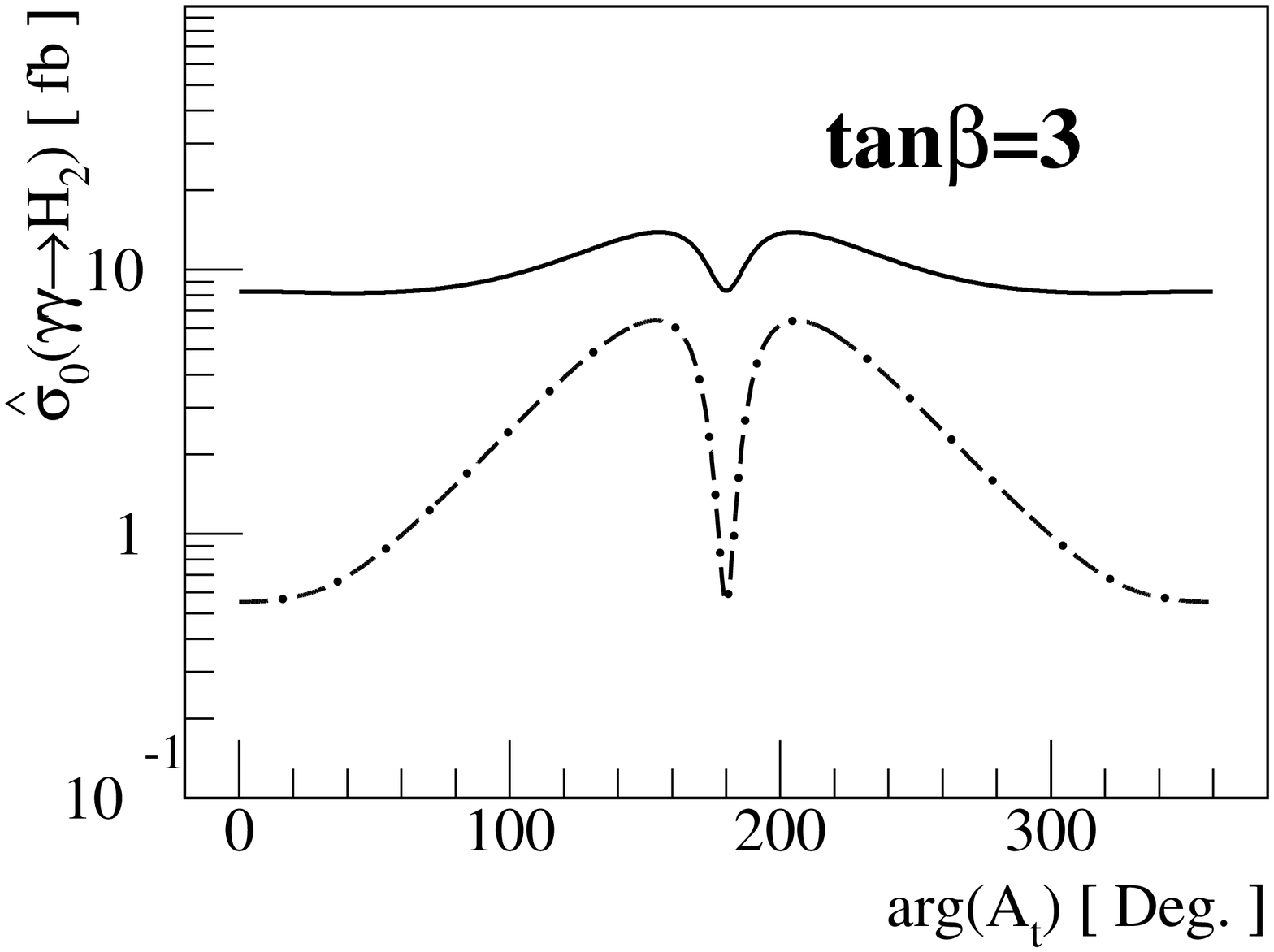}
\epsfxsize=8cm \epsfbox{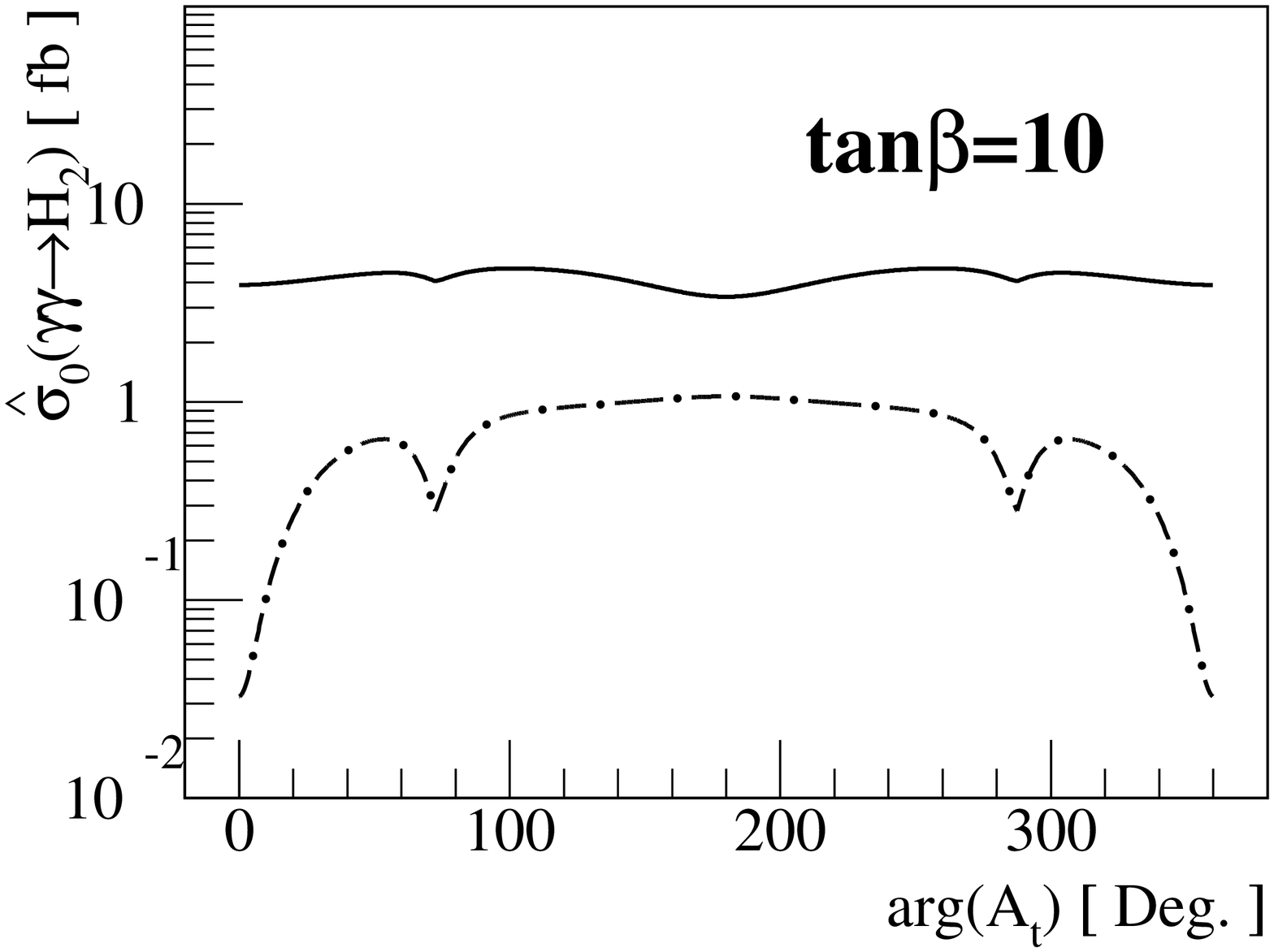}}

~~~~
\caption{ The cross sections for $ \gamma \gamma \rightarrow
H_2$ with chargino loop contributions (solid curve)
and without chargino loop contribution (dash-dotted curve)
in unit of fb as functions of ${\rm arg}(A_t)$ for $M_2=150$ GeV.
The left (right) figure is for $\tan\beta=3$ ($\tan\beta=10$).}
\label{ddd}
\end{figure}

In Fig.~\ref{ddd}, we show the cross section for $\gamma \gamma
\rightarrow H_2$ as a function of ${\rm arg}(A_t)$ for $\tan\beta =3$ (on the
left side) and $\tan\beta = 10$ (on the rigth side), respectively.
The solid (the dash-dotted) curves represents the case with (without) the
chargino loop contributions.
For ${\rm arg}(A_t)=0^\circ$ (or $180^\circ$), the cross section is strongly
enhanced by the chargino loop contributions as discussed in the previous
paragraph. As ${\rm arg}(A_t)$ is turned on, the cross section is
significantly enhanced even without the chargino loop contributions.
This is because all the charged particles including bosons begin to contribute
in the presence of CP violation in the Higgs sector. The dash-dotted curves
strongly depend on ${\rm arg}(A_t)$  for the following reasons. First of
all, the stop masses and the mixing angles depend on ${\rm arg}(A_t)$ very
sensitively.  Note that the stop masses have the $LR$ mixing term
$m^2_{\tilde{t} LR}= m_t(A_t^* e^{-i\xi} - \mu/\tan\beta)$, as shown in
Eq.~(\ref{mLR}) of the Appendix A. Since the mixing between CP-even and
CP-odd neutral Higgs bosons arises from the stop loop [see Eq.~(\ref{m12})],
the $A_t$ phase affects the CP mixing through Im ($A_t \mu$) and the stop
masses in Eq.~(\ref{m12}). Also once CP is broken in the Higgs sector, all
the charged particles including bosons as well as fermions contribute to
$\gamma\gamma \rightarrow H_2$. Therefore the stop loop contribution will
depend on ${\rm arg}(A_t)$. Still the dominant contribution comes from the
chargino loops (see the solid curves in Fig.~2). The net result depends on
${\rm arg}(A_t)$ rather mildly, mainly through the ${\rm arg}(A_t)$ of the
CP-odd and CP-even Higgs mixing.

Also note that the sensitivity of the cross section
$\hat \sigma_0 (\gamma \gamma \rightarrow H_2)$ to ${\rm arg}(A_t)$
decreases as $\tan \beta$ increases. This tendency can be understood by
the strong phase dependences of stop masses, since stop loops contribute to
(i) the mixing of the CP-even and the CP-odd Higgs bosons, and (ii) the
loop diagrams.
The scalar-pseudoscalar mixing is typically characterized by
\[
{\rm Im}(m_{12}^2) \propto h_t^2 {\rm arg}(A_t \mu),  ~~~~
\]
whose $\tan\beta$ dependence is negligible for $3\leq \tan\beta
\leq 10$ [see Eq.~(\ref{m12})]. Also the stop mass eigenvalues are
sensitive to the CP phase ${\rm arg}(A_t)$ when $|A_t|= |\mu|
/\tan\beta$ due to the $LR$ mixing ($\tan \beta = 3$ for our
parameter set $|A_t|=|\mu|/3=0.4$ TeV). The CP mixing would be a
decreasing function of $\tan\beta$ for $\tan\beta \ge 3$, and the
stop masses are less sensitive to CP phase ${\rm arg}(A_t)$ for
the larger $\tan\beta=10$. Therefore, the phase dependence of the
mixing would be a decreasing function of $\tan\beta$ for
$\tan\beta \ge 3$. Another dependence of the cross section on
${\rm arg}(A_t)$ originates from the stop masses in the loop,
which is sensitive to the phase ${\rm arg}(A_t)$ in our choice of
SUSY parameter set. In other words, $\tan\beta$-dependence of the
phase sensitivity comes dominantly from the stop masses as in the
CP-even and CP-odd Higgs mixing. Therefore, the cross section
depends on the phase  ${\rm arg}(A_t)$ less sensitively when $\tan
\beta$ becomes larger for our parameter set. Finally, the heavier
Higgs boson ($H$) is also strongly affected by the CP mixing,
since it can have a large mixing with the CP-odd scalar $A$. The
discussions for $H$ will be similar to those for $A$, and will not
be repeated.

\begin{figure}
\centerline{\epsfxsize=7.0cm \epsfbox{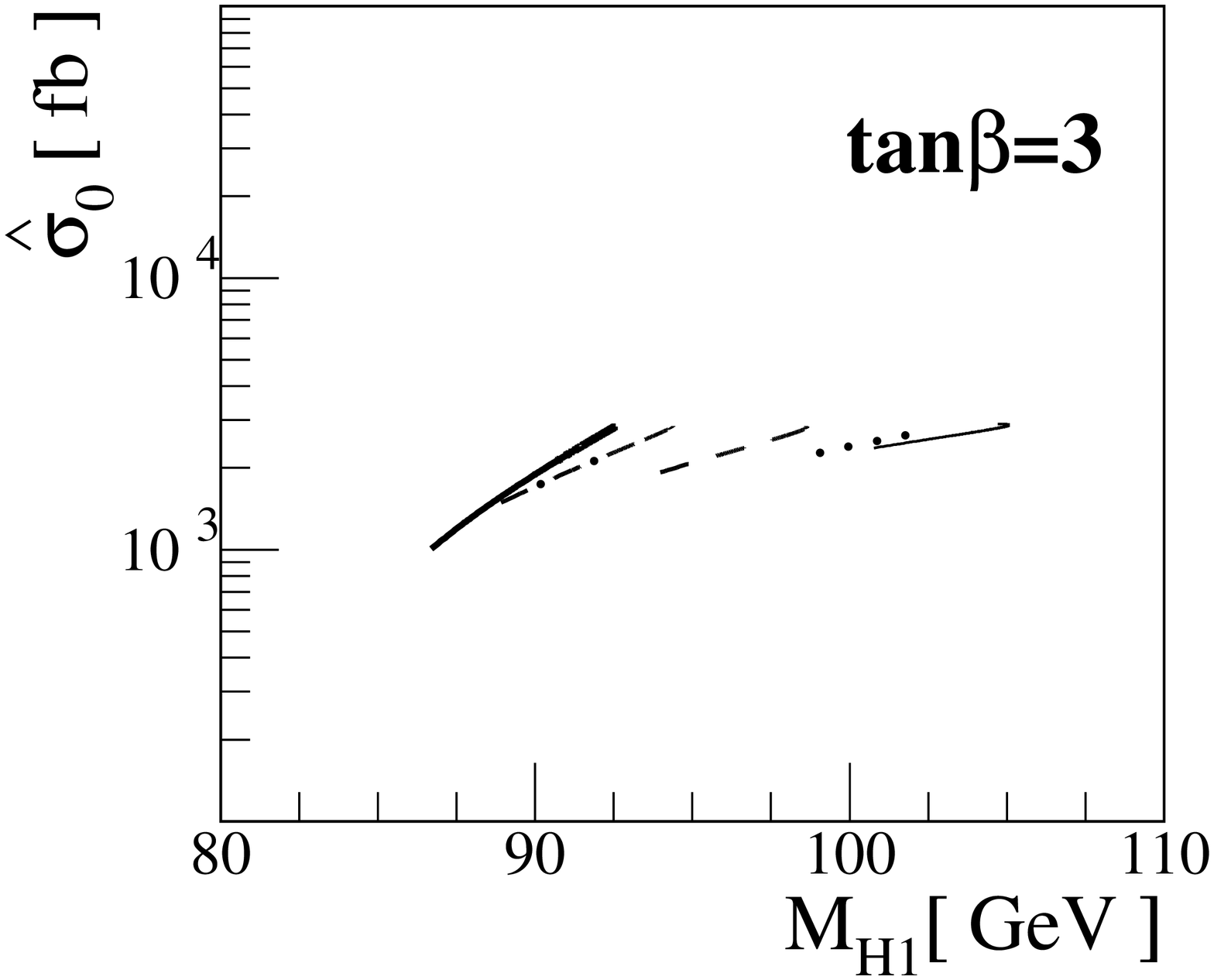}
~~~~~\epsfxsize=7.0cm \epsfbox{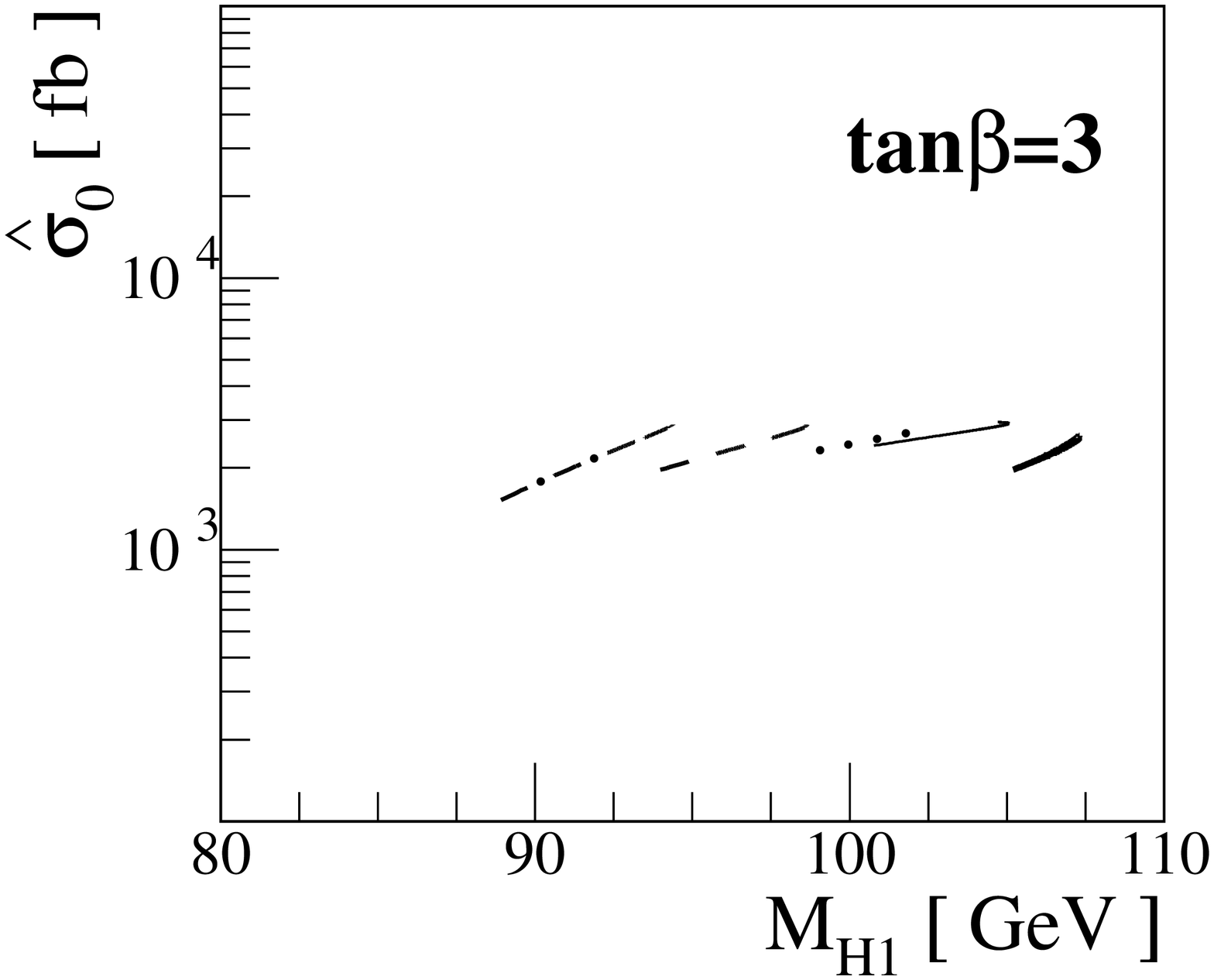}}
\centerline{\epsfxsize=7.0cm \epsfbox{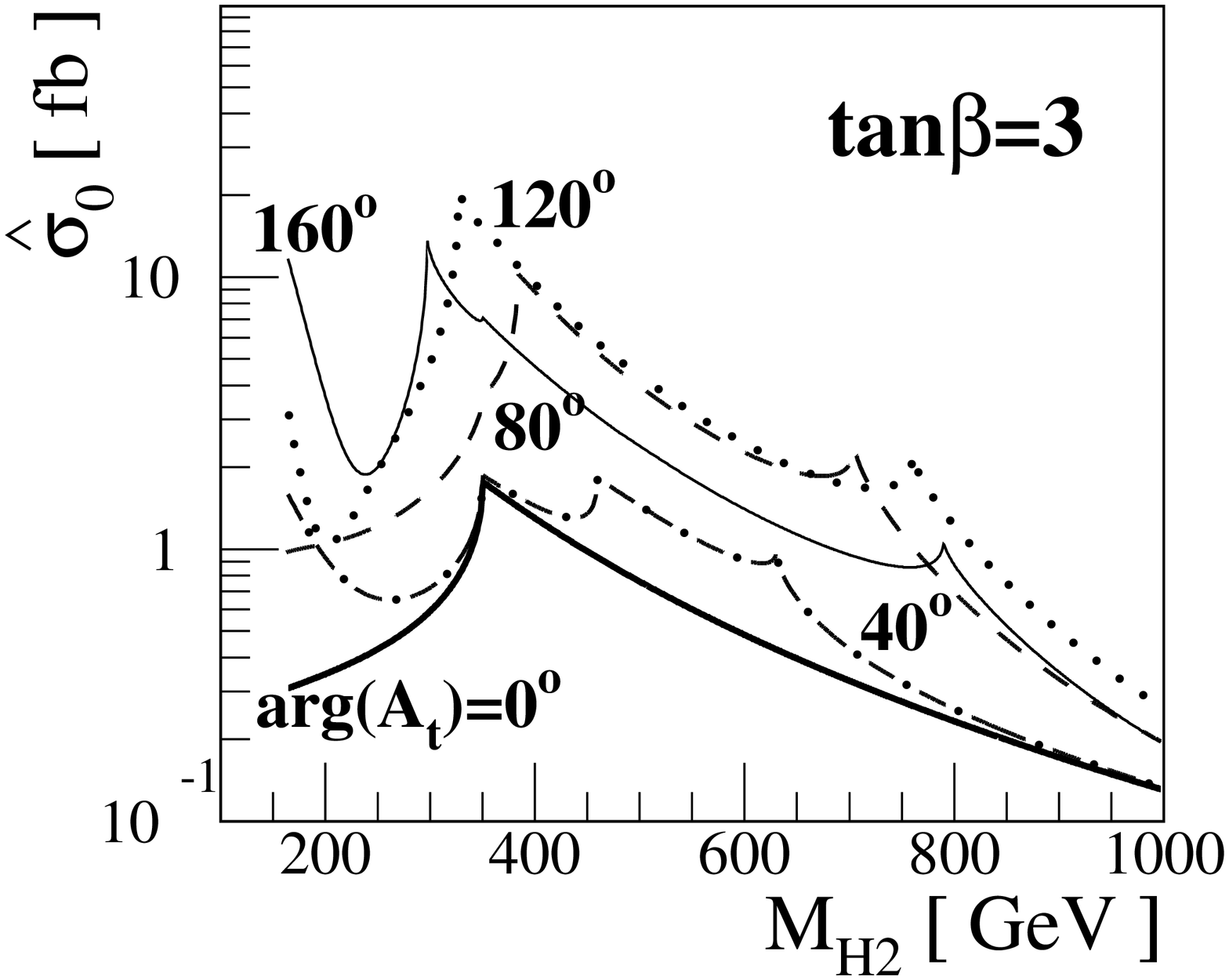}
~~~~~\epsfxsize=7.0cm \epsfbox{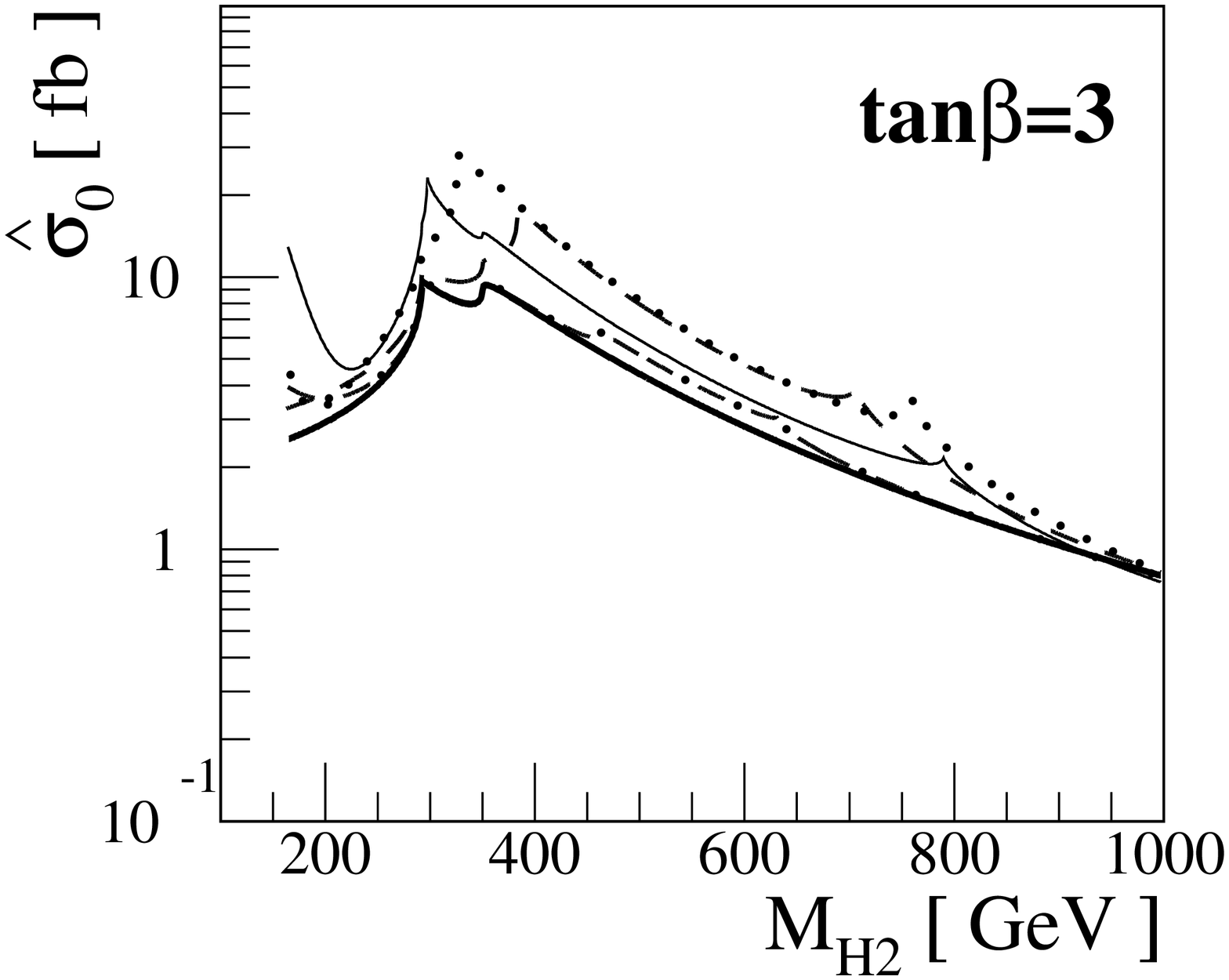}}
\centerline{\epsfxsize=7.0cm \epsfbox{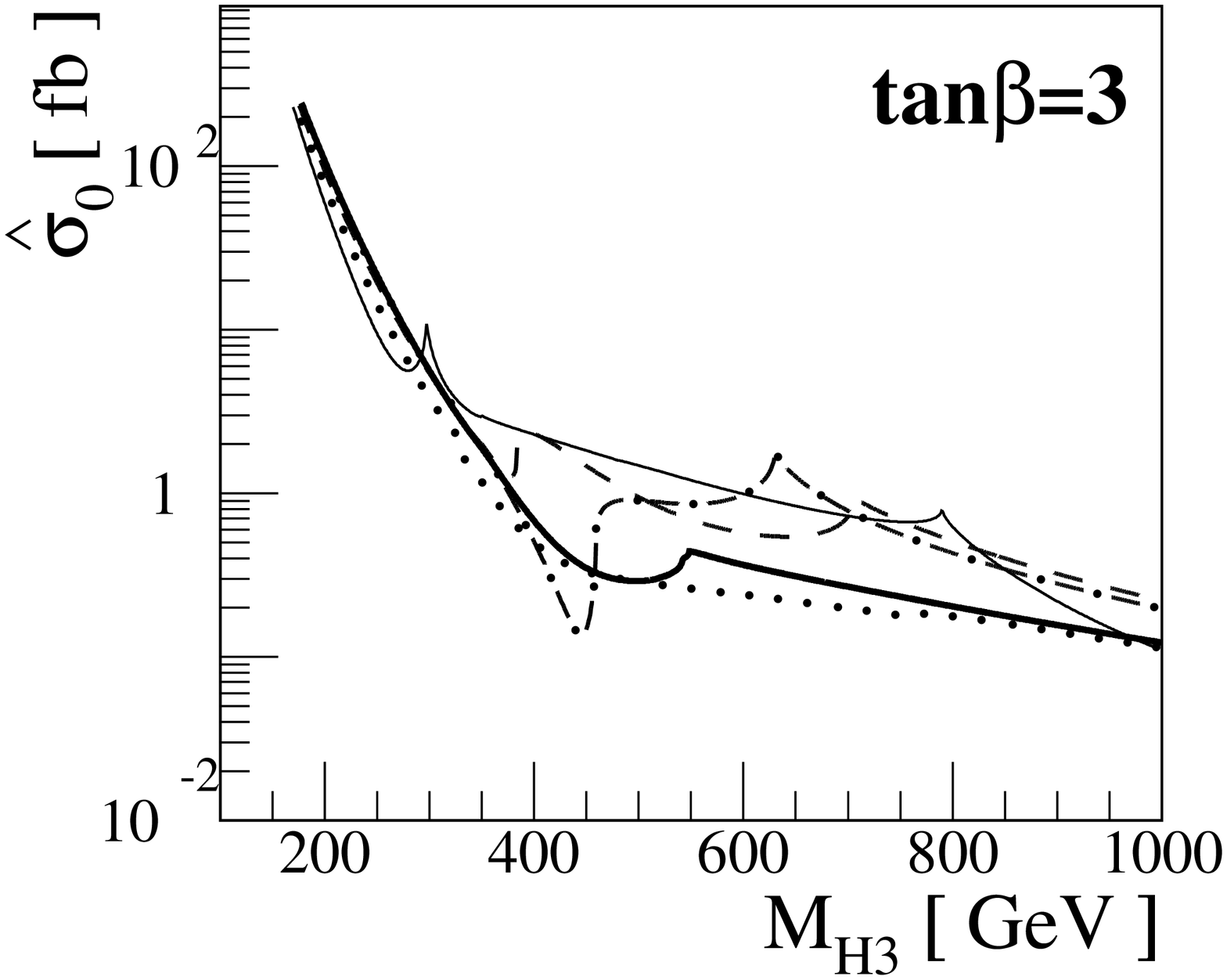}
~~~~~\epsfxsize=7.0cm \epsfbox{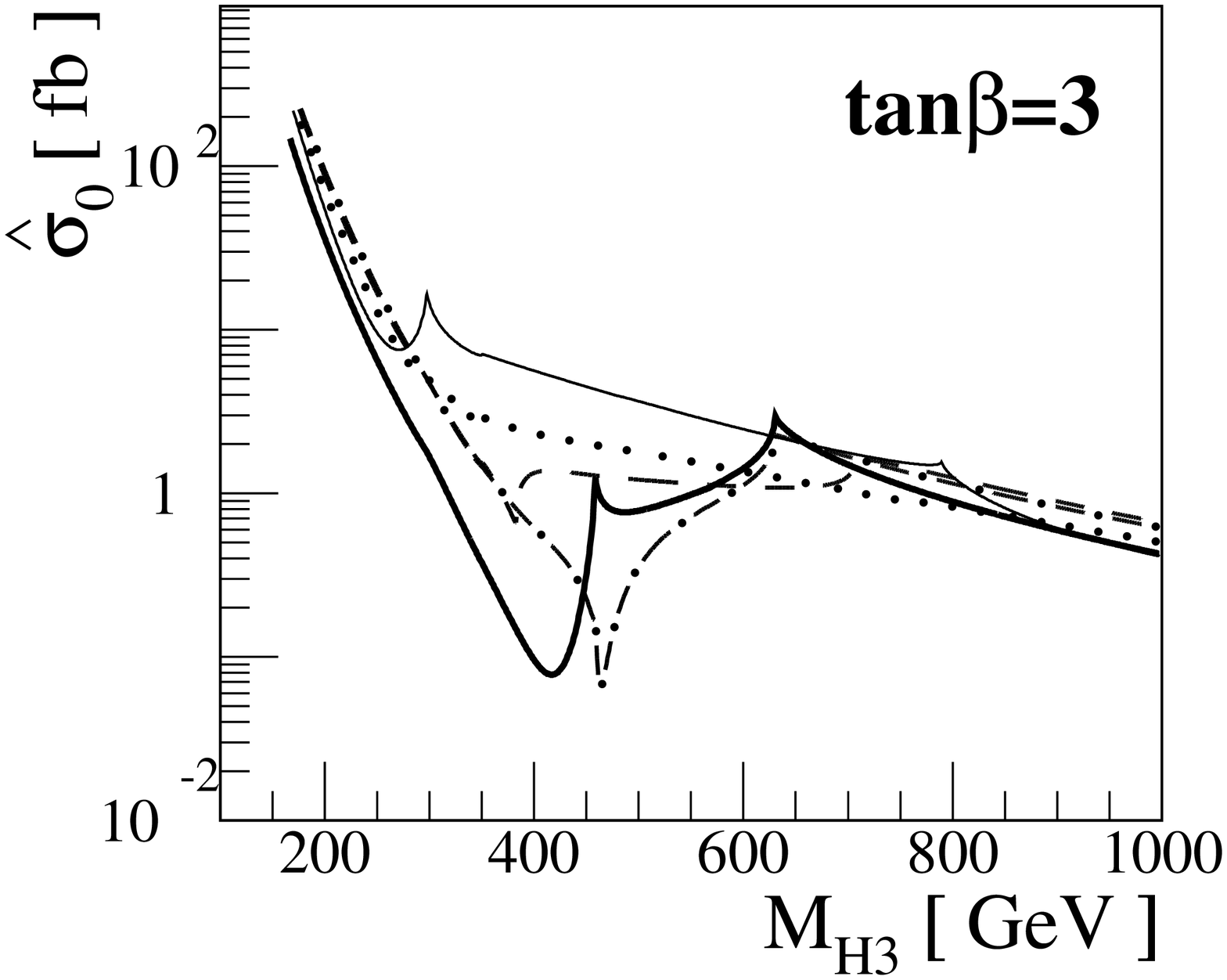}}
~~~~
\caption{The unpolarized cross sections for $\gamma \gamma \rightarrow
H_i$ ($i=1,2,3$) without chargino loop contributions (left column)
and with chargino loop contributions (right column) for $M_2 = 150$ GeV
in units of fb as functions of each Higgs mass for five
different values of the $A_t$ phase;
${\rm arg}(A_t)=0^\circ$ (thick solid curve),
$40^\circ$ (dash-dotted curve),
$80^\circ$ (dashed curve),  $120^\circ$ (dotted curve) and
$160^\circ$(solid curve).}
\label{aaa}
\end{figure}

\begin{figure}
\centerline{\epsfxsize=7cm \epsfbox{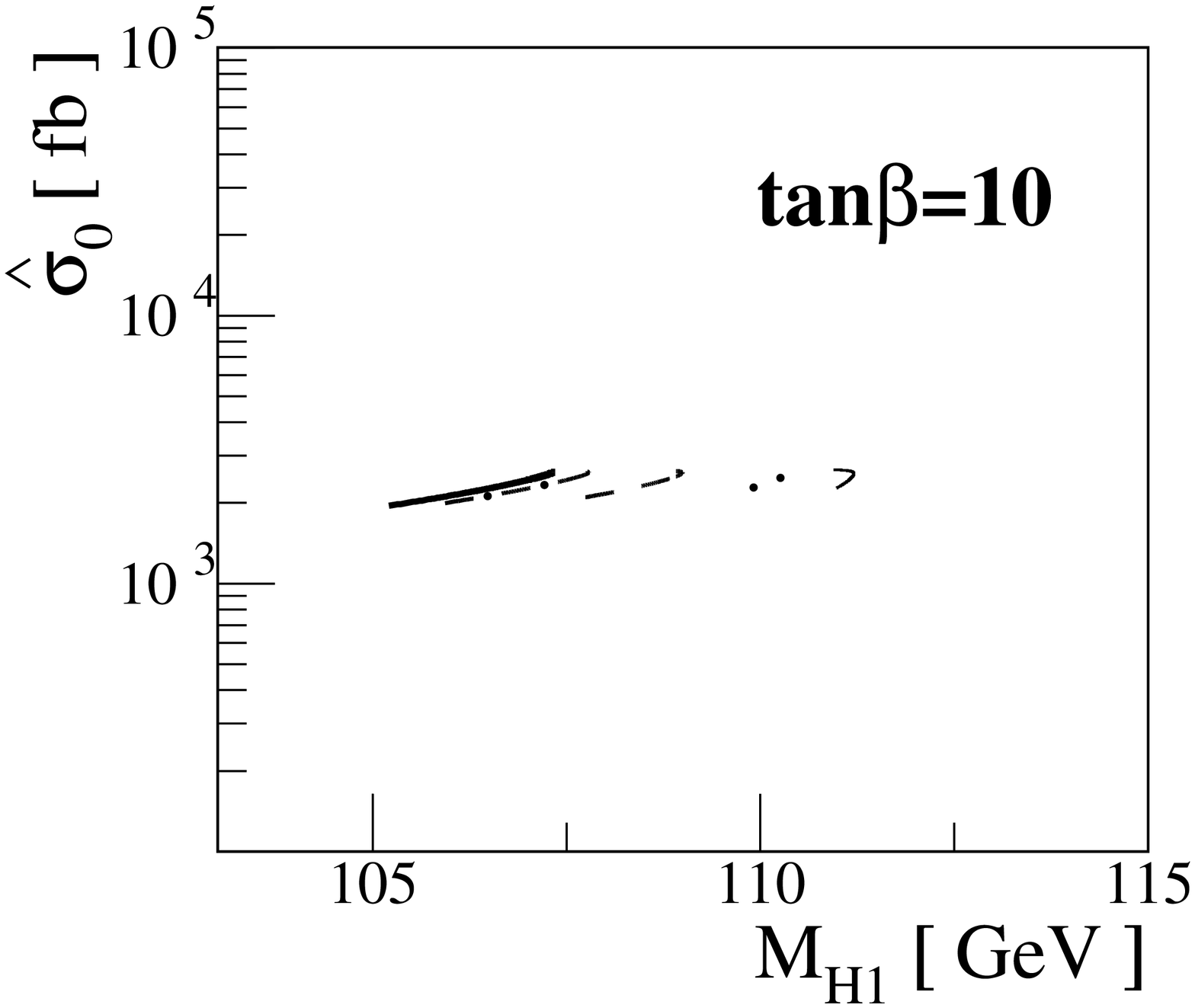}
\epsfxsize=7cm \epsfbox{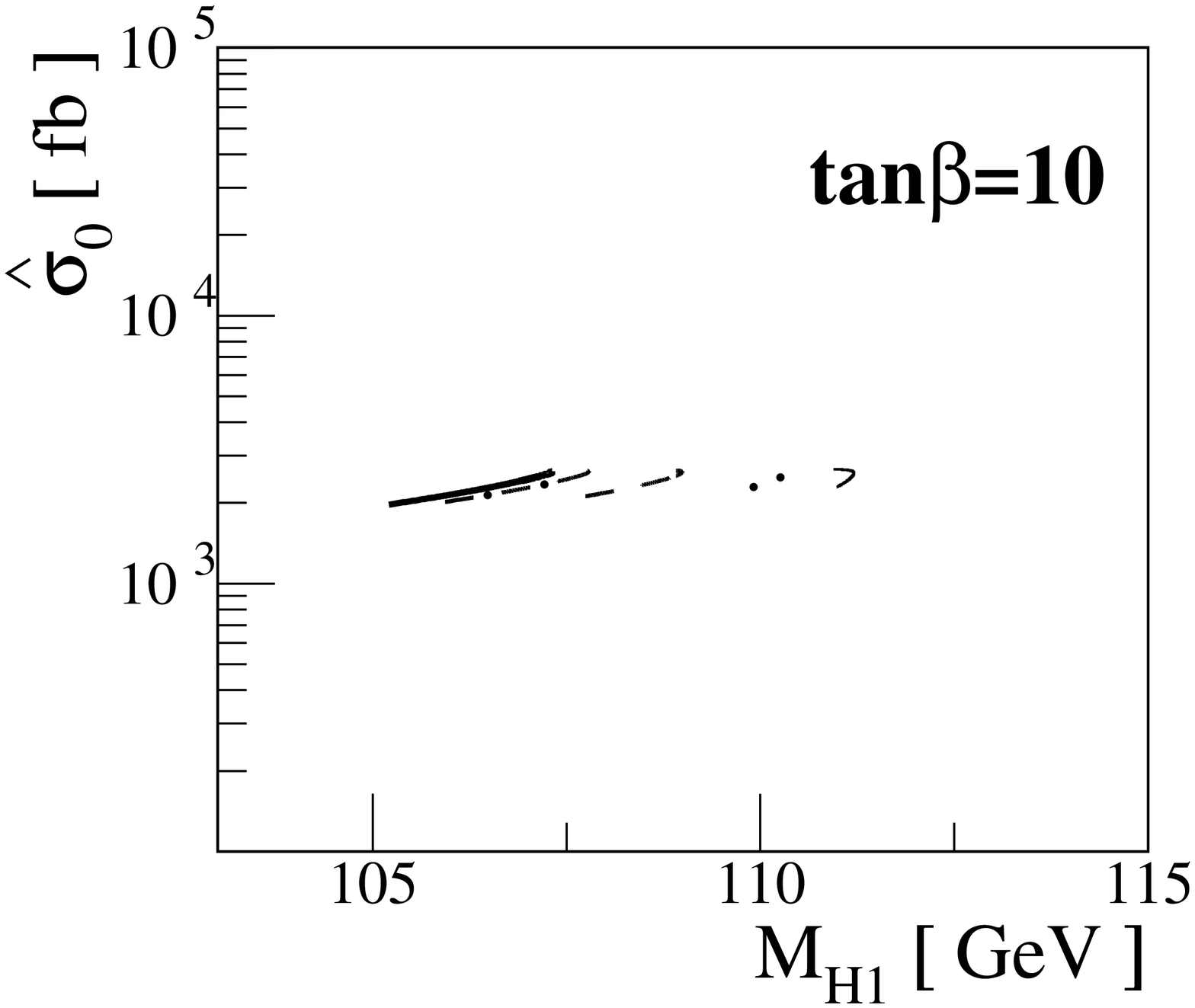}}
\centerline{\epsfxsize=7cm \epsfbox{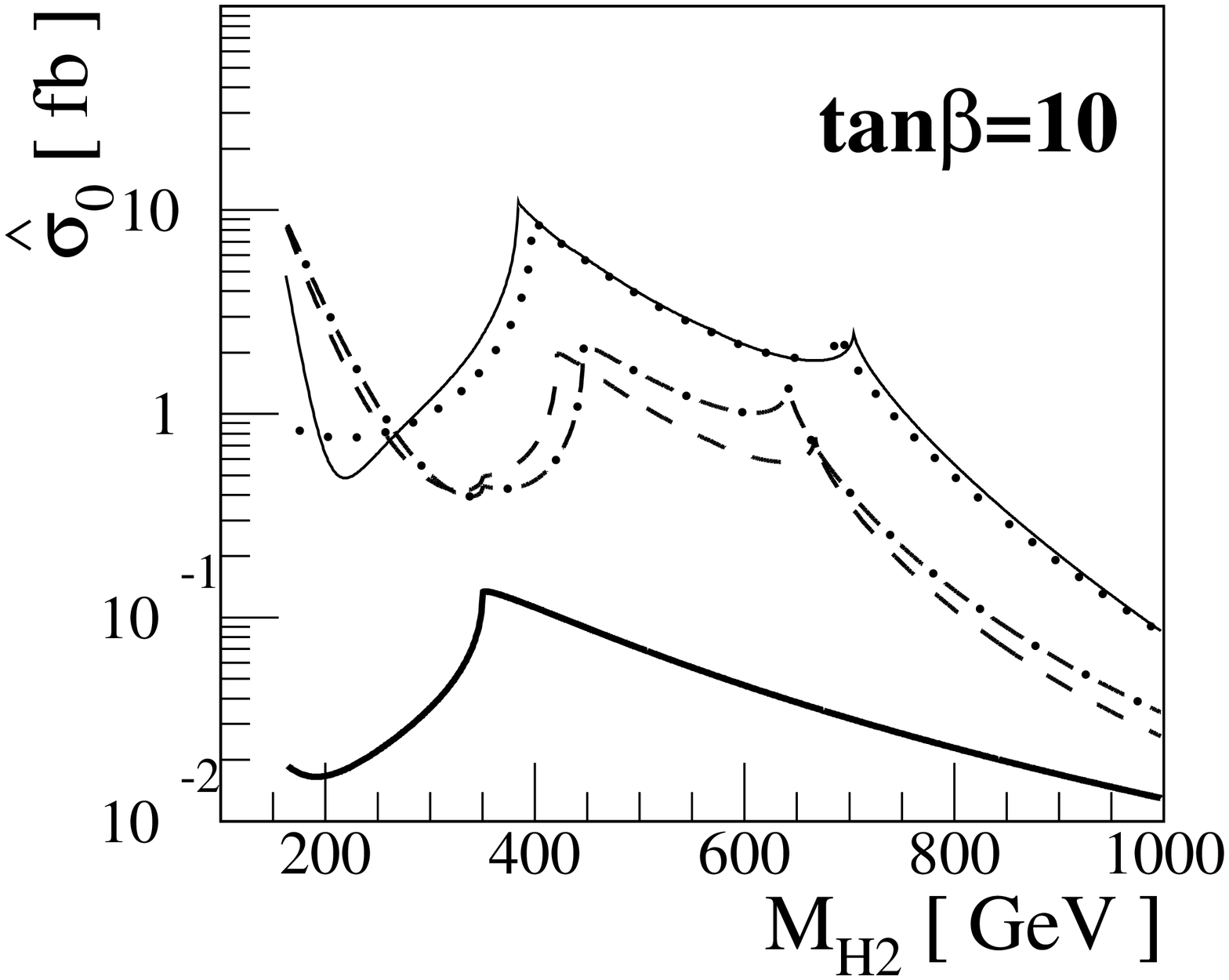}
\epsfxsize=7cm \epsfbox{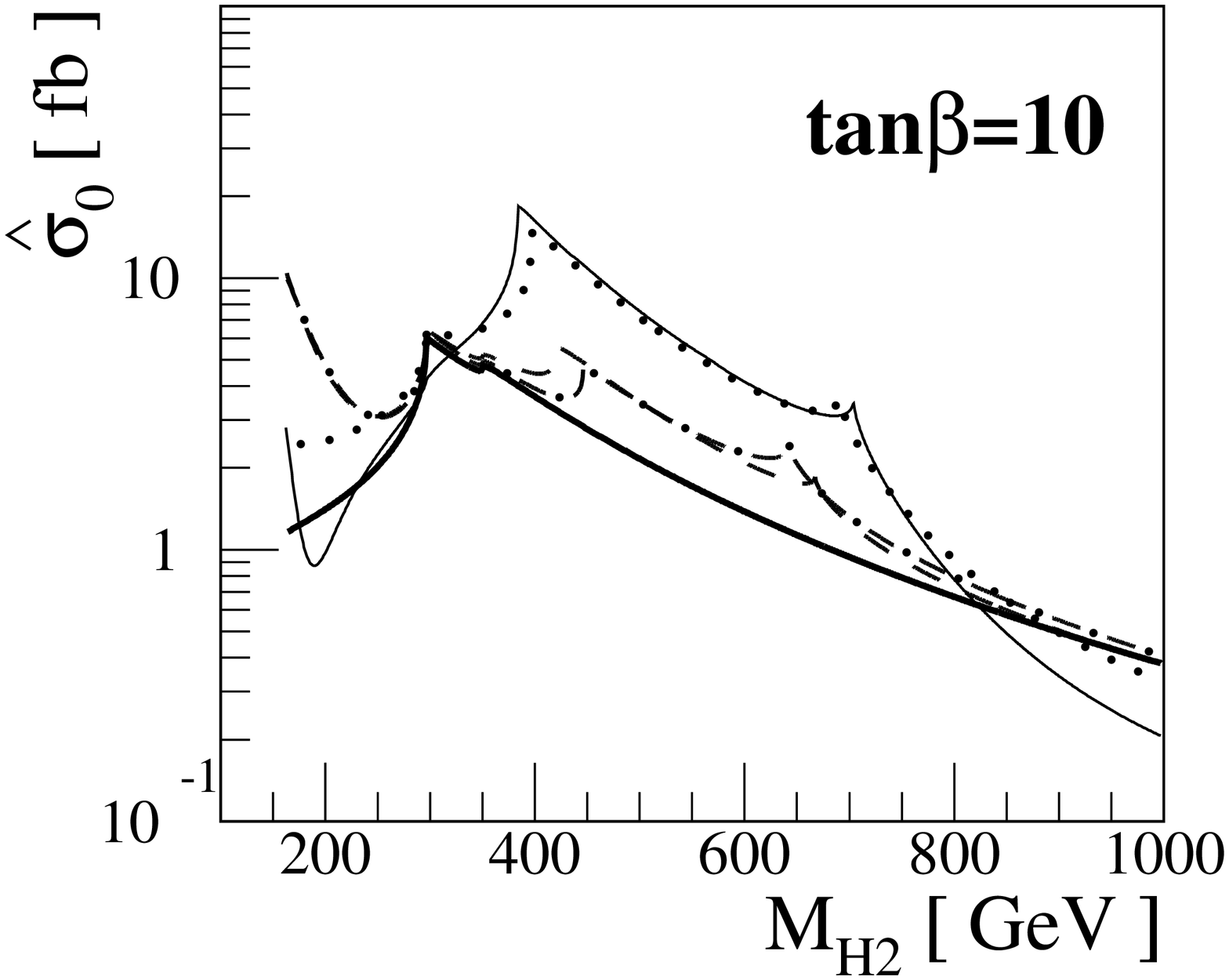}}
\centerline{\epsfxsize=7cm \epsfbox{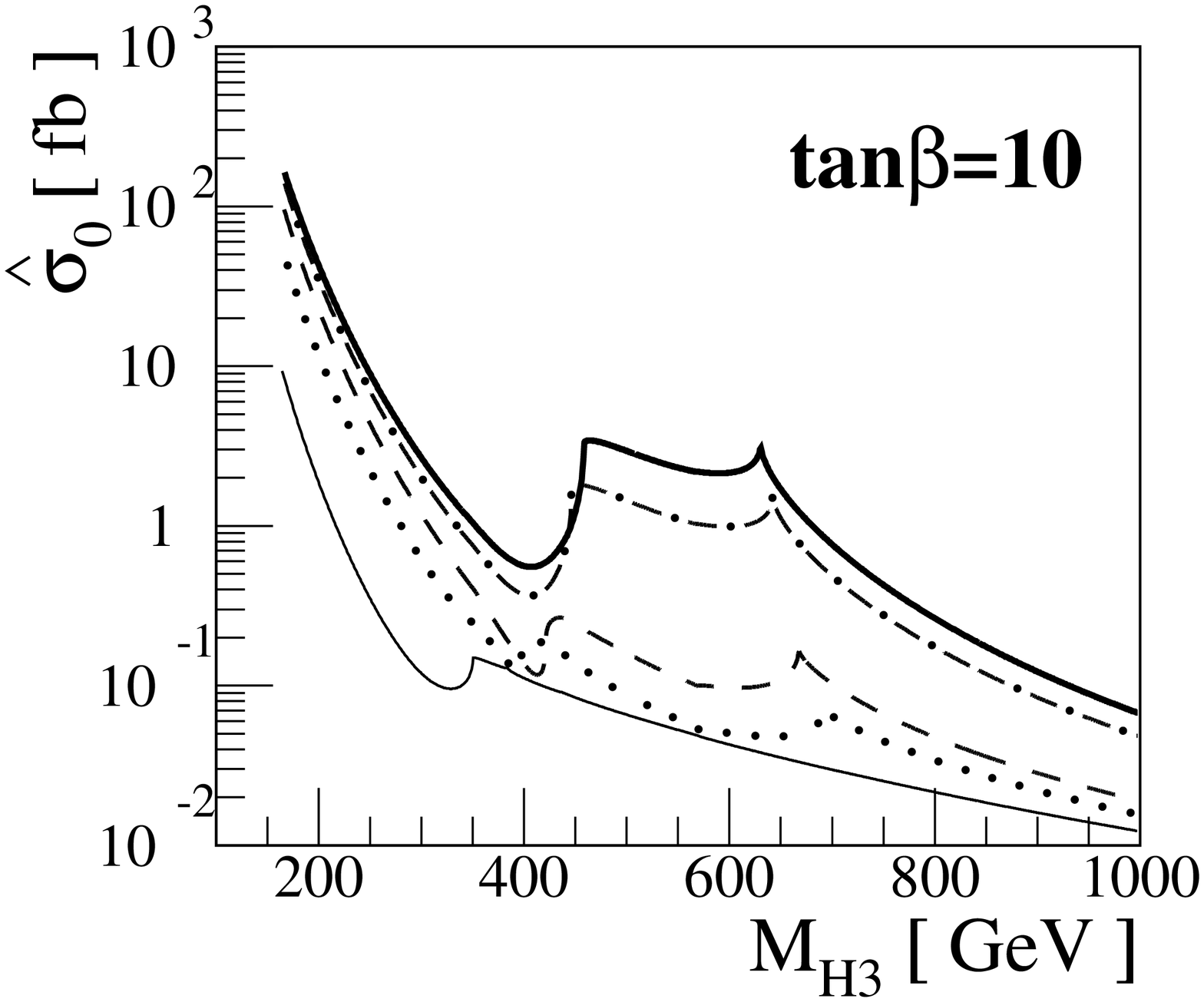}
\epsfxsize=7cm \epsfbox{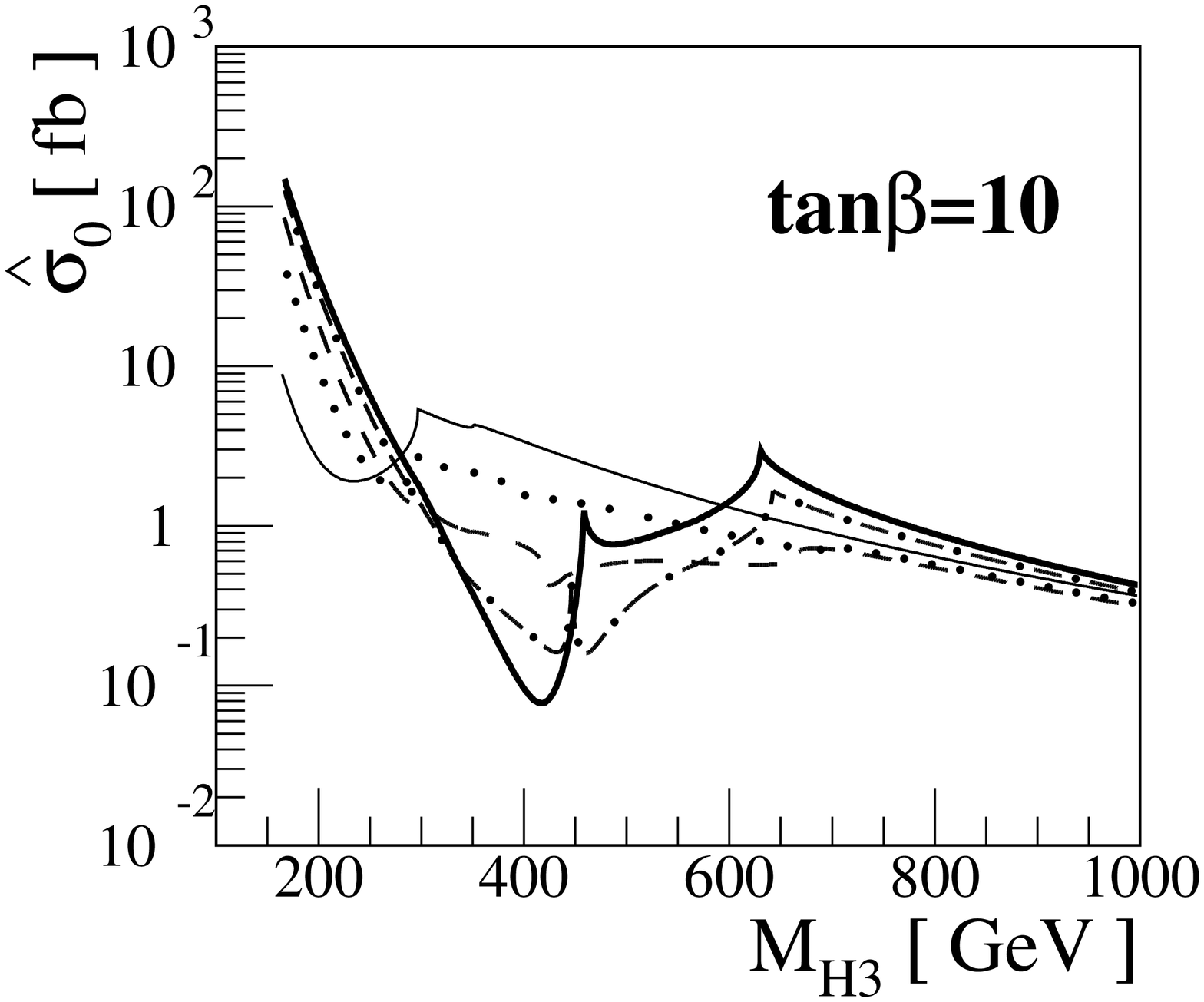}}
~~~~
\caption{The unpolarized cross sections for $\gamma \gamma \rightarrow
H_i$ ($i=1,2,3$) without chargino loop contributions (left column)
and with chargino loop contributions (right column) for
$M_2 = 150$ GeV in units of fb as functions of
each Higgs mass for five  different values of the $A_t$ phase;
${\rm arg}(A_t)=0^\circ$ (thick solid curve),
$40^\circ$ (dash-dotted curve),
$80^\circ$ (dashed curve), $120^\circ$ (dotted curve) and
$160^\circ$ (solid curve).}
\label{bbb}
\end{figure}

In Figs.~\ref{aaa} and \ref{bbb}, we show that the cross sections
$\hat \sigma_0 (H_i)$ ($i=1,2,3$) in units of fb for five different
$A_t$ phases; ${\rm arg}(A_t) = 0^\circ$ (thick solid curve),
$40^\circ$ (dash-dotted curve), $80^\circ $ (dashed curve),
$120^\circ$ (dotted curve) and $160^\circ$ (solid curve)
for $\tan\beta = 3$ (Fig.~\ref{aaa}) and $\tan\beta = 10$ (Fig.~\ref{bbb}).
We present two different cases: without the chargino loop contributions
(the left column) as in Ref.~\cite{SYChoi} and with the chargino loop
contributions (the right column) with $M_2 = 150$ GeV for which
$M_{\tilde{\chi}^-_1} = 146~(148.2)$ GeV for $\tan\beta =3 ~(10)$.
The chargino contributions to $\gamma\gamma \rightarrow H_1$ is negligible,
since $M_{H_1}$ is far below the chargino pair threshold
$2 M_{\tilde{\chi}^-_1}$ for our parameter set. On the other hand, two
heavier Higgs productions are affected by chargino loops by significant
amounts, and we can observe rich structures in the production cross sections
due to the interference of all the charged particles' contributions.  For
example, the production cross sections for $\gamma \gamma \rightarrow H_2$
without the chargino loop contributions (the left columns of Figs.~\ref{aaa}
and \ref{bbb}) have only a single peak at the point $M_{H_2} = 2m_t$ for
${\rm arg}(A_t)=0^\circ$.
If the chargino loop contributions are included (the right columns), the
production cross sections have two comparable peaks at the point
$M_{H_2} = 2M_{\tilde{\chi}^-_1}$ (lighter chargino) and $M_{H_2} = 2m_t$
in the CP-conserving limit. As the CP violating phase arg($A_t$)
increases, the cross section $\hat \sigma_0 (\gamma \gamma \rightarrow H_2)$
starts to get extra contributions from the charged boson loops (involving
sfermions, the charged Higgs-boson and the $W$-boson pairs) due to the mixing
between the CP-odd and the CP-even neutral Higgs bosons.

For a larger $\tan\beta = 10$ (Fig.~4), there appear three qualitative
differences compared to the lower $\tan\beta=3$;
the effect of bottom quark loop contribution, the dominant chargino loop
contributions, and the interchange of the CP-properties of the neutral
Higgs bosons.
\begin{itemize}
\item Since the bottom quark Yukawa coupling (to the CP even Higgs
boson) is proportional to $1/\cos\beta$, the bottom quark
contribution can be significant in the region of large
$\tan\beta$. For ${\rm arg}(A_t)=0^\circ$, the CP-odd Higgs boson
$H_2$ has pseudoscalar couplings to top and bottom quarks, where
the coupling of $H_2$ to top (bottom) quark is proportional to
$\cot\beta$ ($\tan\beta$) [see Eqs.~(\ref{hff}) and (\ref{R}) of
the Appendix B]. Furthermore, there are additional differences
from different electric charges of top and bottom quarks, since
the $\gamma\gamma \rightarrow H_i$ amplitudes depend on $e_q^2$,
which are $(2/3)^2$ vs. $(-1/3)^2$ for (s)top and (s)bottom,
respectively. On the other hand, the loop functions have weaker
$\tan\beta$-dependences. For our parameter set, it turns out that
the bottom quark contribution begins to dominate the top quark
contribution when $\tan\beta \sim 10$, and can be neglected for
$\tan\beta < 10$. \item In the CP conserving limit, the chargino
contribution to $\gamma\gamma \rightarrow A$ is dominant over the
top quark contribution, since the latter is suppressed by
$1/\tan\beta$ relative to the former, even if we assume the mixing
angles in the chargino sector are $O(0.1)$. This is the reason why
the top quark contribution decreases more quickly than the lighter
chargino contributions as $\tan \beta$ increases in
Figs.~\ref{aaa} and \ref{bbb}. \item The final point is the
interchange of the CP properties of the heavier Higgs bosons $H_2$
and $H_3$ for large ${\rm arg}(A_t)$ and large $\tan\beta=10$.
Since there are only fermion contributions to the CP-odd Higgs
production, i.e., two peaks at $M_{H_i}=2M_{\tilde{\chi}^-_1}$ and
$2m_t$, we can find from Fig.~\ref{bbb} that $H_3$ for ${\rm
arg}(A_t)=160^\circ$ has the same CP-odd property as $H_2$ for
${\rm arg}(A_t)=0^\circ$. This can be checked even more easily by
using the polarization asymmetry ${\cal A}_3$ which is +1($-1$)
for a CP-even (CP-odd) Higgs boson, as discussed below in relation
with polarization asymmetries (Figs.~5--7).
\end{itemize}
The importance of chargino loop contributions for $H_3$ production is also
similar to the case of $H_2$ production as discussed above, and we will
not repeat it again.

The number of events is determined by the combination of the
luminosity and the cross section for $\gamma \gamma \rightarrow H_2$.
Although the photon beam luminosity depends on many parameters, if one
only consider the high energy part of the generated photons, the 0.3
conversion factor and the comparable photon spot size to electron beam,
the approximate luminosity of $\gamma \gamma$ collider \cite{ggcol} is
\beq
{\cal L}^{\gamma \gamma} \approx 0.3^2{\cal L}^{e e}_{\rm geom}
\approx 0.1{\cal L}^{e e}_{\rm geom},
\eeq
where ${\cal L}^{e e}_{\rm geom}$ is the luminosity of $e^+ e^-$ collider.
Taking $100~{\rm fb}^{-1}$ as a nominal integrated luminosity in the
$\gamma\gamma$ mode, we can infer from Figs.~\ref{aaa} and \ref{bbb} that
the maximum number of events for the CP-odd Higgs boson is approximately
100 (10) per a year for $\tan\beta=3$ ($\tan\beta=10$), when the
unpolarized cross section does not contain chargino-loop contributions.
However, the chargino-loop contributions enhance the maximum number of
events as approximately 880 (710) for $\tan\beta=3$ ($\tan\beta=10$).
Hence, the chargino loop contributions for the production of the would-be
CP-odd Higgs boson can be significant at the $\gamma \gamma$ collider for
larger $\tan\beta$.

\begin{figure}
\centerline{\epsfxsize=7cm \epsfbox{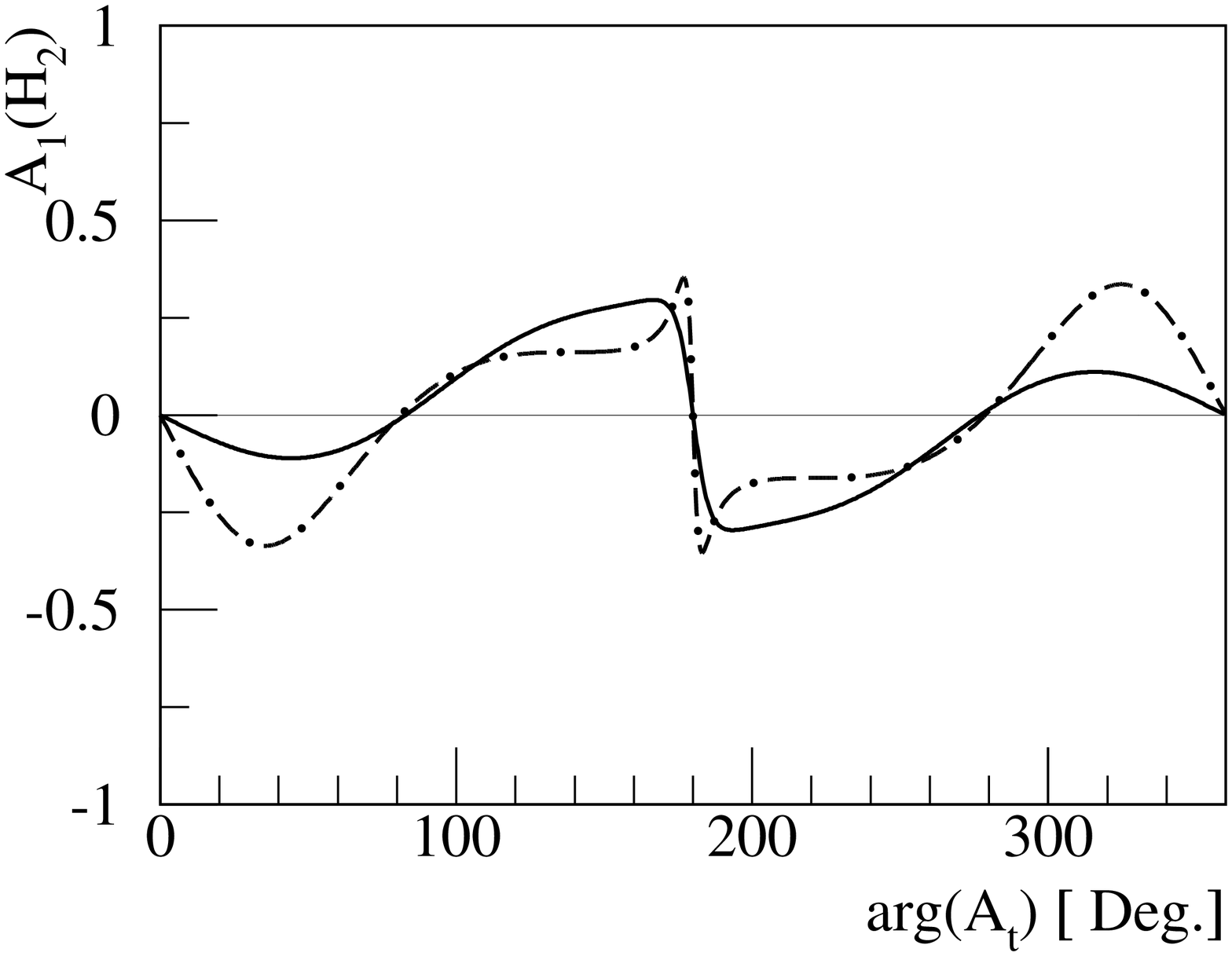}
\epsfxsize=7cm \epsfbox{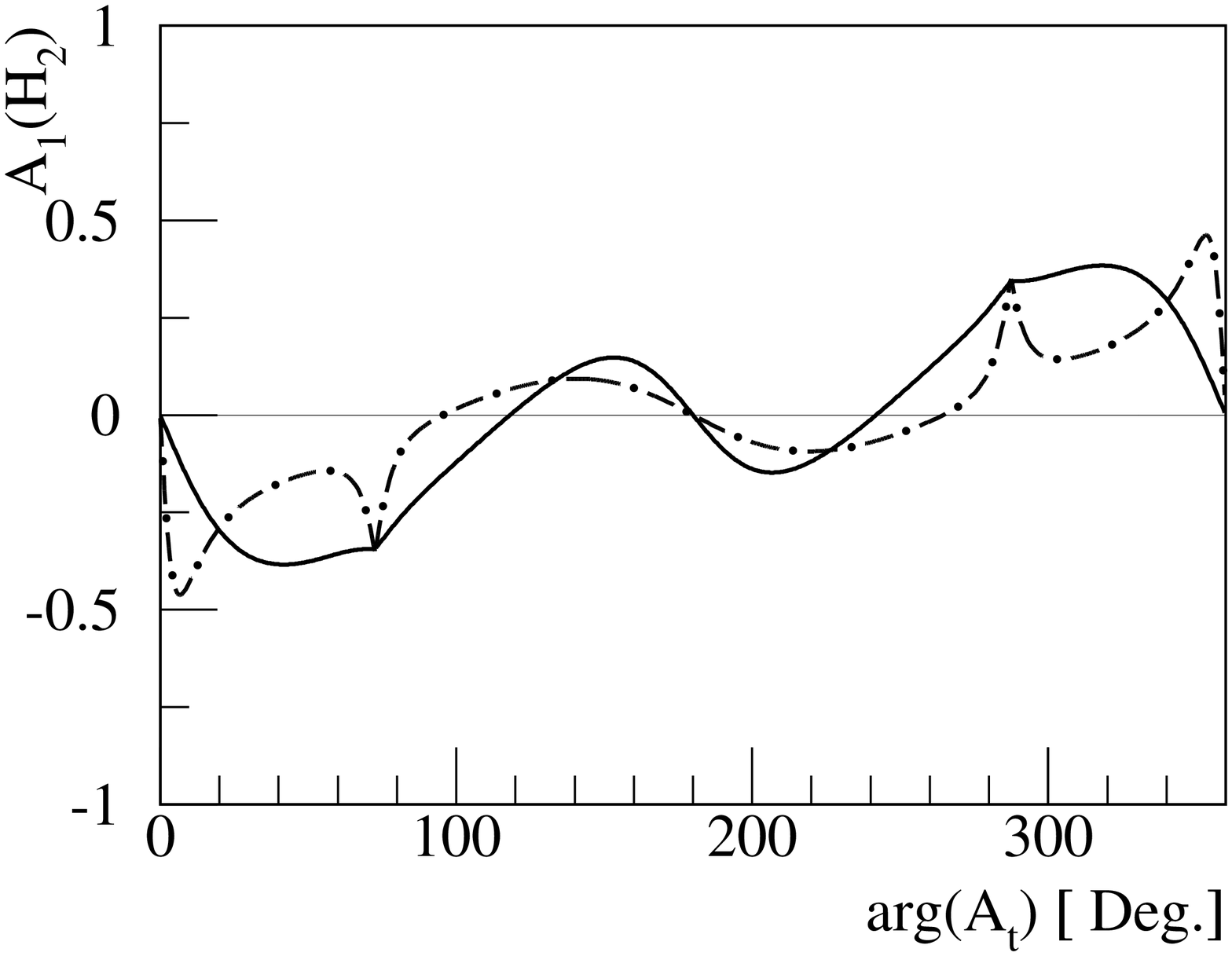}}
\centerline{\epsfxsize=7cm \epsfbox{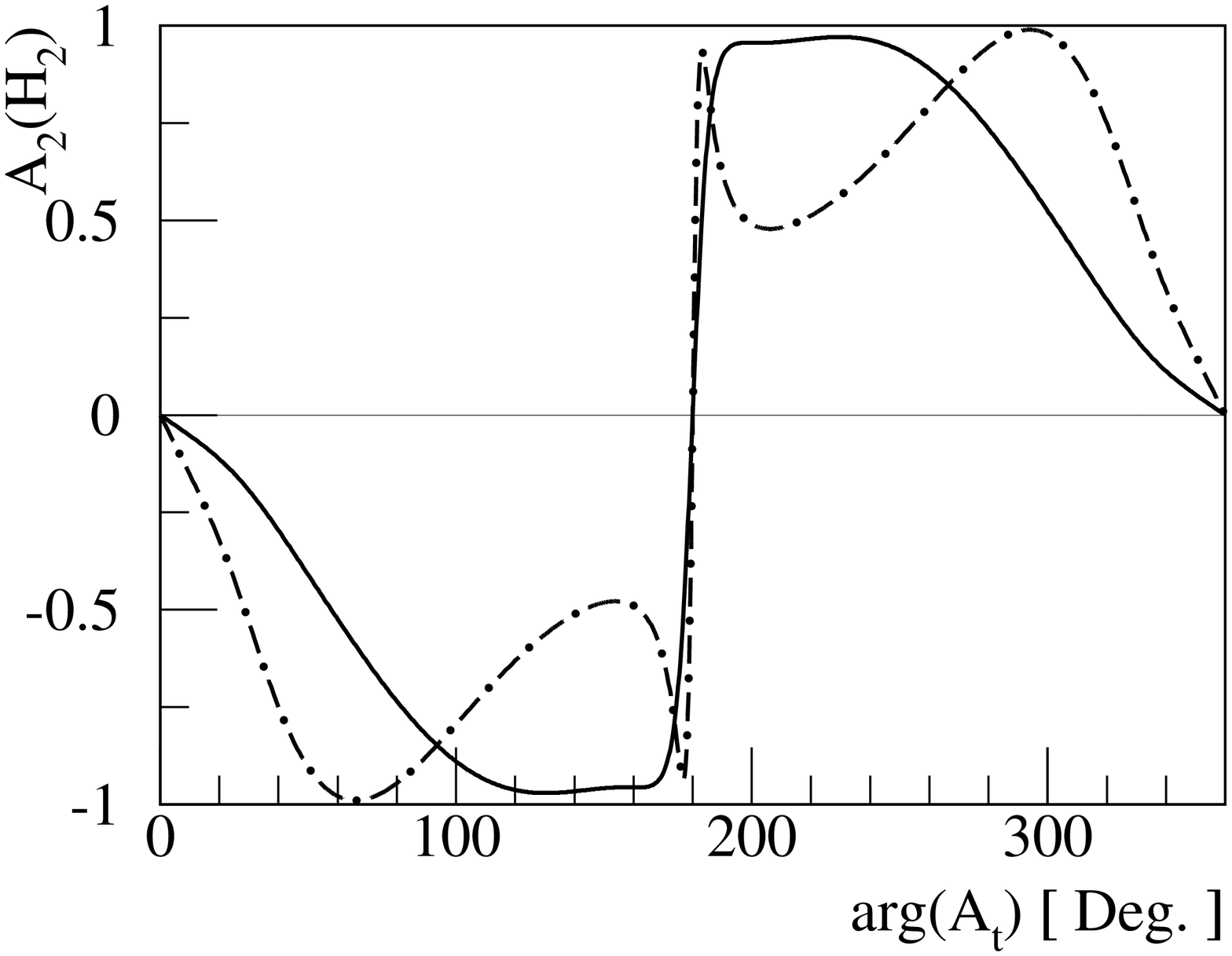}
\epsfxsize=7cm \epsfbox{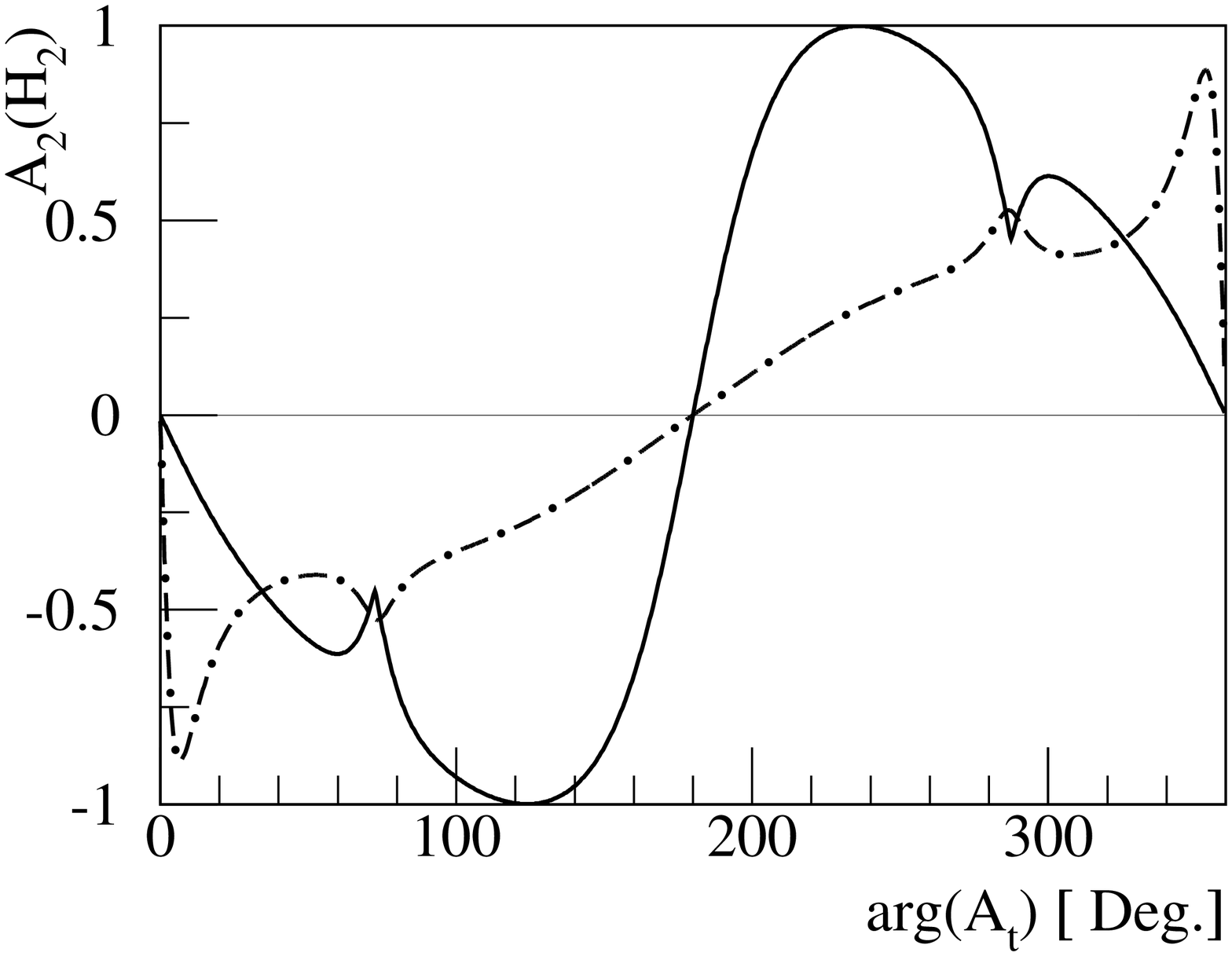}}
\centerline{\epsfxsize=7cm \epsfbox{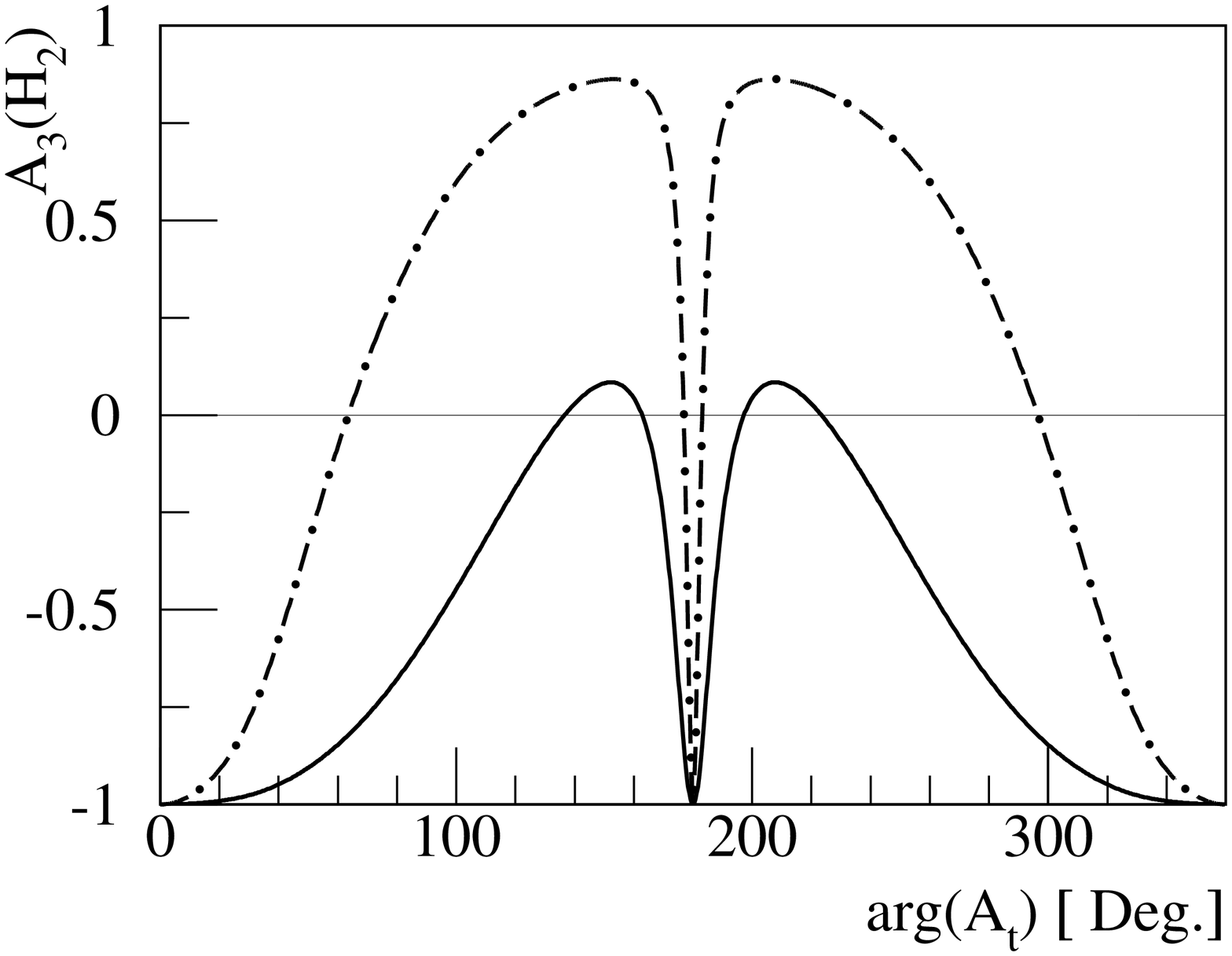}
\epsfxsize=7cm \epsfbox{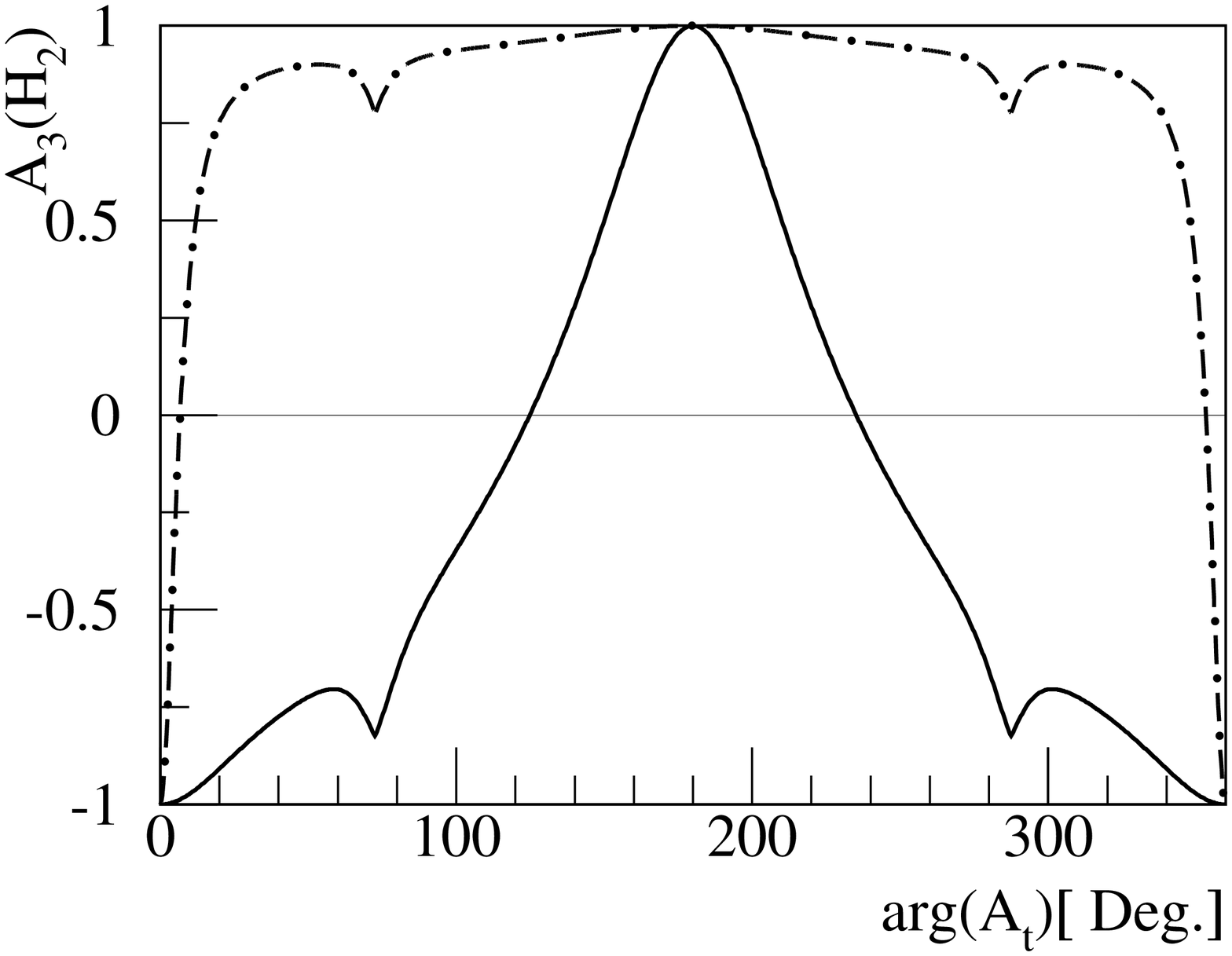}}
~~~~
\caption{ The polarization asymmetries ${\cal A}_1$, ${\cal A}_2$
and ${\cal A}_3$ without (dash-dotted curve) and with (solid curve)
chargino loop contributions as functions of ${\rm arg}(A_t)$.
We take the parameter set (\ref{param}) and $M_{H^+}=300$ GeV.
The left (right) figure is for $\tan\beta=3$ ($\tan\beta=10$).}
\label{eee}
\end{figure}

In Fig.~\ref{eee}, we show three polarization asymmetries of $H_2$
as functions of ${\rm arg}(A_t)$ for $\tan\beta = 3$ (the left column) and
$\tan\beta = 10$ (the right column). As in Figs.~1 and 2, we set $M_{H^+} =
300$ GeV so that $M_{H_2} = 291~(290)$ GeV for $\tan\beta = 3 ~(10)$,
respectively.  The case with (without) the chargino
loop is represented by solid (dash-dotted) curves. We have fixed $M_2 = 150$
GeV as before. The polarization asymmetries ${\cal A}_i (\Phi)$'s satisfy the
following relations:
\bea
{\cal A}_{1,2}(\Phi) = - {\cal A}_{1,2}(360^\circ-\Phi),~~~~~~~
{\cal A}_{3}(\Phi) = + {\cal A}_{3}(360^\circ-\Phi),
\eea
where $\Phi={\rm arg}(A_t \mu) + \xi$ with $\xi=0$. Namely, ${\cal A}_{1,2}$
are CP-odd observables (antisymmetric about $\Phi=180^\circ$) and
${\cal A}_3$ is a CP-even observable (symmetric about $\Phi=180^\circ$).
Note that the chargino loops not only enhance the cross section but also
affect the polarization asymmetries by significant amounts.

\begin{figure}
\centerline{\epsfxsize=6cm \epsfbox{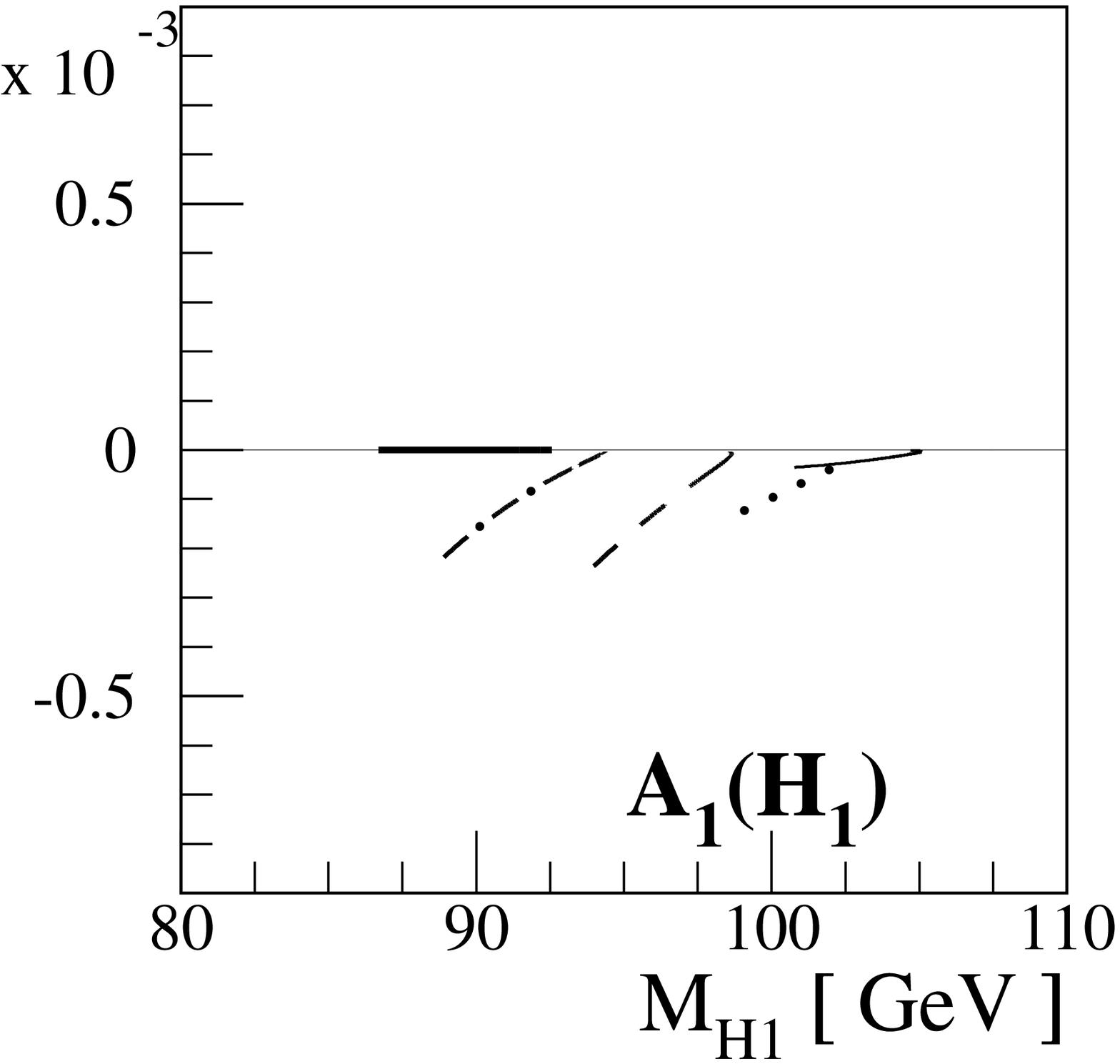}
\epsfxsize=6cm \epsfbox{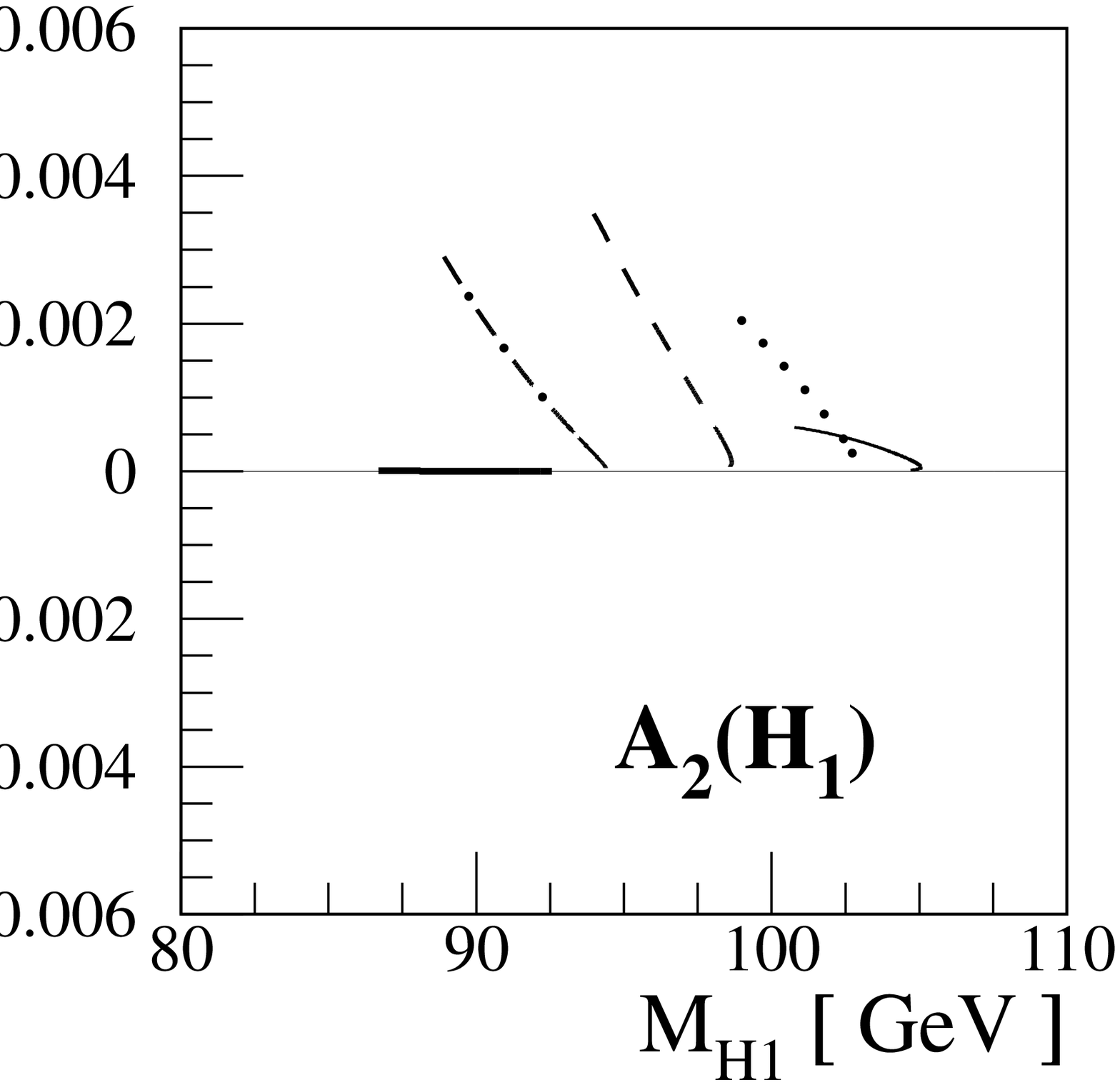}
\epsfxsize=6cm \epsfbox{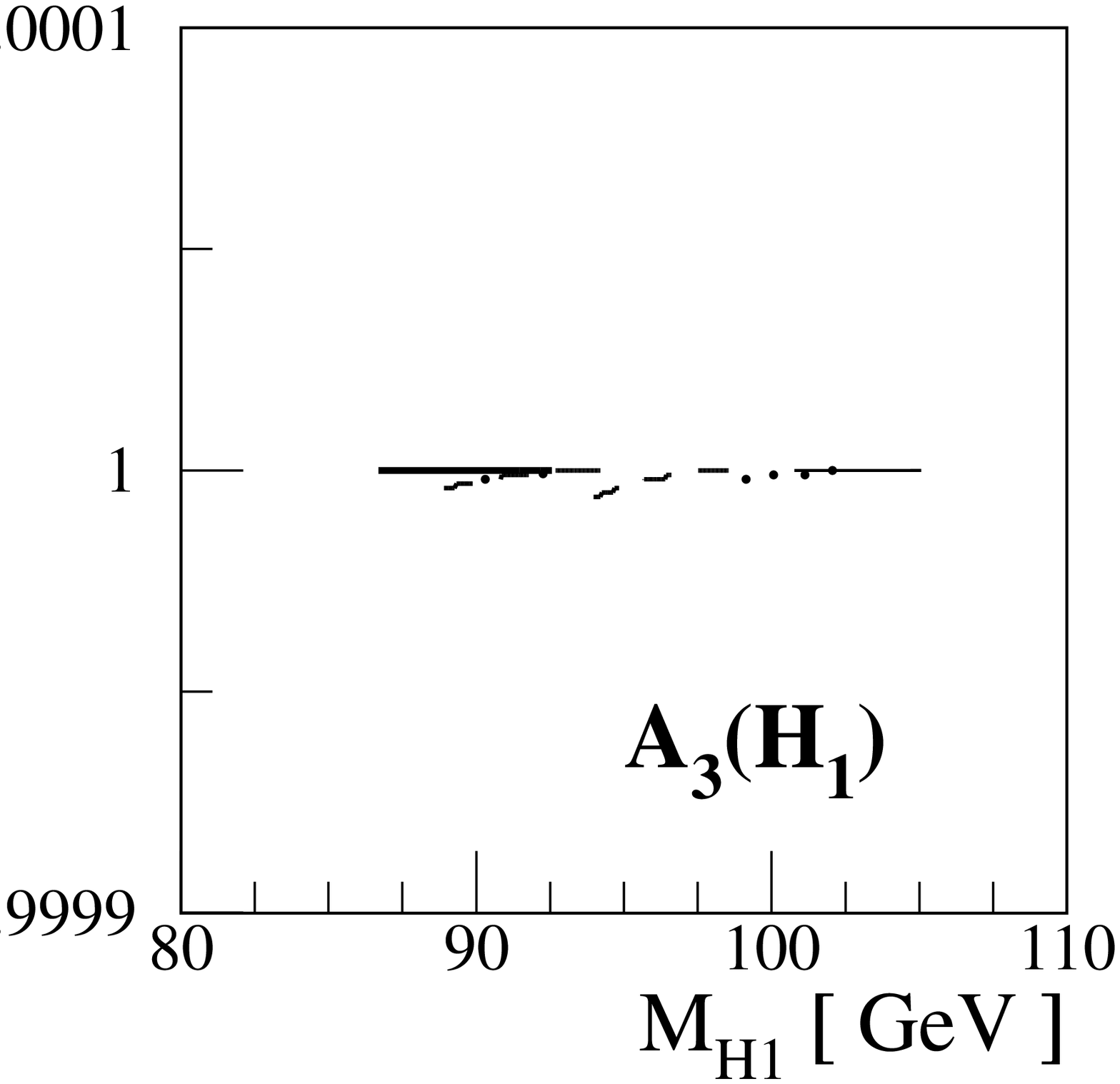}}
\centerline{\epsfxsize=6cm \epsfbox{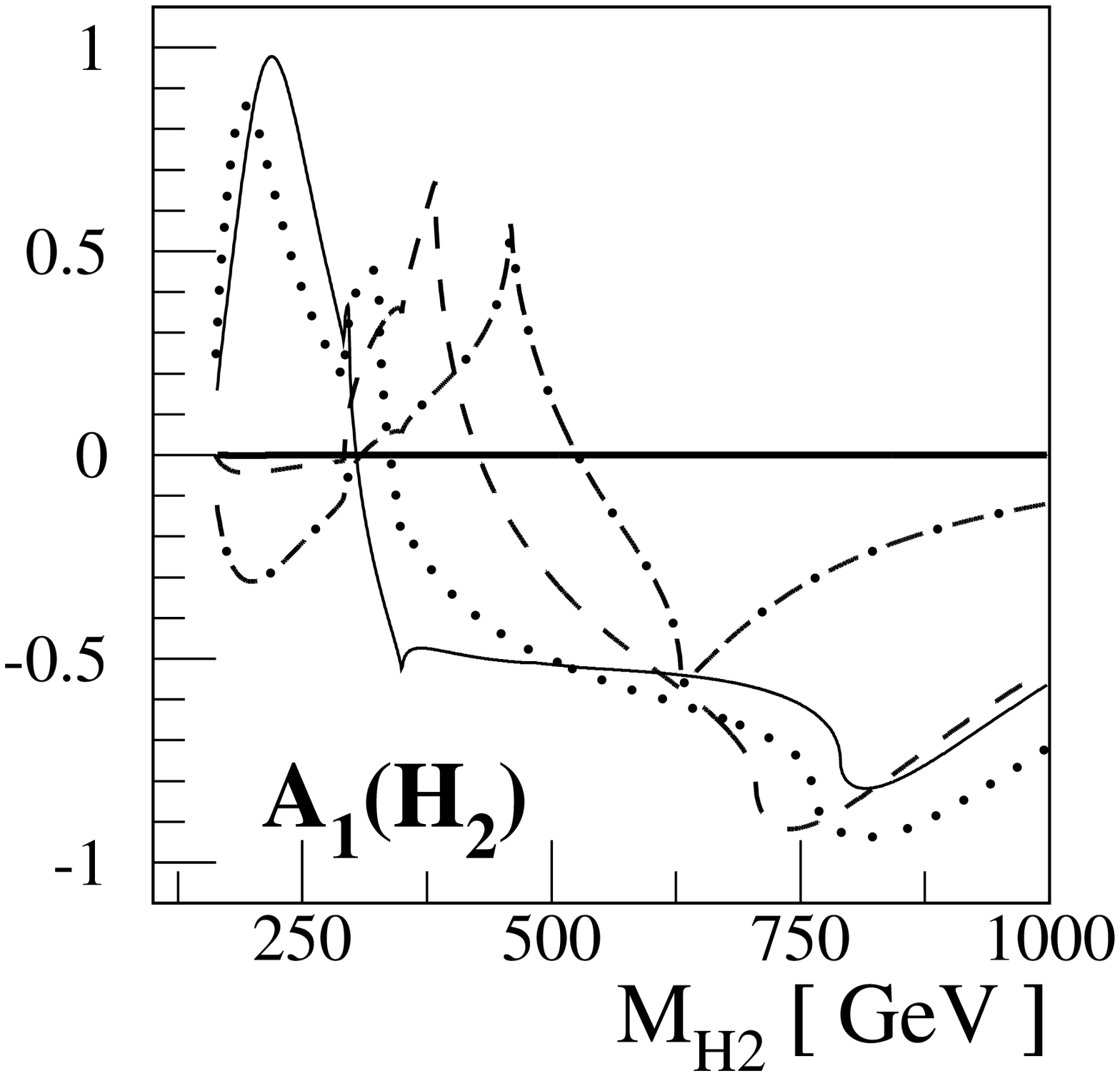}
\epsfxsize=6cm \epsfbox{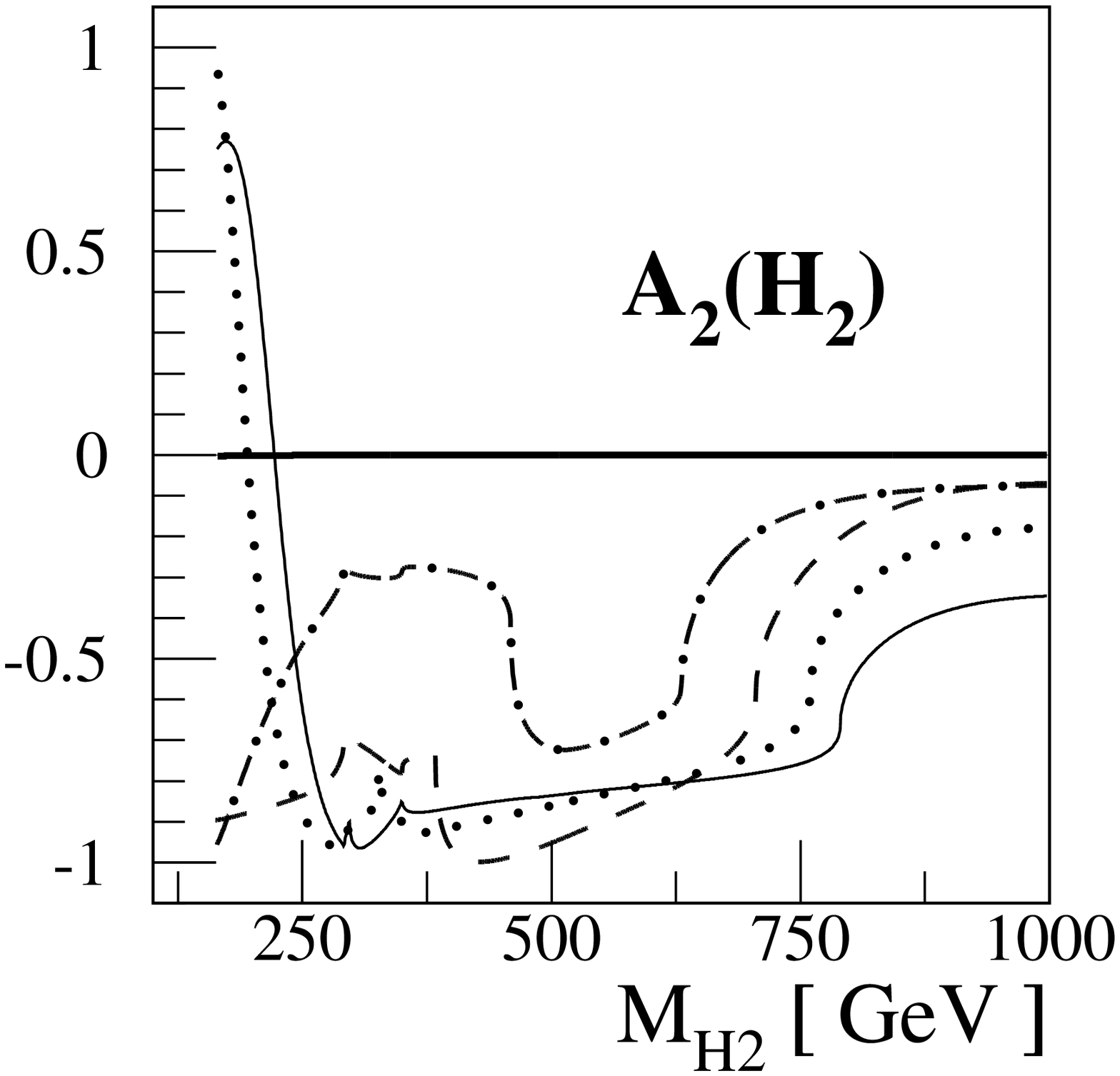}
\epsfxsize=6cm \epsfbox{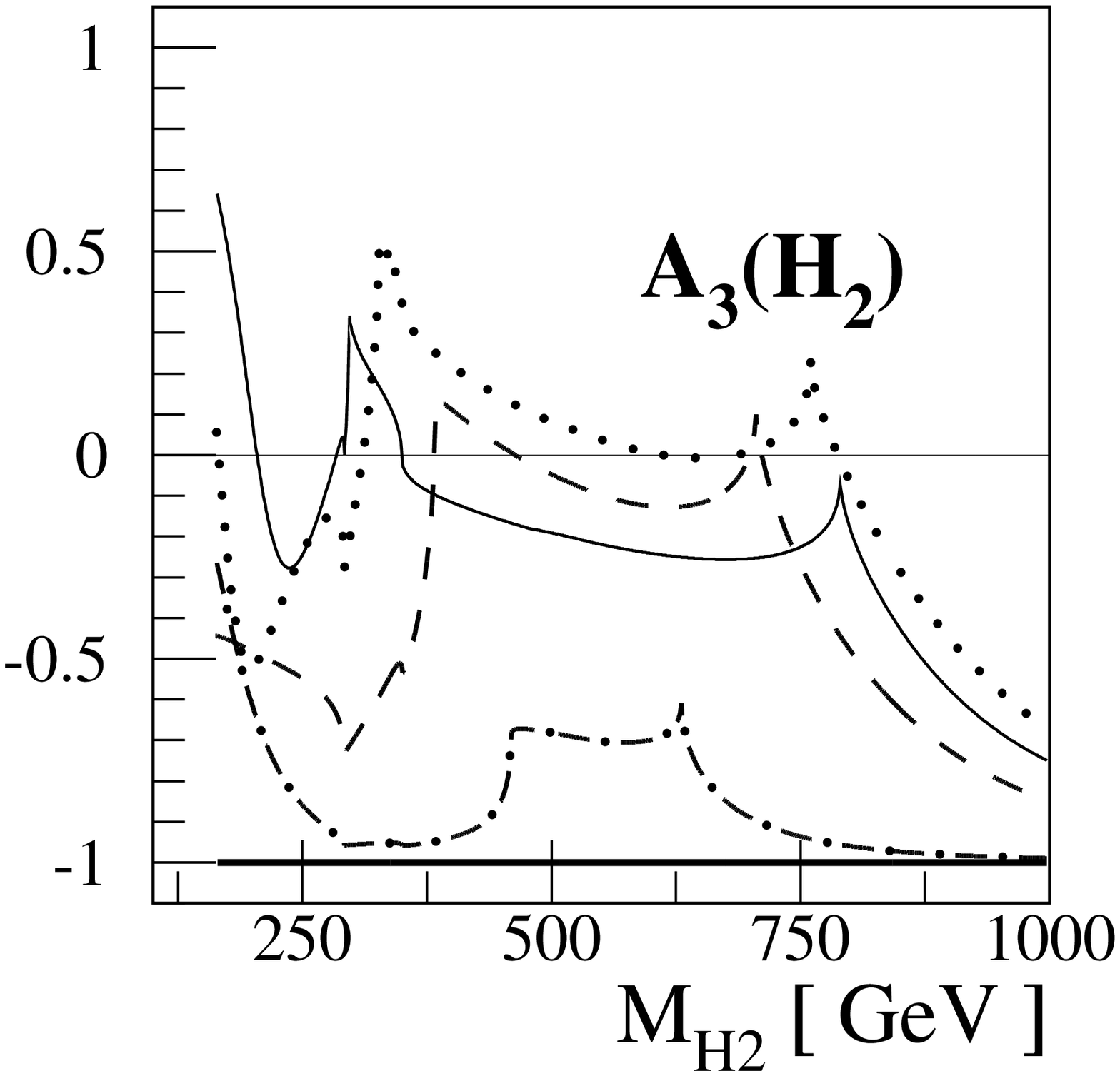}}
\centerline{\epsfxsize=6cm \epsfbox{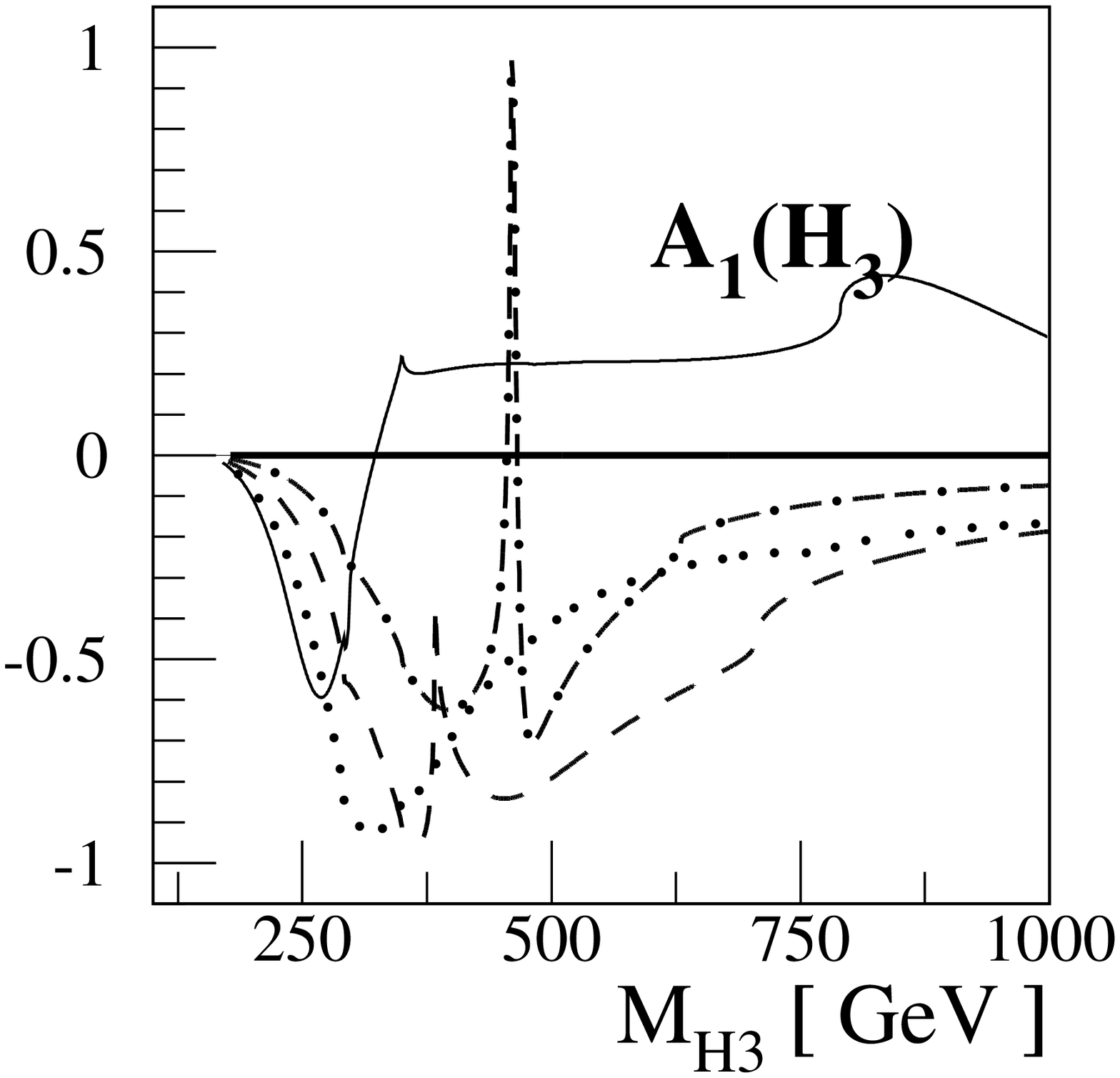}
\epsfxsize=6cm \epsfbox{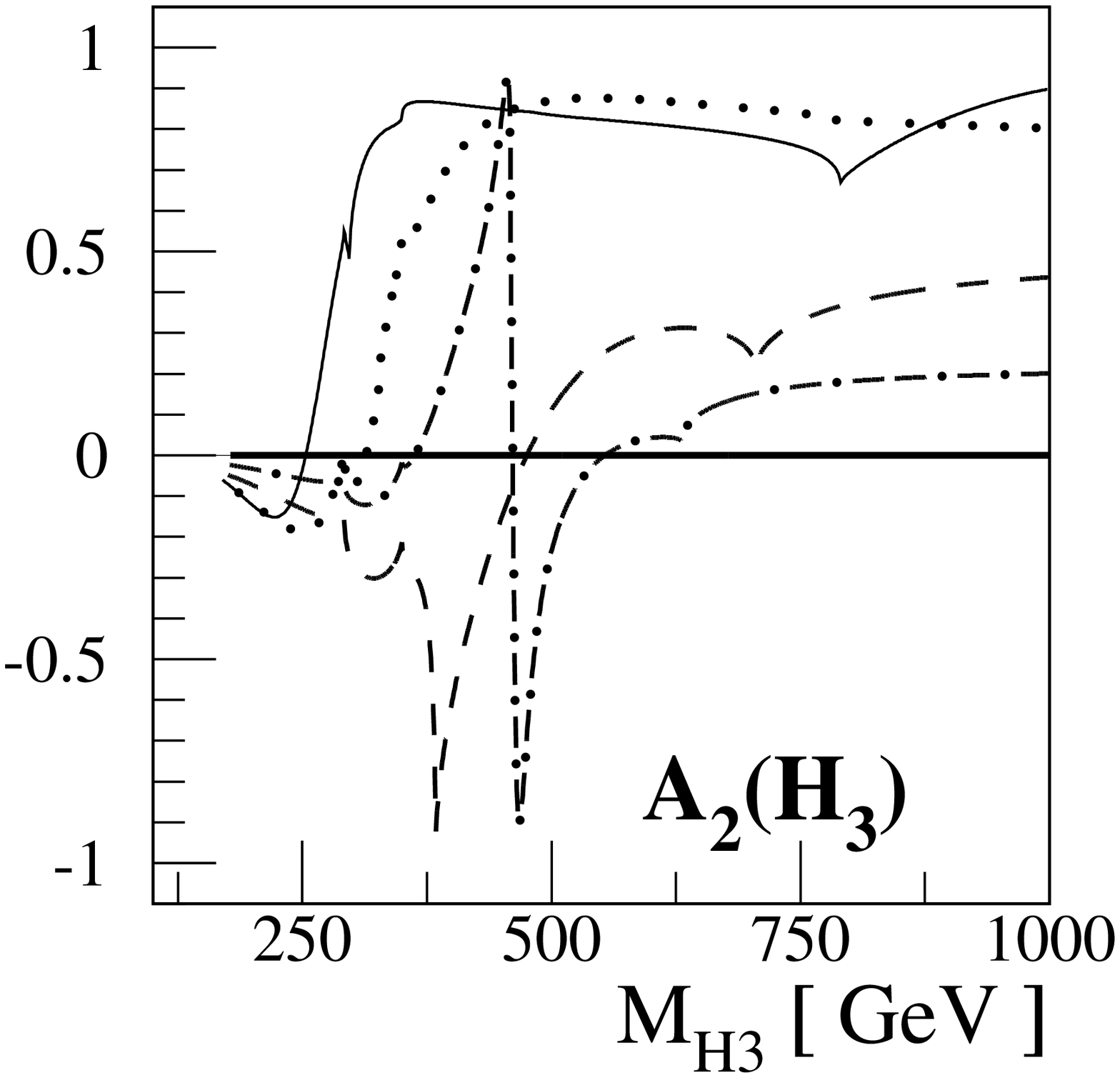}
\epsfxsize=6cm \epsfbox{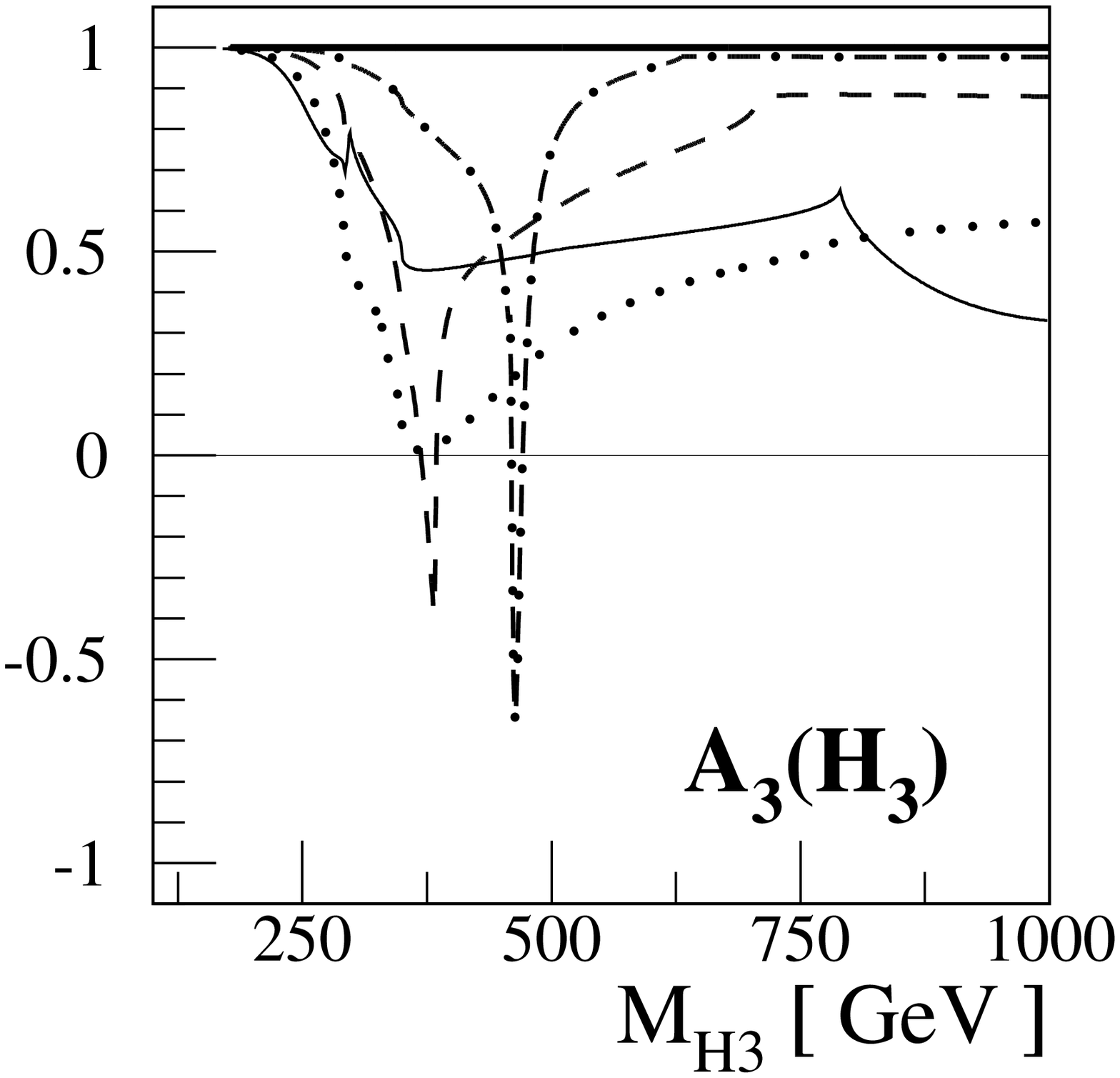}}
~~~~
\caption{The polarization asymmetries ${\cal A}_1$, ${\cal A}_2$
and ${\cal A}_3$ with chargino loop contributions as functions of
each Higgs mass for five different values of the $A_t$ phase with
${\rm arg}(\mu)=0^\circ$; ${\rm arg}(A_t)=0^\circ$ (thick solid curve),
$40^\circ$ (dash-dotted curve), $80^\circ$ (dashed curve),
$120^\circ$ (dotted curve) and $160^\circ$ (solid curve).
We choose the parameter set (\ref{param}) for $\tan\beta =3$.}
\label{fff}
\end{figure}

In Fig.~\ref{fff}, we show the polarization asymmetries ${\cal A}_a (H_j)$
as functions of the neutral Higgs masses for ${\rm arg}(A_t) = 0^{\circ}$,
$40^{\circ}$, $80^{\circ}$, $120^{\circ}$ and $160^{\circ}$ with
$\tan\beta=3$, including all the charged particles in the loops.
The lightest Higgs boson $H_1$ still behaves like a CP-even scalar, since
$-0.03\% \lesssim {\cal A}_1 \le 0$, $0 \le {\cal A}_2 \lesssim 0.4\%$,
and ${\cal A}_3 \simeq 1$.  On the other hand, the heavier $H_2$ and $H_3$
are generically admixtures of CP-even and CP-odd states if the phase of
$A_t$ does not vanish. For $H_2$ and $H_3$, chargino, top and stop loops
give main contributions to the asymmetries above the chargino-pair threshold,
but the chargino and $W^\pm$ loop contributions affect them below the
chargino-pair threshold.

\begin{figure}
\centerline{\epsfxsize=6cm \epsfbox{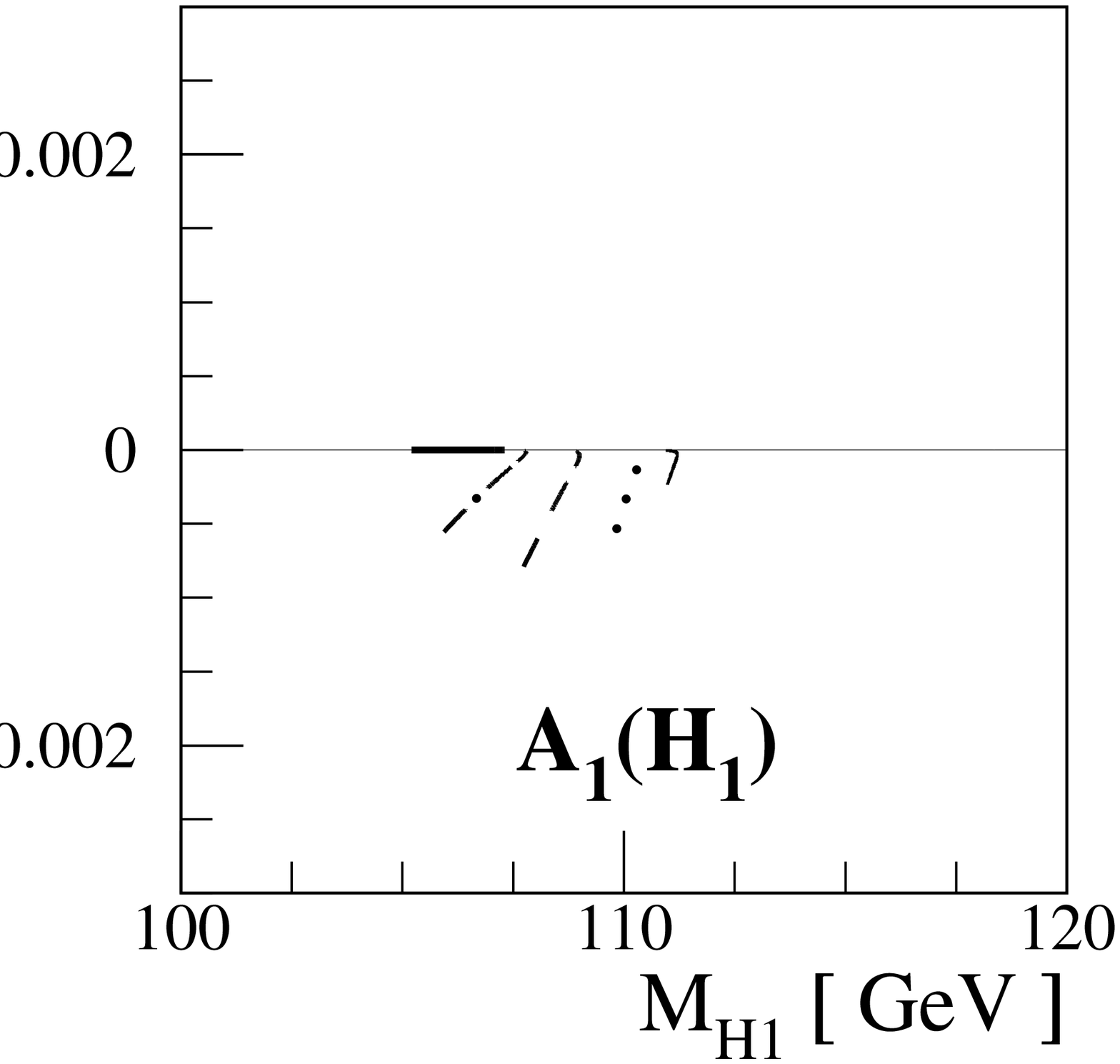}
\epsfxsize=6cm \epsfbox{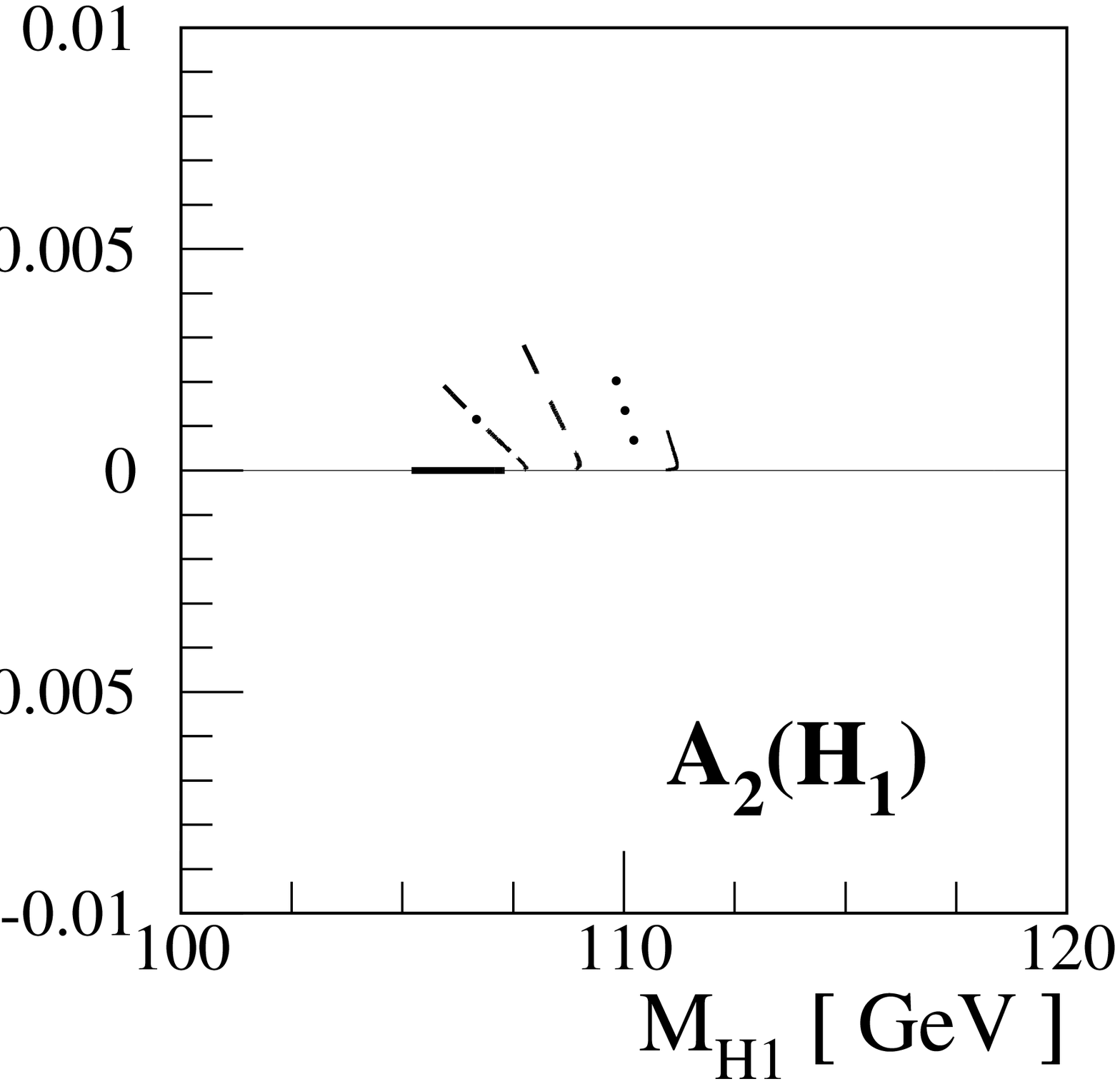}
\epsfxsize=6cm \epsfbox{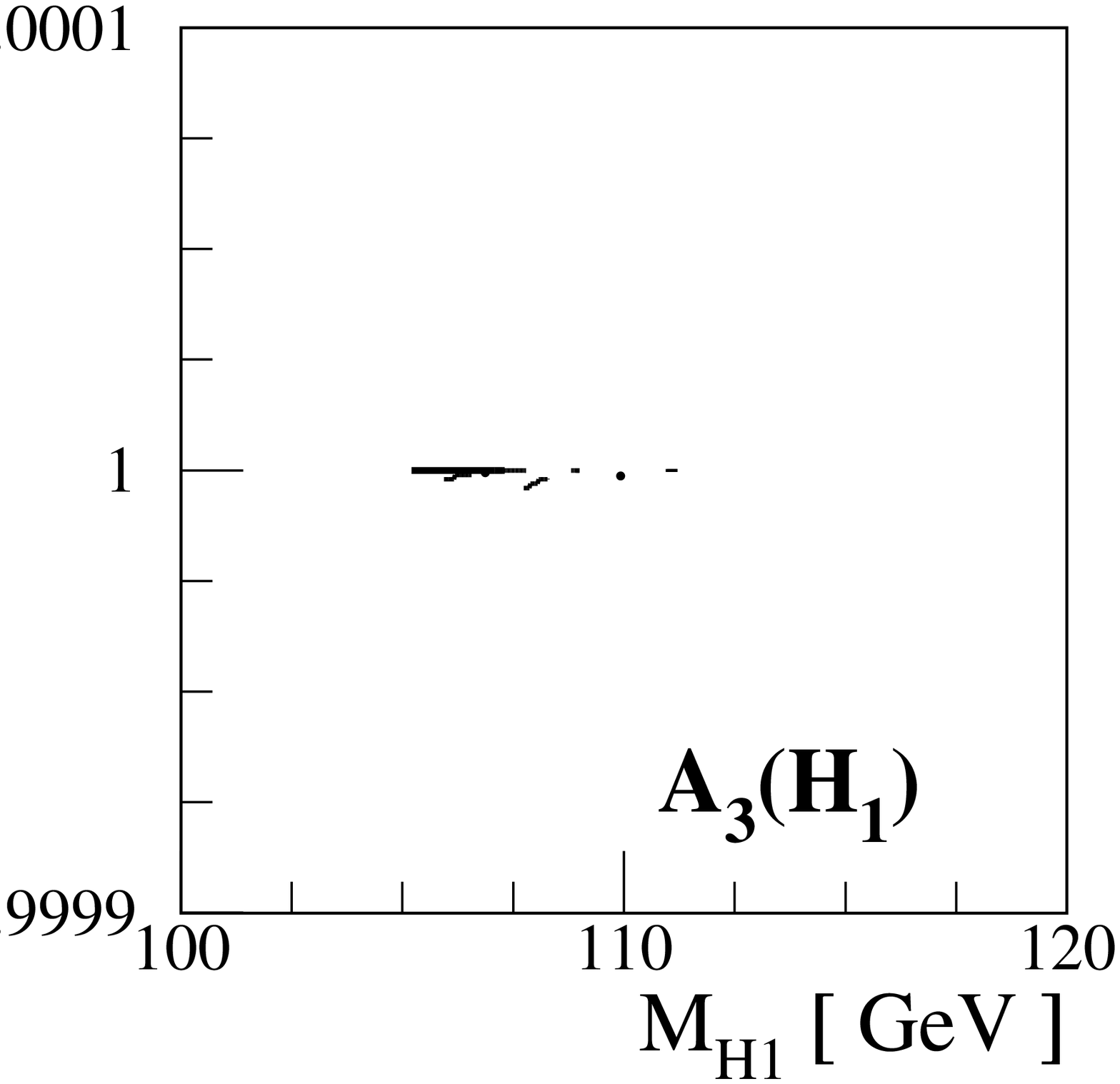}}
\centerline{\epsfxsize=6cm \epsfbox{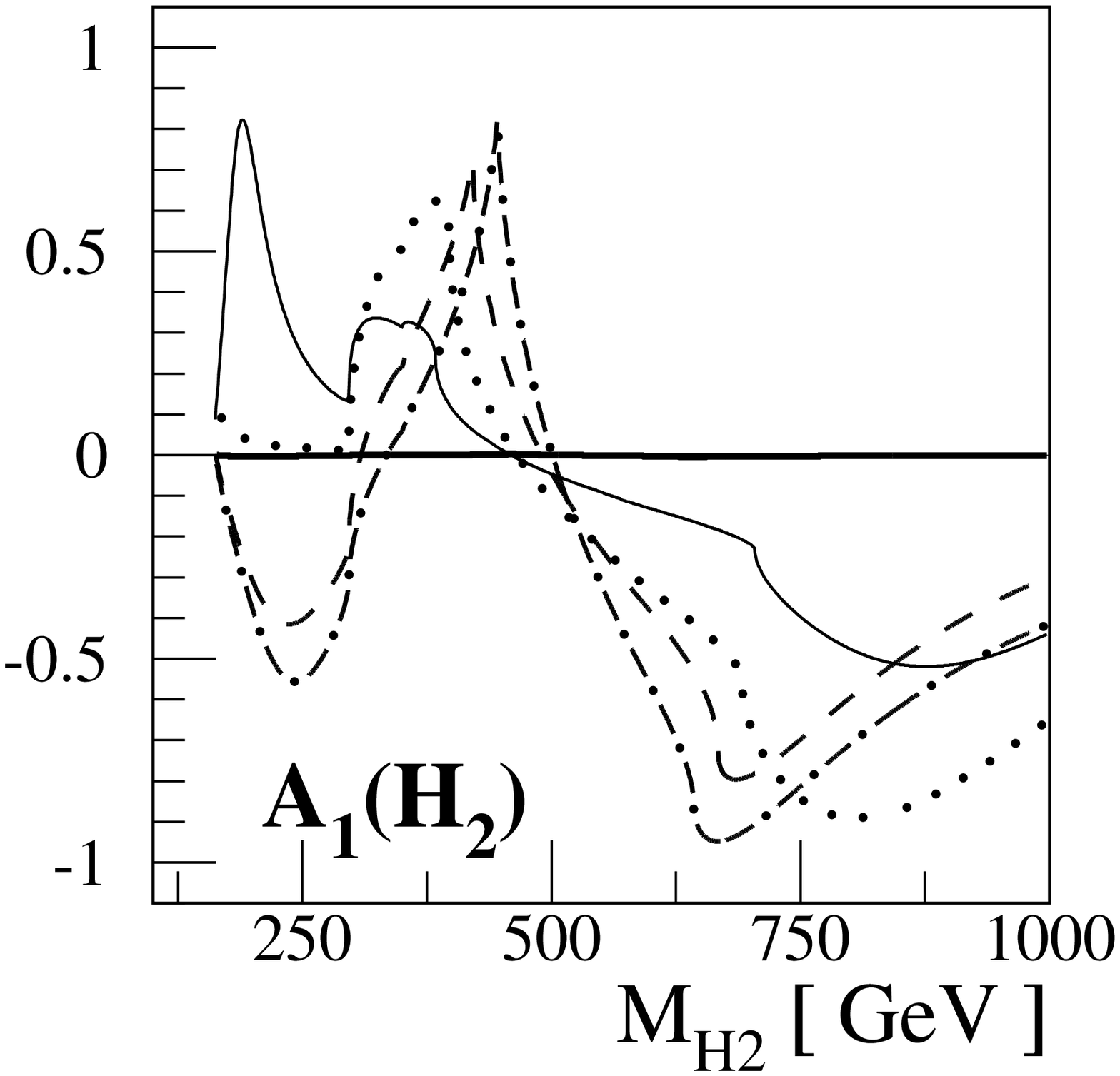}
\epsfxsize=6cm \epsfbox{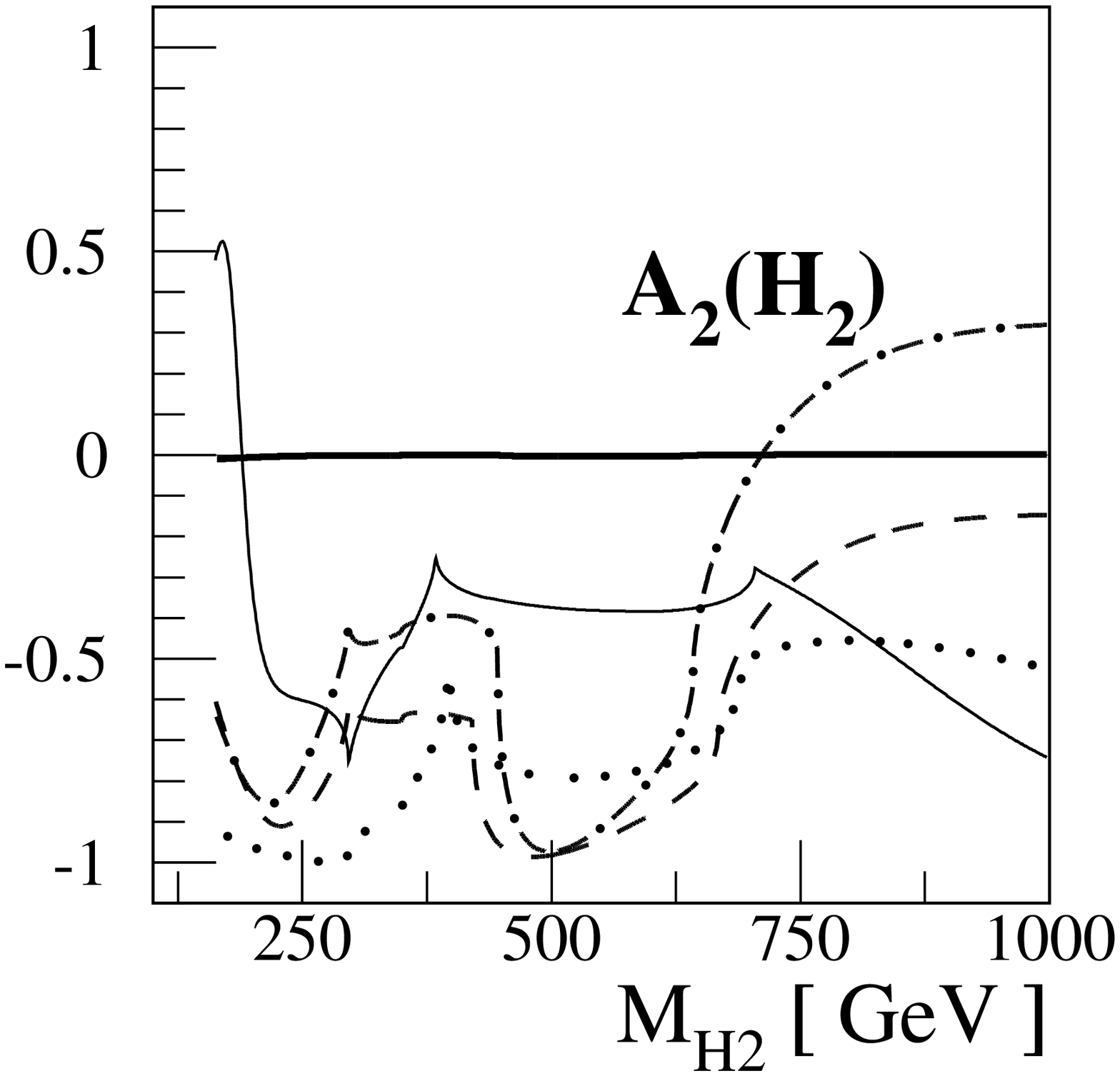}
\epsfxsize=6cm \epsfbox{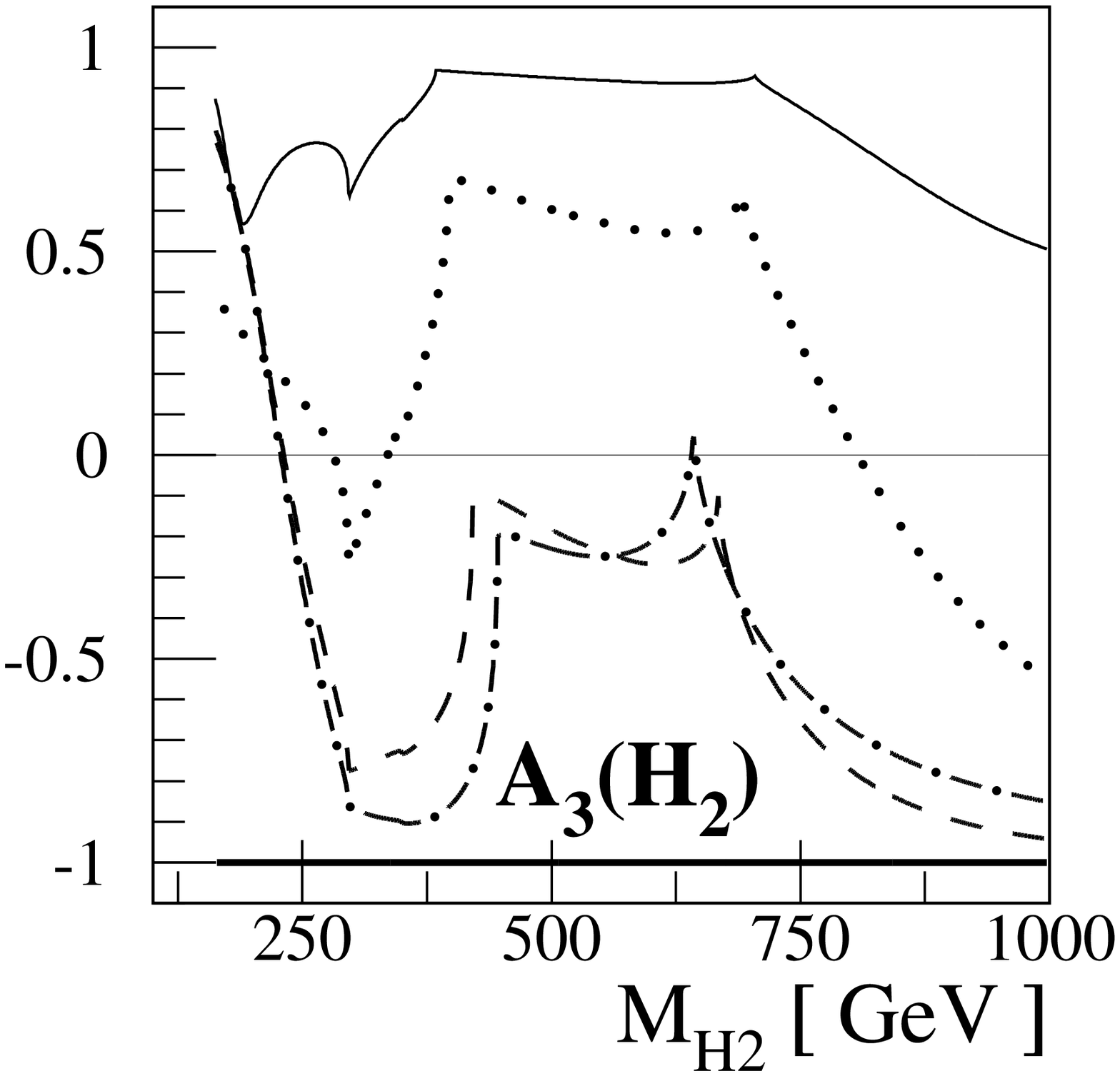}}
\centerline{\epsfxsize=6cm \epsfbox{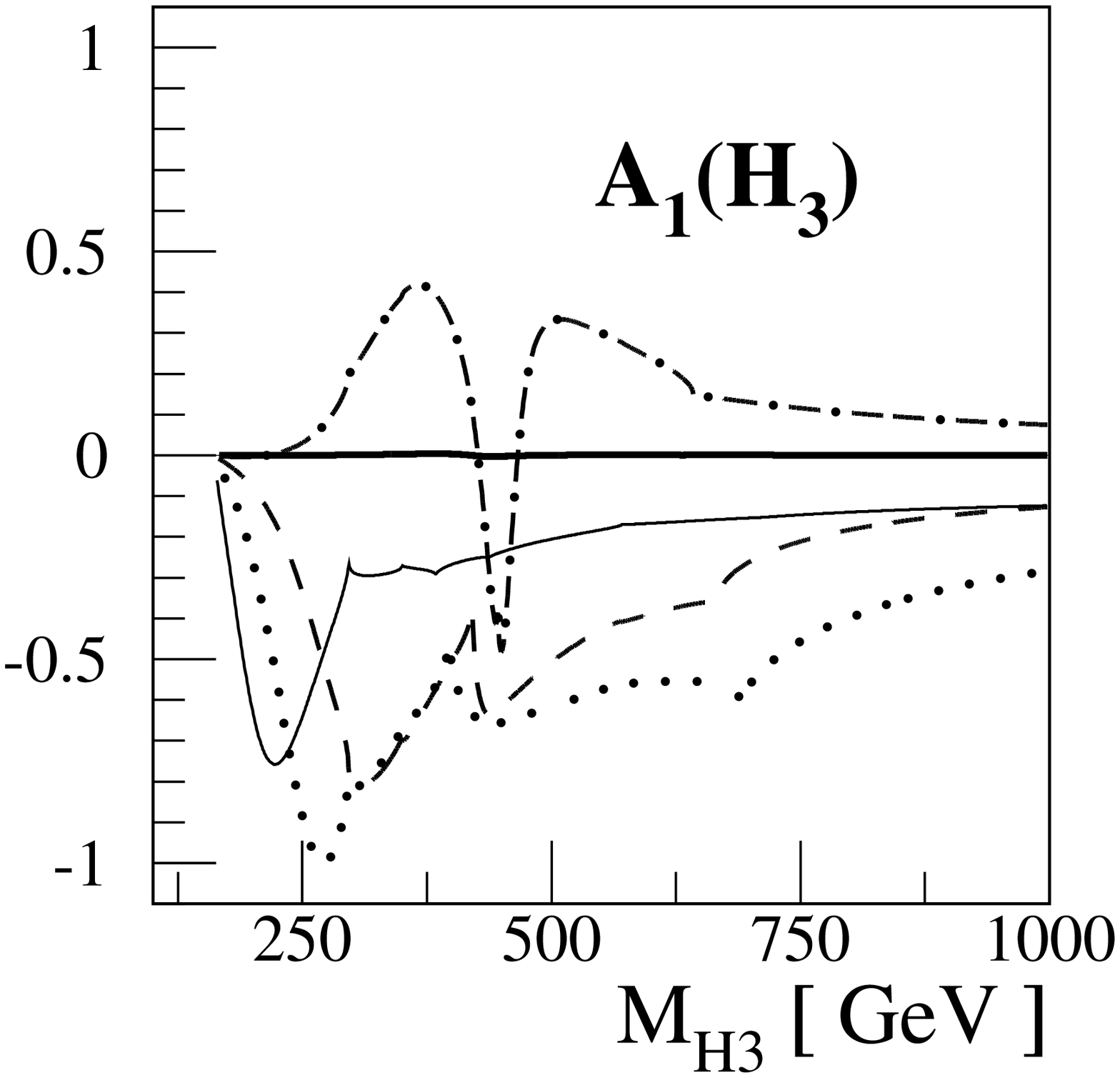}
\epsfxsize=6cm \epsfbox{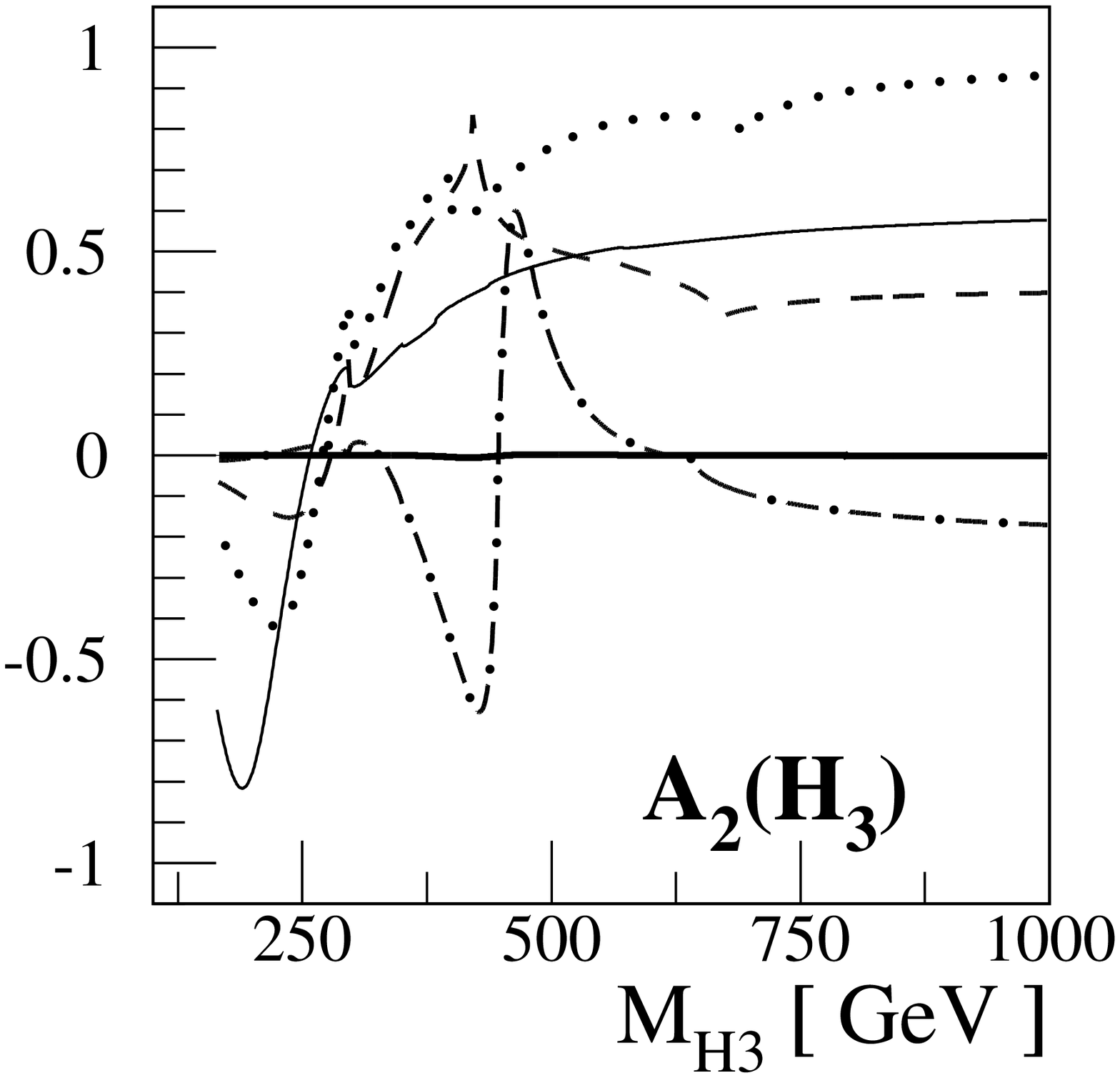}
\epsfxsize=6cm \epsfbox{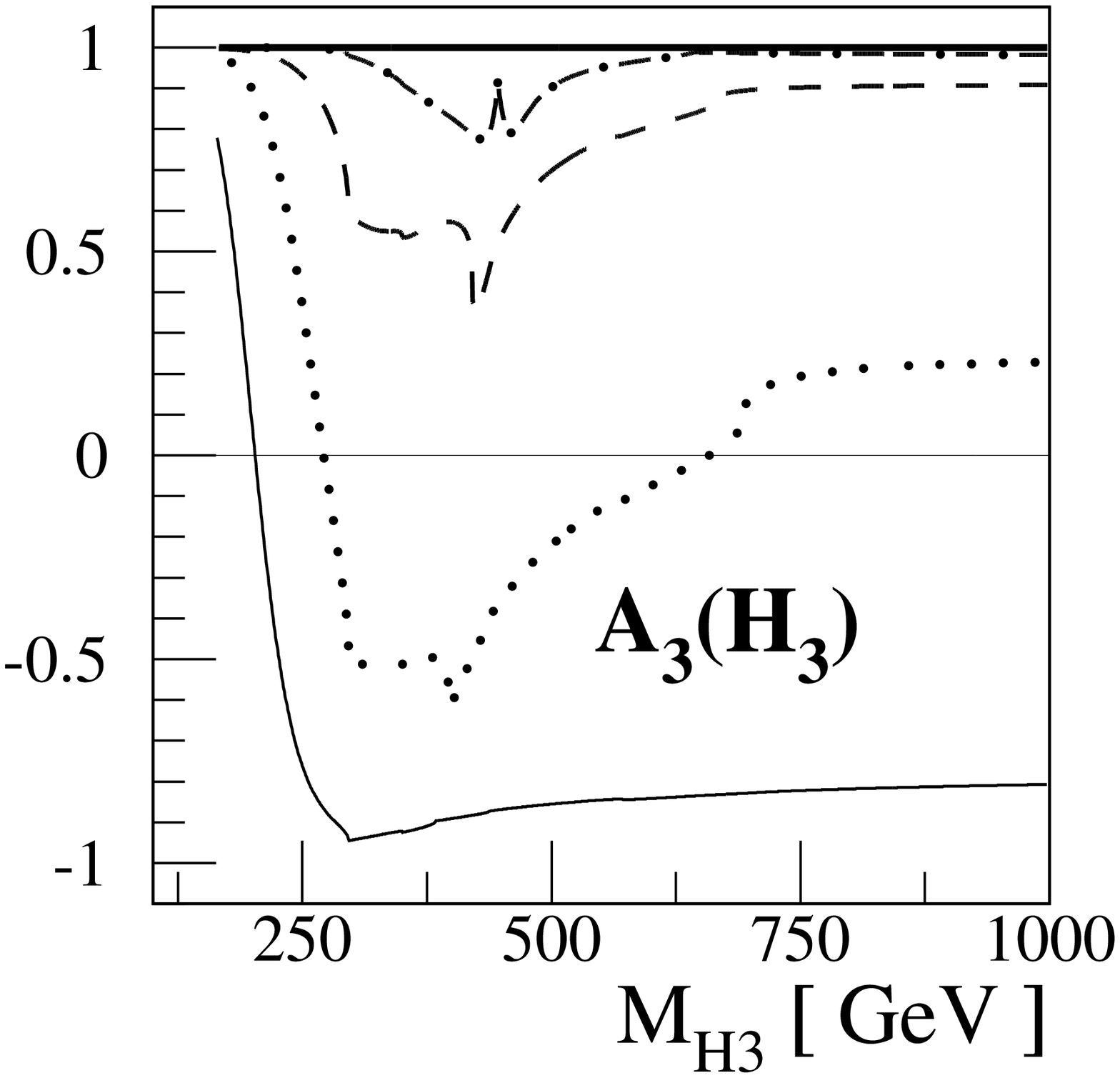}}
~~~~
\caption{The polarization asymmetries ${\cal A}_1$, ${\cal A}_2$
and ${\cal A}_3$ with chargino loop contributions as functions of
each Higgs mass for five different values of the $A_t$ phase;
${\rm arg}(A_t)=0^\circ$ (thick solid curve),
$40^\circ$ (dash-dotted curve), $80^\circ$ (dashed curve), $120^\circ$
(dotted curve) and $160^\circ$ (solid curve). %
We take the parameter set (\ref{param}) for $\tan\beta =10$.}
\label{ggg}
\end{figure}

In Fig.~\ref{ggg}, we present the polarization asymmetries for
$\tan\beta=10$. Again, the lightest Higgs boson $H_1$ behaves like a
CP-even scalar for the larger $\tan\beta$, since
$-0.1\% \lesssim {\cal A}_1 \le 0$, $0 \le {\cal A}_2 \lesssim 0.3\%$,
and ${\cal A}_3 \simeq 1$.  If $\tan\beta$ becomes larger, the top (stop)
loop contribution is accompanied by the bottom (sbottom) contribution to the
polarization asymmetries of the heavier Higgs bosons $H_2$ and $H_3$.
Fig.~\ref{ggg} indicates that as the CP violating phase ${\rm arg}(A_t)$
increases for the case of large $\tan\beta$, the value of the asymmetry
${\cal A}_3$ of $H_2$ approaches that of $H_3$ at ${\rm arg}(A_t)=0^\circ$
and vice versa, i.e., the CP-properties of the heavier Higgs bosons
$H_2$ and $H_3$ are interchanged.

From Figs.~\ref{fff} and \ref{ggg}, the polarization asymmetry
${\cal A}_2 (H_1)$ is the most sensitive CP observable in
detecting the CP violation of the lightest Higgs boson for both
small and large $\tan\beta$, when the chargino contributions are
included. This result is different from the first paper of
Ref.~\cite{SYChoi}, where charginos are neglected by assuming they
are very heavy, and thus ${\cal A}_2$ (${\cal A}_1$) is the most
powerful CP observable for $\tan\beta=3~(\tan\beta=10)$.
Unfortunately the asymmetry itself is very small so that it would
not be easy to find nonzero ${\cal A}_2 (H_1)$.  Still asymmetries
for heavier neutral Higgs bosons can be sizable and thus be used
as the probes of Higgs sector CP violation if they can be produced
with high statistics at NLCs. Therefore, we need to prepare the
colliding photon beams with large linear polarizations as well as
high center of mass energy $\sqrt{s_{\gamma\gamma}}$ in order to
produce neutral Higgs bosons and determine their CP properties in
a model independent manner.

\section{Conclusions}

In this work, we presented a comprehensive analysis of the neutral
Higgs boson productions through $\gamma \gamma \rightarrow
H_{i=1,2,3}$ in the presence of the Higgs sector CP-violation of
the MSSM. In particular, we have included the chargino loop
contributions as well as the contributions from squarks, $W^\pm$
and charged Higgs particles. In many scenarios of SUSY breaking,
charginos are not too heavy that their effects are generically
important. First of all, the production of the would-be CP-odd
$H_2$ boson is enhanced by an order of magnitude when chargino
loop contributions are included even without the Higgs sector CP
violation. If the phase of the $A_t$ parameter is turned on, CP
violation in the Higgs sector become very rich in the structures.
This is also true of the case of the heaviest Higgs boson $H_3$.
Also the polarization asymmetries are affected by the Higgs sector
CP violation.

If the $A_t$ parameter has a large CP violating phase, its effects can appear
in various physical observables: the Higgs sector CP-violation as discussed
in this work, and also the direct CP violation in
$B\rightarrow X_s \gamma$ \cite{baek}, for example.
Since the latter is an indirect signature, it is important to probe SUSY CP
violation in a direct way. Thus it is important to probe CP violation from
the soft SUSY breaking sector such as arg($A_t$) in the Higgs sector CP
violation by using $\gamma\gamma$ colliders as discussed in this work.
In this regards, the $\gamma\gamma$ mode at NLC with high
$\sqrt{s_{\gamma\gamma}}$ and luminosity, and high quality beam
polarizations will be indispensable for this purpose by measuring the
cross sections of $\gamma \gamma \rightarrow H_i$ ($i=1,2,3$) and three
asymmetries ${\cal A}_a (H_j)$ ($a,j=1,2,3$) in the MSSM.

\acknowledgements

We are grateful to S.Y. Choi and Jae Sik Lee for useful communications and
to W.Y. Song for discussions.
This work was supported in part by BK21 Haeksim program of MOE, and by
KOSEF through CHEP at Kyungpook National University.

\appendix
\section{Charginos and scalar tops}
The chargino mass matrix in the $(\tilde{W}^+, \tilde{H}^+)$ basis
\cite{Haber} is
\begin{equation}
{\cal M_C} =
\left(
\begin{array}{cc}
{M}_2 & \sqrt{2} e^{-i \xi} m_W \cos \beta
\\
\sqrt{2} m_W \sin \beta & \mu
\end{array}
\right),
\end{equation}
where ${M}_2 > 0$ and $\mu$ are gaugino and Higgsino masses, and
$e^{+i \xi}$ is the the phase of the up-type Higgs VEV \cite{PiWa}.
Since the mass matrix $X$ is a general complex matrix, it is diagonalized
by a biunitary transformation:
\begin{equation}
\label{M_D}
U^* X V^{-1} \equiv {\rm diag}(M_{\tilde{\chi}^-_1}, M_{\tilde{\chi}^-_2}),
\end{equation}
with $M_{\tilde{\chi}^-_2} \ge M_{\tilde{\chi}^-_1} \ge 0$.
In order for $M_{\tilde{\chi}^-_{i=1,2}}$ to be positive, we define the
unitary matrix $U$ as a product of two unitary matrices
\begin{equation}
\label{H}
U \equiv HU'.
\end{equation}
The angles $\theta_1$ and $\phi_1$ of the unitary matrix
\beq
U' = \left(
\begin{array}{cc}
\cos \frac{\theta_1}{2} & \sin \frac{\theta_1}{2} e^{+i\phi_1} \\
-\sin \frac{\theta_1}{2} e^{-i\phi_1} & \cos \frac{\theta_1}{2}
\end{array}
\right),
\eeq
are given by
\bea
\tan \theta_1 &=& \frac{2 \sqrt{2}m_W \left[ M_2^2 \cos^2\beta
+
|\mu|^2\sin^2\beta + M_2 |\mu| \sin2\beta
\cos(\theta_\mu + \xi) \right]^{1/2}}{ M_2^2
-|\mu|^2-2m_W^2 \cos2\beta},
\\
\tan\phi_1 &=& \frac{ |\mu|\sin(\theta_\mu + \xi)\sin\beta}{M_2 \cos\beta
+ |\mu|\cos( \theta_\mu +\xi)\sin\beta},
\eea
where $\theta_\mu={\rm arg}(\mu)$.  The unitary mixing matrix $V$ is
\begin{equation}
V =
\left(
\begin{array}{cc}
\cos \frac{\theta_2}{2} & \sin \frac{\theta_2}{2} e^{-i\phi_2} \\
-\sin \frac{\theta_2}{2} e^{+i\phi_2} & \cos \frac{\theta_2}{2}
\end{array}
\right),
\end{equation}
where
\begin{eqnarray}
\tan \theta_2 &=& \frac{2 \sqrt{2}m_W \left[ M_2^2 \sin^2\beta
+ |\mu|^2\cos^2\beta + M_2 |\mu| \sin2\beta \cos(\theta_\mu
+ \xi) \right]^{1/2}}{M_2^2 - |\mu|^2 + 2 m_W^2 \cos2\beta},
\\
\tan\phi_2 &=& \frac{M_2 \sin\xi \sin\beta
- |\mu|\sin \theta_\mu \cos\beta}{M_2 \cos\xi \sin\beta
+ |\mu|\cos \theta_\mu \cos\beta}.
\end{eqnarray}
By using the unitary matrix $H = {\rm diag}(e^{i\gamma_1}, e^{i \gamma_2})$,
where  $\gamma_{1,2}$ are the phases of the diagonal
elements of $U'^* X V^{-1}$, we finally obtain
\begin{equation}
U^*XV^{-1} = {\rm diag}(M_{\tilde{\chi}^-_1}, M_{\tilde{\chi}^-_2}).
\end{equation}
And the mass eigenvalues of charginos are
\bea
M_{\tilde{\chi}^-_1,\tilde{\chi}^-_2}^2 &=&
\frac{1}{2}\left( M_2^2 + |\mu|^2 + 2m_W^2 \right)
\mp
\frac{1}{2} \left[ \left( M_2^2 - |\mu|^2 \right)^2 + 4 m_W^4 \cos^2
2\beta \right.
\nl
&&
\left.
~~~~~~+ 4 m_W^2 \left( M_2^2+ |\mu|^2 + 2 M_2 |\mu|
\cos\theta_\mu \sin 2\beta \right) \right]^{1/2}.
\eea
Note that the mass eigenvalues and the mixing angles depend on  the CP
violating phases $\xi$ and $\theta_\mu$.

The stop $(\rm{mass})^2$ matrix ${\cal M}_{\tilde t}^2$ \cite{Bae}
is written as \bea \label{stop mass} {\cal L}_{\rm mass}^{eff} &=&
- ( \tilde{t}^*_L ~ \tilde{t}^*_R ) {\cal M}^2_{\tilde{t}} \left(
\begin{array}{c}
\tilde{t}_L
\\
\tilde{t}_R
\end{array}
\right)
\nonumber
\\
&=&
- ( \tilde{t}^*_L ~ \tilde{t}^*_R )
\left(
\begin{array}{cc}
m_{\tilde{t}L}^2
&m_{\tilde{t}LR}^2
\\
m_{\tilde{t}LR}^{2*}
&m_{\tilde{t}R}^2
\end{array}
\right)
\left(
\begin{array}{c}
\tilde{t}_L
\\
\tilde{t}_R
\end{array}
\right),
\eea
where
\bea
\label{mL}
m_{\tilde{t}L}^2 &=&
M_{\tilde{t}_L}^2 + m_t^2 + m_Z^2
\cos{2\beta}\left(\frac{1}{2}-\frac{2}{3}\sin^2\theta_W\right),
\\
\label{mR}
m_{\tilde{t}R}^2 &=&
M_{\tilde{t}_R}^2 + m_t^2 + m_Z^2
\cos{2\beta}\cdot\frac{2}{3}\sin^2\theta_W,
\\
\label{mLR}
m_{\tilde{t}LR}^2 &=&
m_t \, (A_t^* e^{-i \xi}-\mu \cot\beta).
\eea
The stop mixing angle $\theta_{\tilde t}$ is
\beq
\label{theta t}
\theta_{\tilde t} =
\frac{1}{2} \arctan \left(\frac{2|m_{\tilde{t}LR}^2|}
{m_{\tilde{t}L}^2-m_{\tilde{t}R}^2} \right).
\eeq
The relations between the mass and the weak eigenstates of stops are given by
\bea
\label{mw transf1}
{\tilde{t_1}} &=&
\tilde{t}_L \cos \theta_{\tilde t} +\tilde{t}_R \,e^{-i\beta_{\tilde t}}
\sin \theta_{\tilde t},
\nonumber
\\
\label{mw transf2}
{\tilde{t}_2} &=&
-\tilde{t}_L \, e^{i\beta_{\tilde t}} \sin \theta_{\tilde t}
+\tilde{t}_R \cos\theta_{\tilde t},
\eea
where $\beta_{\tilde t}=-\arg (m_{\tilde{t} LR}^2)$.
The mass eigenvalues of the lighter and heavier stops are
\beq
\label{mt1,2}
m^2_{{{\tilde t}_1},{{\tilde t}_2}}
=
\frac{m_{{\tilde t}L}^2+m_{{\tilde t}R}^2
\mp \sqrt{(m_{{\tilde t}L}^2-m_{{\tilde
t}R}^2)^2+4|m_{{\tilde t}LR}^2|^2}}{2}.
\eeq
Note that $m_{{{\tilde t}_1},{{\tilde t}_2}}^2$ is dependent on the CP
violating phases, $\arg( A_t )$ and $\arg( \mu )$  due to
$m_{{\tilde t}LR}^2$ in Eq.~(\ref{mLR}).

\section{Relevant couplings}
In this sections, we list the couplings relevant to
$\gamma\gamma \rightarrow H_i$ that appear in Table~1.

\begin{itemize}
\item Higgs-fermion-fermion couplings: \beq \label{hff} {\cal
L}_{H\bar{f}f} = - {g m_f \over 2 m_W} \bar{f} \left[ \left( {
v_f^i \over R_\beta^f } \right) - i \gamma_5 \left( {
\bar{R}_\beta^i a_f^i \over R_\beta^f } \right) \right] f H_i ,
\eeq where
\begin{eqnarray}
\label{R}
R_\beta^d & = & \bar{R}_\beta^u = \cos\beta \equiv c_\beta,
~~~~R_\beta^u = \bar{R}_\beta^d = \sin\beta \equiv s_\beta,
\nonumber \\
v_f^d & = & O_{2,i},~~~~v_f^u = O_{3,i},
~~~~a_f^d  =  a_f^u = O_{1,i}.
\end{eqnarray}
Here the matrix $O$ diagonalizes the Higgs mass matrix as in Eq.~(5).
In the presence of Higgs sector CP violation, the Higgs bosons couple with
both CP-even and CP-odd bilinears, $\bar{f} f$ and $\bar{f} \gamma_5 f$,
simultaneously.

\item The Higgs-$W$-$W$ couplings are determined by the gauge couplings:
\beq
{\cal L}_{HW^+ W^-} = g m_W \left( c_\beta O_{2,i} + s_\beta O_{3,i}
\right) H_i W_\mu^+ W^{-\mu} .
\eeq

\item The Higgs-sfermion-sfermion couplings:
\beq
{\cal L}_{H_i \tilde{f}_j \tilde{f}_k} = g^i_{\tilde{f}_j \tilde{f}_k}
\tilde{f}_j^* \tilde{f}_k H_i ,
\eeq
with
\[
 g^i_{\tilde{f}_j \tilde{f}_k} = \tilde{C}^f_{\alpha;\beta\gamma}
O_{\alpha , i} ( U_f )^*_{\beta j} ( U_f )_{\gamma k} .
\]
The matrix $U_f$ diagonalize the sfermion mass matrix:
\[
U_f^{\dagger} M_{\tilde{f}}^2 U_f = {\rm diag}( m_{\tilde{f}_1}^2,
m_{\tilde{f}_2}^2 )
\]
with $m_{\tilde{f}_1} \le m_{\tilde{f}_2}$.
The indices $\alpha$ and $\{\beta, \gamma\}$ label
the three neutral Higgs bosons $(a, \phi_1, \phi_2 )$ and the sfermion
chiralities $\{L, R\}$, respectively. The explicit expressions for
$\tilde{C}^f_{\alpha;\beta\gamma} $ can be found in Ref.~\cite{JSLee}.

\item The $H_i - H^+ - H^-$ couplings are determined by the Higgs
potential.  If we define \beq {\cal L}_{H_i  H^+ H^- } = v C_i H_i
H^+ H^- , \eeq then the couplings $C_i$ are given by \cite{SYChoi}
\[
C_i = \sum_{\alpha=1,2,3} O_{\alpha,i} c_\alpha
\]
with
\begin{eqnarray}
c_1 & = & 2 s_\beta c_\beta {\rm Im}( \lambda_5 e^{2 i \xi} )
- s_\beta^2 {\rm Im}( \lambda_6 e^{i \xi} )
- c_\beta^2 {\rm Im}( \lambda_7 e^{i \xi} ),
\nonumber \\
c_2 & = & 2 s_\beta^2 c_\beta \lambda_1 + c_\beta^3 \lambda_3 - s_\beta^2
c_\beta \lambda_4 - 2 s_\beta^2 c_\beta {\rm Re} ( \lambda_5  e^{2 i \xi} )
\nonumber \\
&&+ s_\beta ( s_\beta^2 - 2 c_\beta^2 ) {\rm Re} ( \lambda_6 e^{i \xi} )
+ s_\beta c_\beta^2 {\rm Re} ( \lambda_7 e^{i \xi} ),
\nonumber \\
c_3 & = & 2 c_\beta^2 s_\beta \lambda_2 + s_\beta^3 \lambda_3 - c_\beta^2
s_\beta \lambda_4 - 2 c_\beta^2 s_\beta {\rm Re} ( \lambda_5  e^{2 i \xi} )
\nonumber \\
&&+ c_\beta  s_\beta^2  {\rm Re} ( \lambda_6 e^{i \xi} )
+ c_\beta ( c_\beta^2 - 2 s_\beta^2 ) {\rm Re} ( \lambda_7 e^{i \xi} ).
\end{eqnarray}

\end{itemize}




\newpage

\newpage

\newpage

\newpage

\newpage

\newpage

\begin{table}
 \caption{The amplitudes $A_i^X$'s and $B_i^X$'s, where $i$ labels three
neutral Higgs bosons, and $X$ labels the species of charged
particles in the triangle loop (with $\tau_{iX} \equiv
M_{H_i}^2/4m_{X}^2$).} \label{table1}
 \begin{tabular}{cc}
$A$'s and $B$'s & Expressions
\\
&
\\
\hline
&
\\
$A_i^f$ & $- 2 ( \sqrt{2} G_F )^{1/2} M_{H_i} N_c e_f^2 \left(
{v_f^i \over R_\beta^f} \right) F_{sf} ( \tau_{if} )$
\\
&
\\
$A_i^{\tilde{f}_j}$ & $ {M_{H_i} N_c e_f^2 g^i_{\tilde{f}_j
\tilde{f}_j } \over 2 m_{\tilde{f}_j}^2} F_0 ( \tau_{i \tilde{f} }
)$
\\
&
\\
$A_i^{W^\pm}$ & $( \sqrt{2} G_F )^{1/2} M_{H_i} \left( c_\beta
O_{2,i} + s_\beta O_{3,i} \right) F_1 ( \tau_{iW} )$
\\
&
\\
$A_i^{H^\pm}$ & ${M_{H_i} v C_i \over 2 m_{H^\pm}^2} F_0 ( \tau_{i
H} )$
\\
&
\\
$A_i^{\tilde{\chi}^\pm_j}$   & $2 {\rm Re}( \kappa_{jj}^i )
{M_{H_i} \over M_{\tilde{\chi}^-_j}} F_{sf} ( \tau_{i
\tilde{\chi}^\pm_j} ) $
\\
&
\\
\hline
&
\\
$B_i^f$  & $  2 ( \sqrt{2} G_F )^{1/2} M_{H_i} N_c e_f^2 \left( {
\overline{R_\beta^f} a_f^i \over R_\beta^f} \right) F_{pf} (
\tau_{if} )$
\\
&
\\
$B_i^{\tilde{\chi}^\pm_j}$ & $- 2 {\rm Im} ( \kappa_{jj}^i )
{M_{H_i} \over M_{\tilde{\chi}^-_j} }F_{pf}( \tau_{i
\tilde{\chi}^\pm_j} )$
\\
 \end{tabular}
 \end{table}

\begin{table}
\caption{Form factor loop functions $F$'s in terms of the scaling function
$f(\tau)$  defined in Eq.~(12).}
\label{table2}
\begin{tabular}{cc}
$F$'s & Definitions
\\
&
\\
\hline
&
\\
$F_{sf} ( \tau )$  & $\tau^{-1} \left[ 1 + ( 1 -  \tau^{-1} ) f ( \tau )
\right]$
\\
&
\\
$F_{pf} ( \tau )$  & $\tau^{-1} f ( \tau )$
\\
&
\\
$F_0 ( \tau )$     & $\tau^{-1} \left[ -1 + \tau^{-1} f ( \tau ) \right]$
\\
&
\\
$F_1 ( \tau )$     & $2 + 3  \tau^{-1} + 3 \tau^{-1} ( 2 - \tau^{-1} )
  f ( \tau )$
\\
\end{tabular}
\end{table}

\end{document}